\documentclass[12pt,letterpaper]{article}
\usepackage[utf8]{inputenc}
\usepackage{color}
\definecolor{coolblack}{rgb}{0.0, 0.18, 0.39}
\definecolor{midnightblue}{rgb}{0.1, 0.1, 0.44}
\definecolor{prussianblue}{rgb}{0.0, 0.19, 0.33}
\definecolor{oxfordblue}{rgb}{0.0, 0.13, 0.28}
\usepackage{hyperref,float}
\hypersetup{
    colorlinks=true,
    linkcolor=blue,
    filecolor=magenta,
    urlcolor=blue,
    citecolor =blue,
}

\usepackage{amsmath}
\usepackage{amsfonts}
\usepackage{amssymb}
\usepackage{graphicx}
\usepackage{lmodern}
\usepackage[round]{natbib}
\usepackage[onehalfspacing]{setspace}
\usepackage{booktabs}
\usepackage{epstopdf}
\usepackage{wrapfig}
\usepackage{cleveref}
\usepackage{textcomp}
\usepackage{pdflscape}
\usepackage{caption}
\usepackage{breqn}
\usepackage{lscape}
\usepackage{pgfplots}
\pgfplotsset{compat=newest}
\usetikzlibrary{plotmarks}
\usepackage{grffile}
\usepackage[left=1.0in,right=1.0in,top=1.0in,bottom=1.0in]{geometry}
\usepackage{adjustbox}
\usepackage{cleveref}
\author{Jozef Barun\'{i}k$^{a,b}$, Mattia Bevilacqua$^{c}$, \& Michael Ellington$^{c}$}
\title{\textbf{Common Firm-level Investor Fears: Evidence from Equity Options}}

\usepackage[toc,page]{appendix}

\newcommand{\spx} {S\&P500}
\newcommand{\vix} {VIX}

\usepackage[normalem]{ulem}
\newcommand\rout{\bgroup\markoverwith{\textcolor{red}{\rule[0.5ex]{2pt}{0.4pt}}}\ULon}

\newcommand{\cf}{\ensuremath{\mathbf{CF} }}
\newcommand{\cfb}{\ensuremath{\mathbf{CF^-}}}
\newcommand{\cfg}{\ensuremath{\mathbf{CF^+}}}

\newcommand{\mf}{\ensuremath{\mathbf{MF} }}
\newcommand{\mfb}{\ensuremath{\mathbf{MF^-}}}
\newcommand{\mfg}{\ensuremath{\mathbf{MF^+}}}

\begin{document}
\begin{titlepage}
\maketitle

\begin{abstract}
\noindent We identify a new type of risk, common firm-level investor fears, from commonalities within the cross-sectional distribution of individual stock options. We define firm-level fears that link with upward price movements as good fears, and those relating to downward price movements as bad fears.  Such information is different to market fears that we extract from index options. Stocks with high sensitivities to common firm-level investor fears earn lower returns, with investors demanding a higher compensation for exposure to common bad fears relative to common good fears. Risk premium estimates for common bad fears range from -5.63\% to -4.92\% per annum.
\end{abstract}

\noindent\textbf{JEL}:
\\
\noindent\textbf{Keywords}:
\\

\begin{spacing}{1}
\noindent\rule{8cm}{0.4pt}\\
\begin{footnotesize}
\noindent $^{\text{a}}$ Institute of Economic Studies, Charles University, Opletalova 26, 110 00, Prague, Czech Republic. \\
$^{\text{b}}$ The Czech Academy of Sciences, IITA, Pod Vodárenskou Věží 4, 182 08, Prague, Czech Republic  \\
$^{\text{c}}$ University of Liverpool Management School, Chatham Building, Chatham Street, L69 7ZH, UK. \\
Email addresses: Jozef Barun\'{i}k, \href{mailto: barunik@fsv.cuni.cz}{barunik@fsv.cuni.cz}, Mattia Bevilacqua \href{mailto:m.bevilacqua@liverpool.ac.uk}{m.bevilacqua@liverpool.ac.uk}, Michael Ellington, \href{mailto:m.ellington@liverpool.ac.uk}{m.ellington@liverpool.ac.uk} \\
\noindent We thank Mykola Babiak for invaluable discussions and comments.  The support from the Czech Science Foundation under the 19-28231X (EXPRO) project is gratefully acknowledged. \\
\noindent \textbf{Disclosure Statement:} Jozef Barun\'{i}k, Mattia Bevilacqua, and Michael Ellington have nothing to disclose. \\
\end{footnotesize}
\end{spacing}

\end{titlepage}


\section{Introduction}\label{sec:Intro}

Option prices contain information regarding uncertainty about future price movements of the underlying asset. Empirical evidence suggests this information is useful for explaining and predicting the cross-section of stock returns \citep[see e.g.][]{bali2009volatility,cremers2010deviations,xing2010does,an2014joint,muravyev2022there}. These studies typically rely on extracting implied volatility, as a proxy for uncertainty for future price movements of the underlying asset, from index options as a gauge for investor fears. Notably, such fears are the expectations of holders of aggregate index funds.

However, relatively little is known about how uncertainty stemming from firm-level options, or as we use interchangeably firm-level investor fears, affects stock returns. Option prices on individual stocks depend on the total volatility of the stock return which incorporates a firm-level component, as well as a market component \citep{campbell2001have}. As such, uncertainty one infers from a cross-sectional distribution of firms carries distinct and more granular information relative to measures of uncertainty about the state of the aggregate stock market, such as the  Chicago Board Options Exchange (CBOE) volatility index (VIX).\footnote{\cite{dew2023cross} present a cross-sectional uncertainty proxy from stock options on individual firms and link it to the business cycle. In fact, in many recent models and empirical work, firm-level uncertainty is the driving force \citep[see e.g.][]{bloom2009impact,gabaix2011granular,acemoglu2012network,herskovic2020firm}.} At the same time, investors fear the prospect of future negative returns, bad fears, more than positive ones, good fears \citep[see e.g.][]{kahneman2013prospect,kilic2019good,bollerslev2020good}. 

This paper is, to the best of our knowledge, the first to study the information content within firm-level good and bad investor fears using a large cross-section of individual equity options for stock returns. Our main contribution is the discovery of a strong common structure in the cross-section of firm-level (good and bad) investor fears that commands a risk premium.  We show, consistent with economic rationale, that stocks with high sensitivities to common firm-level investor fears earn lower returns.  Our results indicate that investors demand a higher compensation for exposure to common bad fears relative to common good fears. The risk premium estimates for common bad fears range from -5.63\% to -4.92\% per annum.

We document a strong factor structure in model-free implied variance measures that we define as firm-level investor fears. We extract implied variance measures from out-of-the-money (OTM) call and put option prices written on a large cross-section of individual stocks in a model-free manner as in \citep{bakshi1997,Bakshi2003}. Then, we decompose these into firm-level good and bad fears in a similar manner to \cite{kilic2019good,barunik2022}. Good fears relate to the prospect of upward price movements that we extract from call options. Bad fears relate to the prospect of downward price movements that we extract from put options. We obtain common factors using principal component analysis (PCA). A single factor we extract from firm-level bad (good) fears explains 75.65\% (83.46\%) of time-variation within the data. 

Such commonalities within firm-level investor fears are inherently different from market fears that we extract from index options.  We use the same model-free approach to compute market fears, and then decompose into good and bad market fears, respectively. The correlation between monthly innovations of common fears and common fears orthogonal to innovations in market fears are in excess of 74\%. We also examine rolling correlations between daily innovations of common fears and market fears.  These statistics reveal low average correlations throughout our sample, with substantially lower values from 2012 onwards. Notably the correlation between common bad fears and bad market fears is always lower than the correlation between common fears and market fears, and common good fears and good market fears, respectively.

Studies do exist on the factor structure in equity options. \cite{engle2015modeling} model the correlation dynamics of implied volatilities and explores the role of the VIX as a common factor in explaining implied volatilities. Their results imply that investors are able to exploit the correlations among implied volatilities for hedging purposes. \cite{christoffersen2018factor} extract factors from equity options for; the short-term level of implied volatilities, the moneyness slope, and the slope of the term-structure. They show that such factors correlate strongly with S\&P500 index options. Both studies use constituents that comprise the Dow Jones Industrial Average (DJIA), and their respective samples end in 2009 and 2010. 

We move beyond this in two ways. First, we use all stocks with available option data. After applying standard filtering \citep[see e.g.][]{carr2011}, we have 526 firms, of which 90\% of are large-cap, with the remaining 10\% being mid-cap stocks. Second, we focus on the pricing implications of innovations to commonalities within firm level investor fears. We exploit the factor structure within firm-level implied variances and extract proxies of aggregate, good and bad common firm-level investor fears.

In contrast to the above, our analysis reveals that co-movement within firm-level investor fears represents a distinct source of information from market-level fears. We emphasize the role of a factor structure present in a large cross-section of option prices that explains the cross-section of stock returns. Our results highlight that investors require compensation for exposure to common bad fears. Differences in firm beta's on common bad fears strongly associate with differences in expected returns. The top common bad fears beta quintile earns average risk-adjusted returns 5.16\% per annum lower than firms in the bottom quintile. We show that risk-adjusted spread portfolio returns maintain statistical and economic significance after controlling for innovations to the VIX, market fears, and an array of firm characteristics in the spirit of \cite{ang2006cross}.

Fama-MacBeth regressions estimate common bad fears risk premiums that range from -5.63\% to -4.92\% per annum, and further substantiates our portfolio sorts. The risk premium estimates of common bad fears from portfolios that control for market fears and other firm characteristics are statistically and economically significant at around -3\% per annum. We show that the \cite{fama2015five} five factors, momentum \citep{jegadeesh1993returns}, common idiosyncratic volatility \citep{herskovic2016common}, liquidity \citep{pastor2003liquidity}, and the variance risk premium \citep{carr2009variance} do not subsume the common bad fears risk premium. These results also hold when considering a battery of anomaly portfolios as test assets. We also show that common firm-level investor fears risk premia are present using alternative factor definitions and the \cite{giglio2021asset} three-pass regression procedure.

Our finding that common good and common bad fears bear different risk premiums resonates well with \cite{farago2018downside} and \cite{bollerslev2020good}. The former develop a model in which there exists a systematic risk factor explicitly relating to bad downside market volatility. They show that a good minus bad realized semivariance measure contains information for the cross-section of stock returns.\footnote{We also consider commonalities among implied variance spreads and show that there is a negative risk premium similar in magnitude to those we report in the main text. This measure aligns more closely with \cite{farago2018downside}. These results are available upon request.} The latter explore the cross-sectional pricing of good and bad realized semivariance measures that relate to positive and negative high frequency price increments. Their results show that firms with high good minus bad realized volatility earn lower returns. A portfolio of assets in the top quintile of good minus bad volatility generating a return 15\% per annum lower than the corresponding bottom quintile portfolio.By contrast, we study the common components within good and bad implied volatilities from individual equity options for stock returns. Our results on common firm-level fears are robust to controlling for various measures of market fears and aggregate volatility.

This paper also relates well with those showing that option prices contain predictive information about stock returns \citep{bali2009volatility,cremers2010deviations,xing2010does}. These studies use deviations between call and put implied volatilities written on index options as proxies for jump risk, or risk neutral skewness. We contribute to these studies by examining the pricing implications of co-movement among implied volatilities of individual equity options. We decompose firm-level implied volatilities into call and put components in order to investigate potential differences in the risk premiums on underlying stock returns. Our asset pricing exercises document a negative and significant risk premium to the common component of firm-level implied volatilities linking to put options, common bad fears. We show that sorting on current loadings to common bad fears generates significant spreads in returns over the next month. Therefore our results are of practical relevance since they reflect an ex ante implementable strategy that one can use to construct hedge portfolios.

Finally, we connect with those studying volatility risk. \cite{ang2006cross} study firm exposures to innovations in the VIX index and show that a negative risk premium is present after controlling for an array of other risk factors and firm characteristics. \cite{cremers2015aggregate} construct volatility and jump risk proxies using data on index option futures and show that volatility and jump risk bear different risk premiums; both of which are negative. \cite{herskovic2016common} considers co-movement in idiosyncratic return volatility and shows that the spread portfolio loading on common idiosyncratic return volatility earns an average return of -5.40\% per annum. 

We investigate the pricing implications stemming from commonalities among ex-ante measures volatility using individual equity options. Our focus is on common firm-level investor fears and the decomposition into good, and bad fears relating to implied volatilities from call and put options, respectively. We show that exposure to common bad fears is where such risk premium is most prevalent. Intuitively this makes sense because investors require compensation for adverse changes to investment opportunities and therefore should accept lower returns in equilibrium for assets that hedge against such changes.

The remainder of this paper proceeds as follows. Section \ref{sec:Measuring} outlines our theoretical background, and how we measure firm-level (good and bad) investor fears. Section \ref{sec:DataCF} describes the data and the common components within investor fears. Sections \ref{sec:Pricing} and \ref{sec:Robust} present our empirical results and robustness analysis, respectively. Finally, Section \ref{sec:Conc} concludes.

\section{Theoretical Background, Investor Beliefs, and Firm-Level Investor Fears}\label{sec:Measuring}

In this Section, we motivate and outline our approach to measuring investor fears, which we proxy from implied variances, using firm-level equity option contracts. First, we motivate our approach by linking it to theory and then discuss investor fears at a general level. Next, we outline how we compute firm-level measures of implied variance and the data we use.  Then, we provide evidence in favour of a common factor within such measures and investigate whether this common factor contains information distinct from what we call market fear; the implied variances that we obtain from S\&P500 index options.

\subsection{Theoretical Background}\label{sec:theory}

Here we outline the economic rationale for why one should expect negative risk prices for exposure to commonalities within firm-level investor fears. We first provide motivation regarding volatility at an aggregate, or market, level. Then, we explain possible mechanisms in the context of existing studies using firm-level volatilities.

Theory provides a several reasons why investors require a premium for exposure to market volatility risk. \cite{merton1973intertemporal} uses an intertemporal capital asset pricing model (ICAPM) to show that investors reduce current consumption to increase precautionary savings in light of uncertainty around market returns. Market volatility hence qualifies as a state variable in traditional multifactor pricing models where risk-averse agents demand stocks to hedge against the risk of deteriorating investment opportunities. This increases the prices of these assets and lowers expected returns. \cite{campbell2018intertemporal} extends the ICAPM framework of \cite{campbell1993intertemporal} by allowing for stochastic volatility. They confirm that returns that positively covary with a variable forecasting future market volatility have low expected returns in equilibrium \citep[][also documents this finding]{campbell1996understanding}.

\cite{farago2018downside} deduce an equilibrium asset pricing model with generalized disappointment aversion that provides an explanation for why such investors price downside volatility more dearly than upside volatility. Specifically, these investors care more about downside losses than upside gains. In doing so, they assign larger weights to outcomes that realize less than the investor's certainty equivalent. This framework yields a systematic risk factor that prices downside market volatility risk, which they confirm empirically.

Regarding commonalities within firm-level volatilities, \cite{herskovic2016common} develop an incomplete markets model with a common idiosyncratic volatility factor driving dispersion in household income growth and also residual stock return volatility. Investors require compensation for changes in current and future cross-sectional consumption growth distribution. Heterogeneous exposures to shocks in common idiosyncratic volatility are the sole driver for differences in the risk premium across stocks. Positive shocks to common idiosyncratic volatility cause loadings to increase thereby lowering expected returns in equilibrium.

\cite{martin2019expected} develop a formula for individual stock returns using risk-neutral variances. They express expected stock returns in terms of the risk-neutral market variance, the risk-neutral variance of the stock, and the value-weighted average of individual stock risk-neutral variances. Although the model has no free parameters, it states a negative relationship between the expected return of the stock and the weighted average of risk-neutral individual stock variances. This term can be thought of as comparable to the common component we extract from firm-level fears and helps provide further rationale for negative risk prices.

\subsection{Investor Beliefs}

Market participants face uncertainty regarding future price movements. The VIX index is a popular gauge of investor fear which tracks market expectations of short-term future price uncertainty using the implied volatilities of S\&P500 equity options. However, following \cite{kahneman2013prospect}'s prospect theory investors care differently about upside gains and downside losses. Those who face downside risks require a relative downside risk premium \citep{ang2006downside} and those who face upside gains may be willing to pay for such an outcome \citep{breckenfelder2012asymmetry}. \cite{kilic2019good}, \cite{barunik2022} and others provide insight on how one can decompose implied volatility (fear) into good components linking to the prospect of upward price movements using call options (good fear), and bad components linking to the prospect of downward price movements using put options (bad fear).

We think of fears as a function of outcomes. Naturally, it is important to be able to measure expectations pertinent to positive and negative outcomes. We use the terms `good fear'' and ``bad fear'' to refer to these two complementary situations. Beliefs linking to the prospect of  good (bad) states - good (bad) fears - reflect the situation where an investor fears uncertainty about price fluctuations, but the uncertainty itself associates with a positive (negative) outcome. The decomposition of implied volatilities into good and bad components represent proxies for good and bad fears. As we discuss above, most of the literature establishes well investor fears, as well as good and bad investor fears, from a market perspective by using information from index options.

Focusing on firm-level investor fears is important for a number of reasons. First, many investors have substantial holdings of individual stocks. Such investors may fail to diversify in an optimal manner or are subject to restrictions by corporate compensation policies, and thus face considerable exposure to firm-level uncertainty regarding future price movements \citep{campbell2001have}. Second, the ability to diversify portfolios containing a subset of large stocks depends on the level of firm-level uncertainty around future price movements of stocks that comprise these portfolios. Third traders who seek to exploit mispricing, or hedge adverse outcomes for stocks within their portfolios, using options face firm-level uncertainty regarding future price movements to a larger degree than they do uncertainty around market movements.

\subsection{Measuring Firm--Level Investor Fears} \label{sec:MFIV}

We use the methods in \cite{bakshi2000spanning} and \cite{Bakshi2003} to extract variance measures from the cross-section of option prices in a model-free manner. We consider the price of a variance contract that pays the squared logarithm of the return at time $t + 1$, which in our case corresponds to a fixed horizon of the next 30 days. Let $s_{i,t}$ denote the natural logarithm of the price $S_{i,t}$ of the $i$th asset at time $t$. The payoff of the variance contract is $r_{i,t+1}^2 = (s_{i,t+1} - s_{i,t})^{2}$ and we define the total implied variance, $\sigma^2_{i,t}$, as the price of the contract:
\begin{equation}\label{eq:MFIV}
\sigma^2_{i,t} \equiv e^{-r_t^f} \mathbb{E}_t^Q \left[r_{i,t+1}^2\right]
\end{equation}
where $\mathbb{E}_t^Q$ is the expectation operator under the risk-neutral measure conditional on time $t$ information and $r_t^f$ is the risk-free rate. \cite{kilic2019good} and \cite{barunik2022} show one can decompose equation (\ref{eq:MFIV}) into two components that relate to the positive and negative returns of the variance contract, respectively. In the absence of arbitrage, the sum of these components is the total implied variance. One obtains the prices of these components from OTM call and put options.

Implied variance measures the expectations of fluctuations to the underlying asset over a given horizon. This reflects investors' fears, which directly relate to uncertainty about the future price movements. Furthermore, \cite{bakshi2000spanning} and \cite{Bakshi2003} show that one can compute $\sigma^2_{i,t}$ from the prices OTM call and put options:
\begin{equation}
\label{eq:iv}
\sigma^2_{i,t} = \underbrace{\int_{S_{i,t}}^{\infty} \frac{2(1-\log(K/S_{i,t}))}{K^2}C(t,t+1,K) dK}_{\sigma^{2,+}_{i,t} }  + \underbrace{\int_{0}^{S_{i,t}} \frac{2(1+\log(S_{i,t}/K))}{K^2}P(t,t+1,K) d K}_{\sigma^{2,-}_{i,t}  },
\end{equation}
where $C(\:\cdot\:)$ and $P(\:\cdot\:)$ denote the prices at time $t$ of a call and put contract with a time to expiration of one period and a strike price of $K$. Call option prices reflect a good state for the stock, while the prices of a put option reflect a bad state for the stock. The two states, most of the time, relate to contrasting investors' beliefs and future expectations \cite{buraschi2006}. OTM puts usually link with hedging and insurance against equity market drops \citep{Han2008}, meanwhile OTM calls more commonly associate with optimistic beliefs \citep{buraschi2006}.

Corresponding to an intuitive measure of expectations of good and bad events for the stock, the payoff from the variance contract can be written as in \cite{kilic2019good} and \cite{barunik2022}:
\begin{equation}\label{eq:MFIV_decomp}
\sigma^2_{i,t} \equiv \underbrace{e^{-r_t^f} \mathbb{E}_t^Q \left[r_{i,t+1}^2 \mathbb{I}_{\{r_{i,t+1}>0\}}\right]}_{\sigma^{2,+}_{i,t}} + \underbrace{e^{-r_t^f} \mathbb{E}_t^Q \left[r_{i,t+1}^2 \mathbb{I}_{\{r_{i,t+1}\le 0\}}\right]}_{\sigma^{2,-}_{i,t}}
\end{equation}
Intuitively, good and bad components of the payoff add to the total, and we can obtain the prices of its components in a model-free manner from a bundle of option prices upon a discretization of equation (\ref{eq:iv}); the Appendix provides details of the procedure we use. The total implied variance is the weighted sum of the option prices, and its components are identifiable by claims that have payoffs relating to the sign of the realized return. Good implied variance is identifiable from call options that pay off when we realize a positive return, and bad implied variance is identifiable through put options paying off upon the realization of a negative return.

Consequently, the first term in Equation (\ref{eq:iv}) refers to positive components and the second term refers to negative components of the payoff of the volatility contract, where $\sigma^{2,+}_{i,t}$ is the good implied variance and $\sigma^{2,-}_{i,t}$ is the bad implied variance. We identify good implied variance by call options that pay off if the return realisation is positive, and bad implied variance by put options that pay off only if the return realisation is negative.

We define $\sigma^{2}_{i,t} $ in Equation (\ref{eq:iv}) as firm-level fears that proxy investors' expectations of future price movements. We refer to $\sigma^{2,+}_{i,t}$ in Equation (\ref{eq:iv}) as firm-level good fears, and $\sigma^{2,-}_{i,t} $ in Equation (\ref{eq:iv}) as firm-level bad fears. Firm-level good fears capture investors' expectations of future upward price movements. Firm-level bad fears capture investors' expectations of future downward price movements.

Importantly, we distinguish whether the information content within firm-level investor fears differs from investor market fears. To do this, we infer investor market fears from index options using the same approach as described here. The only difference is that we replace firm-level options with \spx\:index options. Throughout this paper we use $\mf$, $\mfg$, $\mfb$ to define market fears, good market fears and bad market fears.

\section{Common Firm-level Investor Fears} \label{sec:DataCF}

\subsection{Firm-level Option Data}  \label{Data}

We compute firm-level implied variances using daily data from OptionMetrics over the sample January 03, 2000 to December 31, 2020\footnote{This period allows us to have a good data coverage which was insufficient to compute implied variances prior to January 2000.}. We include all stocks from the time of their IPO and listing, with good options data coverage (i.e. we require stocks to have data spanning more than 5 years of continuous data). We exclude stocks due to: i) bankruptcy; ii) delisting; and iii) mergers and acquisitions.\footnote{Examples of bankruptcies are General Motors, Lehman Brothers and Merrill Lynch; examples of M\&As are Raytheon and United Technologies, Dow Chemical and DuPont, and Walt Disney Company and 21st Century Fox.}

We apply common options filtering rules to further exclude stock options with: i) missing deltas; ii) missing implied volatility; iii) bid prices equal to 0; iv) nil volume; v) nil open interest; vi) negative bid-ask spread; and that vii) violate arbitrage conditions \citep[see, e.g.][]{Bakshi2003,carr2011,christoffersen2012}. Following these filtering criteria, we then remove options with less than 4 contracts on a specific day and are left with 526 firms.\footnote{Most of these data filters are common in the option pricing literature. The volume and open interest constraints ensure that there is genuine interest in the option contract. Options that are close to maturity are removed \citep[see, e.g.][among others]{carr2011,christoffersen2012}. We remove options with a negative bid-ask spread and that violate no-arbitrage constraints, as these option prices are invalid and inconsistent with theory. Finally, we remove ITM contracts, as they tend to be more illiquid than OTM and at-the money options \citep[e.g.][]{christoffersen2012}.} Approximately 90\% of these firms are large-cap; with the remaining 10\% being mid-cap stocks. Most stocks in our sample appear as a constituent of the S\&P500 throughout our sample. Other stocks come from the Russell 1000 for which there is sufficient data coverage. To proxy market investor fears, we use the same filtering criteria for data on \spx index options.

Specifically, each day $t$, our data sample contains daily stock options observations for which we are able to calculate values of the \cite{Bakshi2003} implied variance (and semi-variance) measures. We consider call and put option prices with maturity around 30 days, considering all available strikes for each option. We keep implied variance measures within 23 and 37 days maturity to represent a proxy of investor expectations of the one-month ahead, $t + 1$, fluctuations in the underlying asset.

\begin{table}[!hp]
  \centering
  \caption{\textbf{Descriptive Statistics of Firm-level Implied Variances}\\
  \small{Notes: This table shows the time-series averages of daily cross-sectional means (Mean) and standard deviations (Std) for firm-level implied variance measures.  $ \sigma^2_{i,t}$ refers to total implied variance, while $ \sigma^{2,+}_{i,t} $, $ \sigma^{2,-}_{i,t} $ refer to good and bad implied variances respectively. }}
    \begin{tabular}{lrrr}
    \toprule
    \midrule
    \midrule
          &    $ \sigma^2_{i,t}$   &   $ \sigma^{2,+}_{i,t} $   & $ \sigma^{2,-}_{i,t} $ \\
    \midrule
    Mean  & 0.187 & 0.067 & 0.120 \\
    Std   & 0.182 & 0.055 & 0.129 \\
    Ave. Pairwise covariance & 0.028 & 0.002 & 0.015 \\
     \midrule
    \midrule
    \bottomrule
    \end{tabular}%
  \label{tab:IV_des}%
\end{table}%

Table \ref{tab:IV_des} reports the time-series averages for the cross-sectional means and standard deviations for each of our implied variance measures, as well as the average pairwise covariances. We can see that bad implied variance has a higher time-series average for the cross sectional mean and standard deviation relative to good implied variance, and across all measures, there is a positive average pairwise covariance. Table \ref{tab:IV_des} indicates there is positive co-movement in firm-level investor fears.

\subsection{The Common Factor in Firm-level Investor Fears}

We use principal components analysis (PCA) to estimate the common factor in our implied variance measures. The common factors we extract from firm-level total implied variance, \cf, we refer to as common firm-level investor fears, or common fears for brevity. The common factors we extract from good implied variances, \cfg, we refer to as  good common firm-level investor fears, or common good fears, and those from bad implied variances, \cfb,  we refer to as bad common firm-level investor fears, or common bad fears.  We adopt the expectation-maximization algorithm in \cite{stock2002forecasting,stock2002macroeconomic} and \cite{mccracken2016fred}, using a 252 day window that rolls through our sample in order to eliminate any look ahead bias for our pricing exercise.

\begin{figure}[!hp]
		\centering
		\scalebox{0.85}{\includegraphics{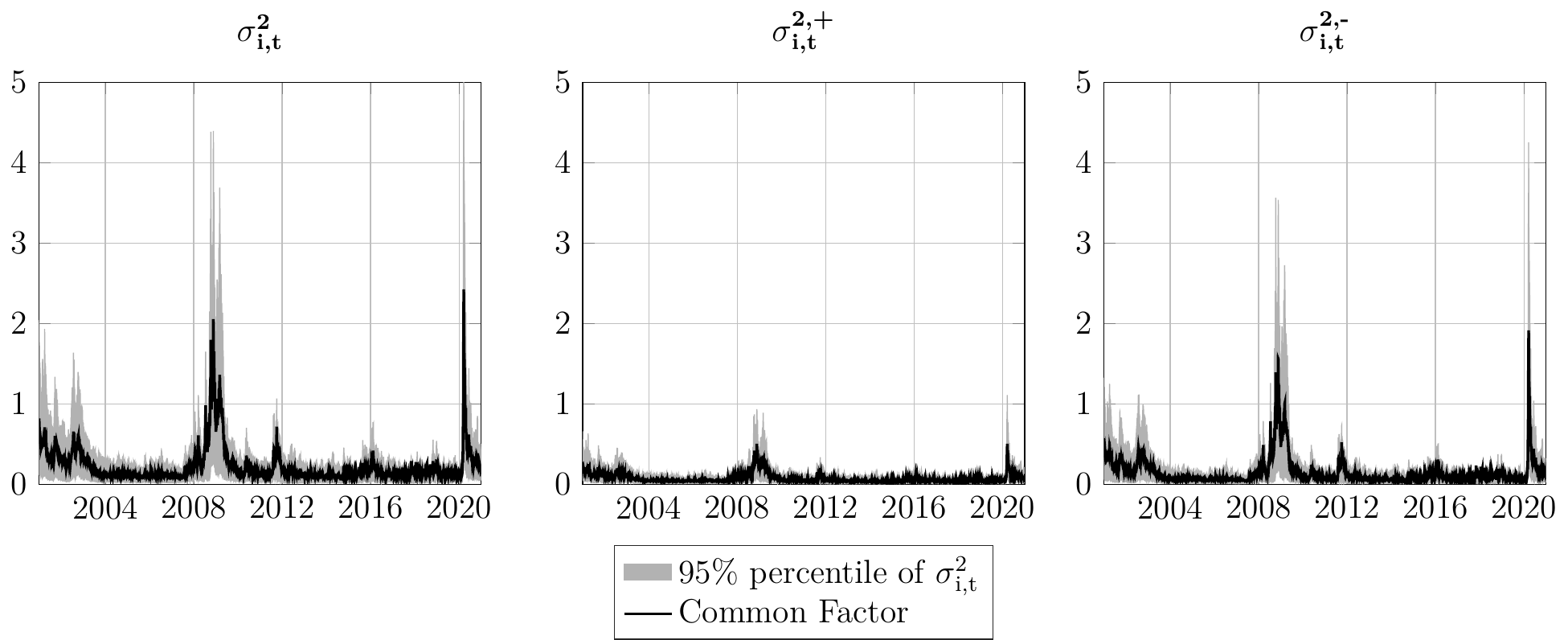}}\\
		\caption{\textbf{Firm-level Implied Variances}\\ \small{Notes: This figure plots the daily 5\% and 95\% percentiles of the cross-sectional distribution for firm-level implied variances from January 03, 2000 to December 31, 2020. The left hand side plot reports firm-level implied variance, the middle plot reports firm-level good implied variances (those we infer from call options), the right hand side plot shows firm-level bad implied variances (those we infer from put options). The solid line is the common factor we extract as the first principal component.}}
      \label{fig:FLIV}
\end{figure}

Figure \ref{fig:FLIV} shows the 5\% and 95\% quantiles of the daily cross-sectional distributions of our firm-level implied variance measures and the common factors we extract from firm-level implied variance measures using PCA. The left hand side plot shows firm-level implied variance, and the middle and right hand side plots show firm-level good and bad implied variances respectively. This shows that firm-level implied variances obey a strong factor structure.

Table \ref{tab:F_dstat} reports descriptive statistics concerning the proportion of variation the common factor explains for our implied variance measures.  We also report the proportion of variation the common factors explain for the full sample January 03, 2000 -- December 30, 2020. On average,  common fears explains 87.1\% of the variation in firm-level implied variances; with common good fears and common bad fears explaining 90.5\% and 83.9\% of variation within firm-level good and bad implied variances respectively. The proportions of variance explained by these common factors range from 69.7\% to 91\%.  The standard deviation of explained variations are between 3.33\% and 6.25\%. Comparing the averages and medians in Panel A to full sample estimates in Panel B shows the rolling sample estimates, pertinent to our pricing exercise, are explain a higher proportion of variation when looking at our sample through a rolling window. This provides substantial evidence in favour of a common factor driving firm level investor fears.

\begin{table}[!htbp]
  \centering
  \caption{\textbf{Proportion of Variation Explained by Common Factor from Firm-level Implied Variances} \\
  \small{Notes: This table shows the percentage of variation that the common factor we extract from firm-level implied variances explains in the dataset. Panel A shows the time-series average, median, minimum (Min), maximum (Max), and standard deviation (Std) for the common factors we extract using a 252-day window rolling through the sample January 03,2000 to December 31, 2020 daily.  Panel B shows the percentage variation the common factor explains using the full sample January 03, 2000 to December 31, 2020.}}
    \begin{tabular}{lrrr}
    \toprule
    \midrule
    \midrule
    \textbf{A:} Rolling Sample & \multicolumn{1}{l}{\cf} & \multicolumn{1}{l}{\cfg} & \multicolumn{1}{l}{\cfb} \\
    \midrule
    Mean (\%)  & 87.13 & 90.51 & 83.94 \\
    Median (\%) & 88.75 & 90.98 & 85.68 \\
    Min  (\%)  & 75.56 & 82.04 & 69.68 \\
    Max  (\%)  & 95.06 & 95.51 & 93.89 \\
    Std  (\%)  & 4.86 & 3.33 & 6.25 \\
    \midrule
    \midrule
    \textbf{B:} Full Sample & \multicolumn{1}{l}{\cf} & \multicolumn{1}{l}{\cfg} & \multicolumn{1}{l}{\cfb} \\
    \midrule
    \% variation & 77.75& 83.46 & 75.65 \\
     \midrule
    \midrule
    \bottomrule
    \end{tabular}%
  \label{tab:F_dstat}%
\end{table}%

\subsection{Is the Information within Common Firm-level Investor Fears different to Investor Market Fears?}

\begin{figure}[!hp]
		\centering
		\scalebox{0.8}{\includegraphics{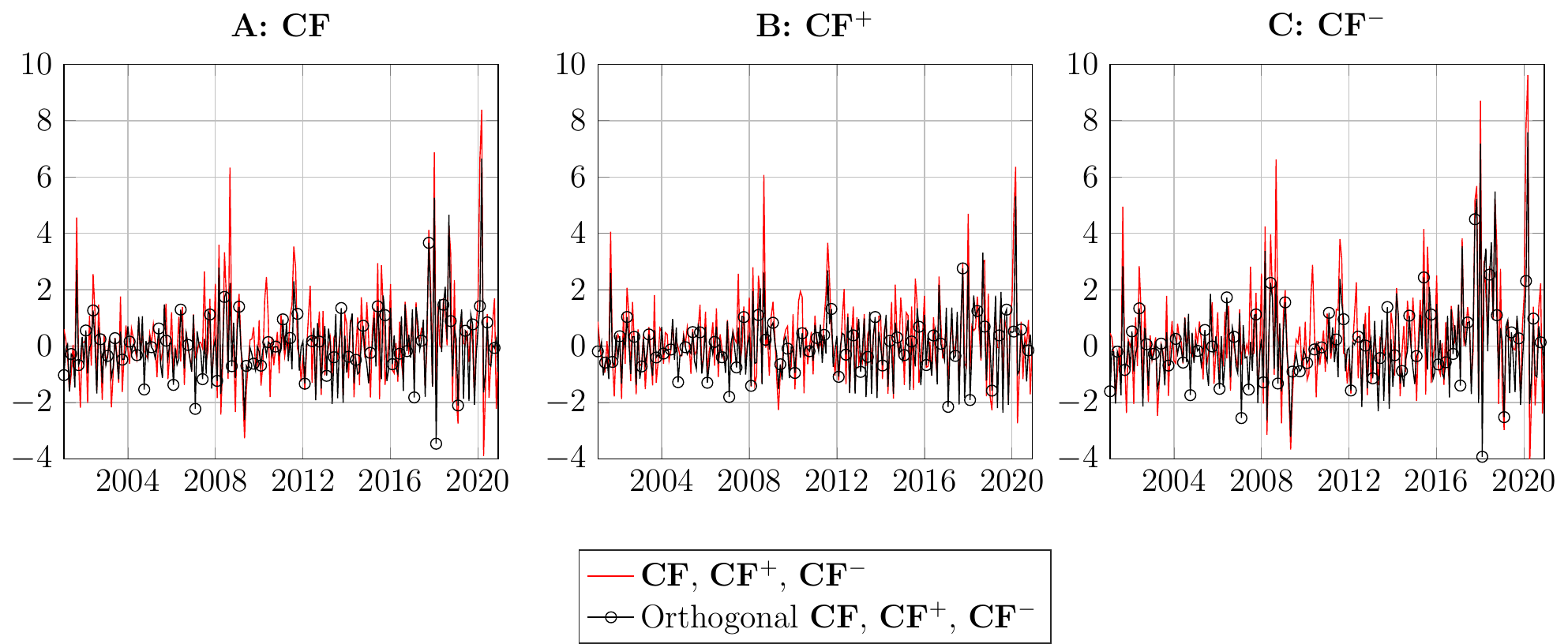}}\\
		\caption{\textbf{Monthly Innovations to Common Factors of Firm-level Implied Variances}\\ \small{Notes: This figure plots innovations to Common Fears (solid line) and Orthogonal Common Fears (dot markers). Common Fears are the common factors we extract from firm-level implied variances. Orthogonal Common Fears are the common factor we extract from firm-level implied variances that are orthogonal to market fears. Market fears are analogous to firm-level fears in that they are implied variances (total, good, and bad) we infer from \spx\ index options. Orthogonal Common Fears are the residuals from a regression with Common Fears as the dependent variable and Market Fears as the independent variable. Panel A shows these for Common Fears. Panels B and C show these for Common Good Fears and Common Bad Fears, respectively.}}
      \label{fig:CFIVS}
\end{figure}

Here we investigate whether common firm-level fears contain different information to market fears. Figure \ref{fig:CFIVS} shows monthly innovations to the common factors we extract from firm-level implied variances using the entire sample. Panel A shows innovations to common fears, and Panels B and C show innovations to common good fears and common bad fears respectively. Alongside these plots, we also report the innovations orthogonal to those from the corresponding market fears measures we compute from index options. Orthogonal innovations are the residuals we compute by regressing common fears on market fears. The correlations between the plots in Panel A is 74\%, and in Panels B and C the correlations are 76\% and 79\% respectively.

\begin{figure}[!hp]
		\centering
		\scalebox{0.9}{\includegraphics{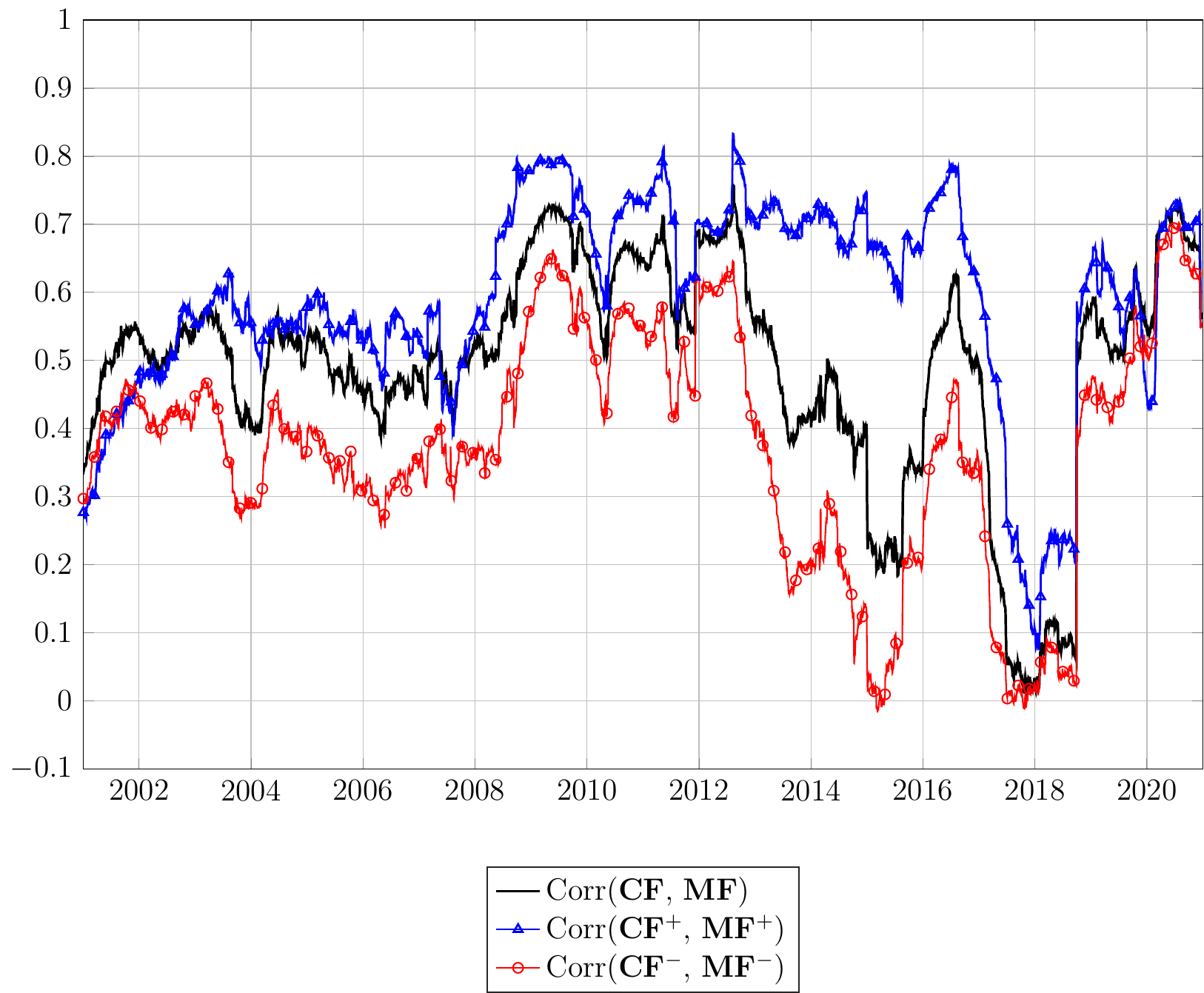}}\\
		\caption{\textbf{Correlations between Common Fears and Market Fears}\\ \small{Notes: This figure plots the daily correlations between Common Fears and Market Fears from January 03, 2001 to December 31, 2020. The former is the common factor in firm-level implied variances and the latter is the implied variance from \spx index options. We compute correlations using a rolling 1-year window. The solid line shows the correlations between Common Fears and Market Fears (total implied-variances), the line with triangle markers corresponds to the correlations between Common Good Fears and Good Market Fears, the line with o markers corresponds to the correlations between Common Bad Fears and Bad Market Fears.}}
      \label{fig:Corrs}
\end{figure}

To further explore differences, we plot the correlations between daily innovations to common firm-level fears and the corresponding market fears using a 1-year rolling window in Figure \ref{fig:Corrs}.  There are three takeaway points from these plots. First, correlations vary substantially throughout the sample.  These range from highs of around 0.8 for common good fears and good market fears, to lows of -0.05 for common bad fears and bad market fears. Second, correlations  between common firm-level fears and market fears appears to rise during periods of financial and economic turbulence, as well as bear markets.  Notably correlations are substantially lower from 2012 onwards. Finally, the correlation between innovations to common bad fears and bad market fears is almost always lower than common fears and market fears, and common good fears and good market fears.

Overall, this implies that shocks to common fears in the firm-level domain are distinct to their corresponding fears at the market index level.  In particular, innovations to common bad fears are notably different to bad market fears. We now move on to pricing exercises to understand the compensation investors demand for bearing such exposure.

\newpage
\section{Pricing Common Fears}\label{sec:Pricing}

To examine the pricing implications for our measures of common fears, we obtain data on stock prices from the Center for Research in Security Prices (CRSP).  We take all common stocks listed on the New York Stock Exchange (NYSE), NASDAQ and AMEX. To construct our sample, we adopt the following filtering criteria. First,  we omit stocks in month $t$+1 if their market capitalization at the end of month $t$ is in the bottom 30\% percentile of the cross-sectional distribution at the end of month $t$. Next, we omit stocks in month $t$+1 whose price at the end of month $t$ is less than \$5. Then, we winsorize by removing stocks in month $t$+1 if their returns lie in the top and bottom 5\% percentiles of returns at the end of month $t$. As we explain below, our pricing exercise uses rolling regressions using 1-year of data. Therefore we require stocks to have this data available to estimate loadings. This means that the average number of stocks in our sample is 1045 with a minimum of 883 and a maximum of 1385.\footnote{We assess the robustness of our findings to these filtering criteria by relaxing the stringencies.  In this case, we omit stocks in month $t$+1 if their market capitalization at the end of month $t$ is in the bottom 20\% percentile of the cross-sectional distribution at the end of month $t$. Next, we omit stocks in month $t$+1 whose price at the end of month $t$ is less than \$1. Then, we winsorize by removing stocks in month $t$+1 if their returns lie in the top and bottom 5\% percentiles of returns at the end of month $t$. the average number of stocks in this sample is 1360, the minimum and maximum number of stocks are 1118 and 1937, respectively.  These results are qualitatively similar, to those we report in the main text and are available upon request.} 

Our pricing exercises are similar to \cite{ang2006cross}, \cite{cremers2015aggregate}, and \cite{herskovic2016common} who follow standard approaches within the asset pricing literature; namely portfolio sorts and Fama-MacBeth regressions.  For portfolio sorts, we obtain factor loadings from daily data using a 1-year rolling window from regressions of the $i$th stock's excess return

\begin{eqnarray}
r_{i,t} &=& \beta^{0}_{i} + \beta^{\Delta \cf}_{i}\Delta \cf_{t} + \beta^{\Delta \text{VIX}}_{i} \Delta\text{VIX}_{t}+ \epsilon_{i,t} \label{eq:load1} \\
r_{i,t} &=& \beta^{0}_{i} + \beta^{\Delta \cfg}_{i}\Delta \cfg_{t} + \beta^{\Delta \text{VIX}}_{i} \Delta\text{VIX}_{t}+ \epsilon_{i,t} \label{eq:load2} \\
r_{i,t} &=& \beta^{0}_{i} + \beta^{\Delta \cfb}_{i}\Delta \cfb_{t} + \beta^{\Delta \text{VIX}}_{i} \Delta\text{VIX}_{t}+ \epsilon_{i,t} \label{eq:load3}
\end{eqnarray}

\noindent in which $\Delta \cf_{t}  $, $\Delta \cfg_{t}  $ , and $ \Delta \cfb_{t} $, are the respective innovations to common fears, common good fears and common bad fears; $ \Delta\text{VIX}_{t} $ are innovations to the VIX index. We also consider replacing $\Delta\text{VIX}_{t} $ with market fears, $\Delta \mf,\: \Delta \mfg,\: \Delta \mfb $, that we infer from S\&P500 index options  in a similar manner to the firm-level implied variances \citep[similar to][]{herskovic2016common}.\footnote{We also consider replacing $\Delta\text{VIX}_{t}$ with the market risk premium, market capitalization, and trading volume in the spirit of \cite{ang2006cross,cremers2015aggregate}. We report these results in the Appendix.}

Using these loadings we consider portfolio sorts for value-weighted portfolios. For single portfolio sorts, at the end of month $t$, we sort stocks into quintile portfolios from their respective loadings on $\beta^{\Delta \cf}_{i}$,  $\beta^{\Delta \cfg}_{i}$, $\beta^{\Delta \cfb}_{i}$, compute returns over the next month, $t+1$, and repeat this process.  For sorts that control for firm characteristics, we first sort stocks into quintiles using the characteristic of interest, and then on one of our common fears risk proxies, $\beta^{\Delta \cf}_{i}$,  $\beta^{\Delta \cfg}_{i}$, $\beta^{\Delta \cfb}_{i}$ \citep[see e.g.][]{ang2006cross,herskovic2016common}; this neutralizes the portfolios from the characteristic of interest. For double sorts, we first sort stocks into quintiles on the characteristic of interest, and then on one of our common fears risk proxies, $\beta^{\Delta \cf}_{i}$,  $\beta^{\Delta \cfg}_{i}$, $\beta^{\Delta \cfb}_{i}$.

Along with the excess expected returns for quintile portfolios, we also report risk-adjusted returns which are the alphas from the Fama-French 5 factor model,  $\alpha^{\text{FF5}}$, and the alphas from the Fama-French 5-factor model that accounts for the momentum factor.\footnote{These data, along with additional test assets are from Kenneth French's Data Library.} Reporting risk-adjusted returns allows us to control for other factors known to affect stock returns.


Table \ref{tab:single_sorts} shows results from single portfolio sorts using loadings on common fears in Panel A, common good fears in Panel B, and common bad fears in Panel C.  We report the expected excess return and risk-adjusted returns relative to the respective Fama-French 5 factor model and the Fama-French 5-factor model plus the momentum factor, for quintile portfolios,  and the long--short portfolio that buys the portfolio with high loadings (portfolio 5) and sells the portfolio with low loadings (portfolio 1).

In general, and consistent with economic rationale, excess and risk-adjusted returns are almost monotonically decreasing. However, it is only when looking at portfolio sorts on common bad fears that the spread portfolio excess and risk-adjusted returns are statistically significant. The expected excess return is -0.53\% per month which corresponds to an annualized return of -6.36\%. Looking at alphas from the multi-factor pricing models, this portfolio earns an economically meaningful annualized return of -5.04\% from the Fama-French 5 factor model, and -5.15\% when adding the momentum factor.

\begin{table}[!hp]
  \centering
  \caption{\textbf{Single Portfolio Sorts from Loadings on Common Fears, Common Good Fears, and Common Bad Fears}\\
  \small{Notes: This table shows value-weighted portfolios that we sort on loadings to: i) common fears in Panel A; ii) common good fears in Panel B; and iii) common bad fears in Panel C.  In each Panel, we report monthly excess returns for quintile portfolios and the long-short portfolio that goes long the portfolio of stocks with a high loading on common fears and short the portfolio of stocks with low loadings to common fears. We also report the risk adjusted returns from the Fama-French 5 factor model and the Fama-French 5-factor model accounting for Momentum.}}
    \begin{tabular}{lrrrrrr}
    \toprule
    \midrule
    \midrule
       & \multicolumn{1}{l}{low $\beta^{\Delta\cf}$} &       &       &       \multicolumn{2}{r}{High $\beta^{\Delta\cf}$} &  \\
    \midrule
    \textbf{A:} Common Fears  & 1     & 2     & 3     & 4     & 5     & 5-1 \\
    \midrule
    mean (\%) & 1.14  & 1.11  & 0.90  & 0.96  & 0.79  & -0.35 \\
    $t$-stat & 3.29  & 3.61  & 3.29  & 3.70  & 3.44  & -1.74 \\
    $\alpha^{\text{FF5}}$  & 0.32  & 0.41  & 0.25  & 0.37  & 0.19  & -0.13 \\
    $t$-stat & 4.42  & 2.93  & 2.41  & 2.66  & 1.74  & -0.90 \\
    $\alpha^{\text{FF5 + MOM}}$ & 0.32  & 0.40  & 0.24  & 0.36  & 0.18  & -0.15 \\
    $t$-stat & 4.38  & 2.77  & 2.40  & 2.58  & 1.74  & -1.02 \\
    \midrule
    \midrule
    \textbf{B:} Common Good Fears & 1     & 2     & 3     & 4     & 5     & 5-1 \\
    \midrule
    mean (\%) & 1.13  & 0.99  & 1.20  & 0.79  & 0.76  & -0.37 \\
    $t$-stat & 3.16  & 3.11  & 4.27  & 3.01  & 3.17  & -1.53 \\
    $\alpha^{\text{FF5}}$  & 0.35  & 0.23  & 0.56  & 0.27  & 0.14  & -0.21 \\
    $t$-stat & 2.91  & 3.71  & 3.87  & 1.99  & 0.87  & -0.92 \\
    $\alpha^{\text{FF5 + MOM}}$  & 0.35  & 0.23  & 0.55  & 0.27  & 0.12  & -0.23 \\
    $t$-stat & 3.08  & 3.80  & 3.71  & 1.92  & 0.84  & -1.09 \\
    \midrule
    \midrule
    \textbf{C:} Common Bad Fears & 1     & 2     & 3     & 4     & 5     & 5-1 \\
    \midrule
    mean (\%) & 1.28  & 0.92  & 1.00  & 0.91  & 0.76  & \textbf{-0.53} \\
    $t$-stat & 3.87  & 3.01  & 3.79  & 3.47  & 3.13  & \textbf{-2.82} \\
    $\alpha^{\text{FF5}}$  & 0.55  & 0.16  & 0.33  & 0.36  & 0.13  & \textbf{-0.42} \\
    $t$-stat & 3.29  & 1.63  & 3.24  & 2.47  & 1.59  & \textbf{-2.32} \\
    $\alpha^{\text{FF5 + MOM}}$ & 0.54  & 0.16  & 0.32  & 0.35  & 0.12  & \textbf{-0.43} \\
    $t$-stat & 3.22  & 1.54  & 3.24  & 2.42  & 1.48  & \textbf{-2.46} \\
    \midrule
    \midrule
    \bottomrule
    \end{tabular}%
  \label{tab:single_sorts}%
\end{table}%

Table \ref{tab:single_sortsVIX} analogous results to those in Table \ref{tab:single_sorts} that control for innovations to the VIX index, $\Delta$VIX. We can see that excess returns monotonically decrease as loadings to common fears increase, and risk-adjusted returns are almost monotonically decreasing. A similar story emerges here in that the spread portfolios are statistically significant for common bad fears. The economic significance for risk-adjusted returns for the spread portfolio sorting on common bad fears reduces to -0.21\% per month (-2.52\% annualized) and -0.23\% (-2.76\% annualized) relative to the Fama-Fench 5 factor model and Fama-French 5 factor model plus momentum respectively.

\begin{table}[!hp]
  \centering
  \caption{\textbf{Portfolio Sorts from Loadings on Common Fears, Common Good Fears, and Common Bad Fears Controlling for $\Delta$VIX}\\
  \small{Notes: This table shows value-weighted portfolios that we sort on loadings to: i) common fears in Panel A; ii) common good fears in Panel B; and iii) common bad fears in Panel C whilst controlling for innovations to the VIX index ($\Delta$\vix).  In each Panel, we report monthly excess returns for quintile portfolios and the long-short portfolio that goes long the portfolio of stocks with a high loading on common fears and short the portfolio of stocks with low loadings to common fears. We also report the risk adjusted returns from the Fama-French 5 factor model and the Fama-French 5-factor model accounting for Momentum.}}
    \begin{tabular}{lrrrrrr}
    \toprule
    \midrule
    \midrule
       & \multicolumn{1}{l}{low $\beta^{\Delta\cf}$} &       &       &       \multicolumn{2}{r}{High $\beta^{\Delta\cf}$} &  \\
    \midrule
    \textbf{A:} Common Fears  & 1     & 2     & 3     & 4     & 5     & 5-1 \\
    \midrule
    mean (\%) & 1.25  & 1.06  & 0.96  & 0.92  & 0.79  & \textbf{-0.46} \\
    $t$-stat & 3.64  & 3.46  & 3.55  & 3.75  & 3.51  & \textbf{-2.62} \\
    $\alpha^{\text{FF5}}$  & 0.42  & 0.29  & 0.24  & 0.26  & 0.18  & -0.23 \\
    $t$-stat & 4.18  & 4.53  & 3.98  & 4.99  & 2.42  & -1.57 \\
    $\alpha^{\text{FF5 + MOM}}$ & 0.42  & 0.29  & 0.24  & 0.25  & 0.17  & \textbf{-0.25} \\
    $t$-stat & 4.49  & 4.46  & 3.83  & 5.17  & 2.74  & \textbf{-2.00} \\
    \midrule
    \midrule
    \textbf{B:} Common Good Fears & 1     & 2     & 3     & 4     & 5     & 5-1 \\
    \midrule
    mean (\%) & 1.16  & 1.04  & 1.05  & 0.93  & 0.79  & -0.38 \\
    $t$-stat & 3.26  & 3.34  & 3.82  & 3.70  & 3.70  & -1.74 \\
    $\alpha^{\text{FF5}}$   & 0.39  & 0.26  & 0.32  & 0.25  & 0.18  & -0.21 \\
    $t$-stat & 3.87  & 4.65  & 6.16  & 4.56  & 1.92  & -1.24 \\
   $\alpha^{\text{FF5 + MOM}}$ & 0.39  & 0.26  & 0.31  & 0.25  & 0.16  & -0.23 \\
    $t$-stat & 4.04  & 4.68  & 6.09  & 4.60  & 2.03  & -1.50 \\
    \midrule
    \midrule
    \textbf{C:} Common Bad Fears & 1     & 2     & 3     & 4     & 5     & 5-1 \\
    \midrule
    mean (\%) & 1.23  & 1.06  & 0.96  & 0.91  & 0.82  & \textbf{-0.41} \\
    $t$-stat & 3.68  & 3.59  & 3.55  & 3.59  & 3.43  & \textbf{-2.66} \\
    $\alpha^{\text{FF5}}$ & 0.40  & 0.29  & 0.25  & 0.26  & 0.19  & -0.21 \\
    $t$-stat & 5.53  & 5.01  & 3.95  & 4.54  & 2.51  & -1.78 \\
    $\alpha^{\text{FF5 + MOM}}$ & 0.41  & 0.29  & 0.24  & 0.25  & 0.18  & \textbf{-0.23} \\
    $t$-stat & 5.82  & 4.97  & 3.59  & 4.92  & 2.72  & \textbf{-2.24} \\
    \midrule
    \midrule
    \bottomrule
    \end{tabular}%
  \label{tab:single_sortsVIX}%
\end{table}%

In Table \ref{tab:single_sortsVIX2} we report portfolio sorts on common fears loadings that control for corresponding market fears. Here, both excess returns and risk-adjusted returns decrease monotonically as loadings to common fears increase. All spread portfolios earn negative returns with statistically significant returns from the spread portfolio using common bad fears. The risk-adjusted return controlling for the Fama-French 5 factors and momentum is significant at 1\% levels with a $t$-statistic of -2.24. The economic significance of this risk-adjusted return reduce slightly to -0.18\% per month (-2.16\% annualized) relative to the result in Table \ref{tab:single_sortsVIX}. In the Appendix, we report results from conditional double sorts in Tables \ref{tab:double_sorts} and \ref{tab:double_sorts2}. These results first sort on loadings to innovations in the VIX index and market fears respectively. Overall, these results show monotonically decreasing returns for portfolios that load on common fears relative to controlling for the market. These results show that such spread portfolios are significant for the majority of quantiles at conventional levels; with the remaining quantiles at 10\% levels. There is no pattern in quintile portfolios sorting on innovations in the VIX index or market fears.

\begin{table}[!hp]
  \centering
  \caption{\textbf{Portfolio Sorts from Loadings on Common Fears, Common Good Fears, and Common Bad Fears Controlling for Market Fears}\\
  \small{Notes: This table shows value-weighted portfolios that we sort on loadings to: i) common fears in Panel A; ii) common good fears in Panel B; and iii) common bad fears in Panel C whilst controlling for innovations market fears. Market fears are implied volatilities we compute from index options following \cite{bakshi2000spanning} and \cite{Bakshi2003}. In each Panel, we report monthly excess returns for quintile portfolios and the long-short portfolio that goes long the portfolio of stocks with a high loading on common fears and short the portfolio of stocks with low loadings to common fears. We also report the risk adjusted returns from the Fama-French 5 factor model and the Fama-French 5-factor model accounting for Momentum.}}
    \begin{tabular}{lrrrrrr}
    \toprule
    \midrule
    \midrule
       & \multicolumn{1}{l}{low $\beta^{\Delta\cf}$} &       &       &       \multicolumn{2}{r}{High $\beta^{\Delta\cf}$} &  \\
    \midrule
    \textbf{A:} Common Fears  & 1     & 2     & 3     & 4     & 5     & 5-1 \\
    \midrule
    mean (\%) & 1.25  & 1.05  & 0.94  & 0.93  & 0.82  & \textbf{-0.44} \\
    $t$-stat & 3.56  & 3.48  & 3.32  & 3.77  & 3.79  & \textbf{-2.33} \\
    $\alpha^{\text{FF5}}$  & 0.40  & 0.30  & 0.20  & 0.25  & 0.24  & -0.16 \\
    $t$-stat & 4.03  & 5.30  & 2.51  & 4.18  & 3.61  & -1.16 \\
    $\alpha^{\text{FF5 + MOM}}$ & 0.41  & 0.30  & 0.19  & 0.24  & 0.23  & -0.18 \\
    $t$-stat & 4.38  & 5.12  & 2.80  & 4.32  & 4.06  & -1.56 \\
    \midrule
    \midrule
    \textbf{B:} Common Good Fears & 1     & 2     & 3     & 4     & 5     & 5-1 \\
    \midrule
    mean (\%) & 1.12  & 1.11  & 0.99  & 0.97  & 0.81  & -0.31 \\
    $t$-stat & 3.08  & 3.52  & 3.53  & 3.85  & 3.95  & -1.38 \\
    $\alpha^{\text{FF5}}$  & 0.33  & 0.32  & 0.24  & 0.29  & 0.23  & -0.09 \\
    $t$-stat & 2.76  & 4.75  & 4.50  & 4.70  & 2.80  & -0.51 \\
    $\alpha^{\text{FF5 + MOM}}$  & 0.33  & 0.31  & 0.23  & 0.28  & 0.22  & -0.11 \\
    $t$-stat & 3.05  & 4.65  & 4.53  & 4.93  & 3.07  & -0.73 \\
    \midrule
    \midrule
    \textbf{C:} Common Bad Fears & 1     & 2     & 3     & 4     & 5     & 5-1 \\
    \midrule
    mean (\%) & 1.22  & 1.03  & 0.97  & 0.91  & 0.81  & \textbf{-0.41} \\
    $t$-stat & 3.64  & 3.41  & 3.55  & 3.56  & 3.63  & \textbf{-2.46} \\
    $\alpha^{\text{FF5}}$ & 0.39  & 0.26  & 0.25  & 0.23  & 0.23  & -0.16 \\
    $t$-stat & 5.56  & 4.35  & 4.62  & 3.11  & 3.82  & -1.47 \\
    $\alpha^{\text{FF5 + MOM}}$ & 0.40  & 0.26  & 0.24  & 0.22  & 0.22  & \textbf{-0.18} \\
    $t$-stat & 6.17  & 4.27  & 4.61  & 3.47  & 4.21  & \textbf{-2.01} \\
    \midrule
    \midrule
    \bottomrule
    \end{tabular}%
  \label{tab:single_sortsVIX2}%
\end{table}%


To complement this analysis, we also report Fama-MacBeth two-pass regressions using the portfolios we sort on loadings to common fears, common good fears, and common bad fears.  Our procedure estimates portfolio betas using the full-sample and then a single cross-sectional regression to estimate risk premia. For common fears, common good fears, and common bad fears, we construct a factor mimicking portfolio following the approach in \cite{ang2006cross}. This is because our factors are not directly observable, and hence not tradable.  Our base assets for the mimicking portfolios use the portfolios we sort on common fears, common good fears, and common bad fears. This is because they all have different sensitivities to the factor. This approach provides us with a daily factor mimicking portfolio that contains no look-ahead bias due to how we extract the common fears factors, and how we construct the base assets. We take the average daily return in each month to convert the factor mimicking portfolio returns to monthly in each month by taking the average to monthly by taking the average return over month.\footnote{Converting the daily returns to monthly by cumulating returns over the month does not change the statistical significance of our results, however it provides inflated point estimates of risk premia.  Looking at the end of month returns produces qualitatively similar results, but again results in inflated point estimates of risk-premia.}

In Table \ref{tab:FMB_decile}, we present the risk premia estimates for decile portfolios that sort on loadings to common fears, $\beta^{\Delta\cf}$, in columns 1--6; common good fears, $\beta^{\Delta\cfg} $, in columns 7--12; and common bad fears, $\beta^{\Delta\cfb} $, in columns 13--18. We show results for decile portfolios to have a large enough cross-section to estimate these models. Available on request are results for quintile portfolios using a smaller number of asset pricing factors that convey the same message as those we present here.

\begin{table}[!hp]
  \centering
  \caption{\textbf{Fama-MacBeth Analysis: Decile Common Fears Portfolios} \\
  \small{This table shows the Fama-MacBeth two-pass regression analysis for decile portfolios we construct on loadings to: common fears, $\beta^{\Delta\cf}$, in columns 1--6; common good fears, $\beta^{\Delta\cfg} $, in columns 7--12; and common bad fears, $\beta^{\Delta\cfb} $, in columns 13--18. $\lambda_{\cf}$ denotes the corresponding risk premium estimates for common fears, common good fears and common bad fears.  Results in columns 1, 7, 13 control for the Fama-French 5 factor model. MKT is the market risk premium; SMB is small-minus-big; HML is high-minus-low; RMW is robust-minus-weak; CMA is conservative minus aggressive). Meanwhile columns 2--5, 8--11, and 14--17 control for the variance risk premium (VRP), momentum (MOM), common idiosyncratic volatility (CIV) \citep{herskovic2016common}, and liquidity (LIQ) \citep{pastor2003liquidity}, respectively. Columns 6, 12, 18 add both CIV and LIQ to the Fama-French 5 factor model. Below risk premia estimates we report Newey and West $t$-statistics with 12 lags that adjust for errors in variables as in \cite{shanken1992estimation}. $ \bar{R}^2 $ is the adjusted R-squared.}
  }
  \adjustbox{max height=9.70in ,max width=\textwidth, keepaspectratio}{
    \begin{tabular}{lrrrrrrrrrrrrrrrrrr}
    \toprule
    \midrule
    \midrule
          & \multicolumn{6}{c}{Decile $\beta^{\Delta\cf} $ Portfolios} & \multicolumn{6}{c}{Decile $\beta^{\Delta\cfg} $ Portfolios} & \multicolumn{6}{c}{Decile $\beta^{\Delta\cfb} $ Portfolios} \\
    \midrule
          & 1     & 2     & 3     & 4     & 5     & 6     & 7     & 8     & 9     & 10    & 11    & 12    & 13    & 14    & 15    & 16    & 17    & 18 \\
    \midrule
   $\lambda_{0}$ & 0.21  & -1.39 & 0.28  & 0.11  & 0.08  & 0.09  & -0.35 & -0.40 & -0.58 & -0.64 & -1.05 & -3.45 & 0.33  & 0.72  & -0.11 & 0.76  & 0.40  & 0.33 \\
    $t$-stat & 0.24  & -0.78 & 0.30  & 0.13  & 0.10  & 0.11  & -0.56 & -0.56 & -0.92 & -0.34 & -1.11 & -1.01 & 0.63  & 1.30  & -0.19 & 1.35  & 0.78  & 0.55 \\
    $\lambda_{\cf}$ & -0.22 & -0.69 & -0.21 & -0.27 & -0.39 & -0.40 & -0.26 & -0.27 & -0.29 & -0.35 & -0.49 & -1.27 & \textbf{-0.41} & \textbf{-0.45} & \textbf{-0.45} & \textbf{-0.47} & \textbf{-0.42} & \textbf{-0.41} \\
    $t$-stat & -1.07 & -1.21 & -1.04 & -1.19 & -1.36 & -1.41 & -1.05 & -0.97 & -1.25 & -0.61 & -1.46 & -1.20 & \textbf{-3.54} & \textbf{-3.93} & \textbf{-3.76} & \textbf{-4.11} & \textbf{-3.69} & \textbf{-3.29} \\
    $\lambda_{MKT}$ & 0.51  & 2.36  & 0.42  & 0.63  & 0.65  & 0.63  & 1.43  & 1.55  & 1.88  & 1.77  & 2.67  & 5.76  & 0.75  & 0.12  & 1.29  & 0.04  & 0.37  & 0.45 \\
    $t$-stat & 0.45  & 1.06  & 0.36  & 0.55  & 0.59  & 0.55  & 1.60  & 1.66  & 1.79  & 0.72  & 2.02  & 1.25  & 0.98  & 0.14  & 1.80  & 0.04  & 0.43  & 0.55 \\
    $\lambda_{SMB}$ & 1.40  & 1.29  & 1.40  & 1.38  & 1.33  & 1.33  & 0.56  & 0.42  & 0.21  & 0.53  & -0.38 & -1.28 & 0.24  & 0.41  & 0.03  & 0.67  & 0.94  & 0.96 \\
    $t$-stat & 1.71  & 1.54  & 1.72  & 1.68  & 1.61  & 1.60  & 0.51  & 0.44  & 0.16  & 0.46  & -0.30 & -0.64 & 0.49  & 0.86  & 0.05  & 1.41  & 1.64  & 1.51 \\
    $\lambda_{HML}$ & 3.06  & -0.30 & 3.02  & 2.74  & 2.70  & 2.76  & 0.88  & 0.77  & 0.57  & 0.85  & -0.85 & -2.46 & 2.33  & 2.66  & 2.59  & 2.02  & 1.78  & 1.75 \\
    $t$-stat & 1.54  & -0.07 & 1.54  & 1.32  & 1.35  & 1.32  & 0.71  & 0.68  & 0.42  & 0.66  & -0.46 & -0.82 & 2.25  & 2.48  & 2.34  & 1.93  & 1.71  & 1.72 \\
    $\lambda_{RMW}$ & -0.41 & 1.78  & -0.36 & -0.32 & -0.08 & -0.05 & 0.46  & 0.30  & -0.11 & 0.51  & -0.84 & -1.55 & -0.41 & -0.62 & -0.51 & -0.09 & -0.32 & -0.37 \\
    $t$-stat & -0.53 & 0.74  & -0.47 & -0.41 & -0.10 & -0.07 & 0.31  & 0.26  & -0.06 & 0.37  & -0.48 & -0.66 & -0.92 & -1.45 & -1.16 & -0.17 & -0.68 & -0.69 \\
    $\lambda_{CMA}$ & 0.94  & 1.24  & 0.89  & 0.79  & 0.90  & 0.94  & 0.93  & 0.94  & 0.75  & 0.95  & 0.63  & 0.51  & -0.40 & -0.06 & -0.35 & -0.04 & 0.27  & 0.30 \\
    $t$-stat & 1.62  & 2.21  & 1.39  & 1.19  & 1.50  & 1.55  & 1.20  & 1.19  & 0.87  & 1.18  & 0.78  & 0.63  & -0.82 & -0.14 & -0.76 & -0.10 & 0.61  & 0.59 \\
    $\lambda_{VRP}$ &       & -0.21 &       &       &       &       &       & 0.08  &       &       &       &       &       & 0.22  &       &       &       &  \\
    $t$-stat &       & -1.00 &       &       &       &       &       & 0.32  &       &       &       &       &       & 1.18  &       &       &       &  \\
    $\lambda_{MOM}$ &       &       & -0.21 &       &       &       &       &       & 0.07  &       &       &       &       &       & -0.70 &       &       &  \\
    $t$-stat &       &       & -1.65 &       &       &       &       &       & 0.05  &       &       &       &       &       & -0.67 &       &       &  \\
    $\lambda_{CIV}$ &       &       &       & 0.15  &       & -1.26 &       &       &       & 0.18  &       & 0.78  &       &       &       & -0.80 &       & 1.05 \\
    $t$-stat &       &       &       & 0.47  &       & -0.30 &       &       &       & 0.27  &       & 0.83  &       &       &       & -2.13 &       & 0.14 \\
    $\lambda_{LIQ}$ &       &       &       &       & 0.24  & 0.27  &       &       &       &       & -0.14 & -0.22 &       &       &       &       & -0.16 & -0.20 \\
    $t$-stat &       &       &       &       & 1.00  & 0.96  &       &       &       &       & -0.65 & -0.84 &       &       &       &       & -1.20 & -0.69 \\
    \midrule
    $\bar{R}^2$ & 0.938 & 0.979 & 0.922 & 0.943 & 0.979 & 0.971 & 0.890 & 0.856 & 0.886 & 0.857 & 0.919 & 0.982 & 0.756 & 0.980 & 0.781 & 0.872 & 0.939 & 0.911 \\
    \midrule
    \midrule
    \bottomrule
    \end{tabular}%
    }
  \label{tab:FMB_decile}%
\end{table}%

Results in columns 1, 7, 13 control for the Fama-French 5 factor model. Meanwhile columns 2--5, 8--11, and 14--17 control for the variance risk premium (VRP), momentum (MOM), common idiosyncratic volatility (CIV) \citep{herskovic2016common}, and liquidity (LIQ) \citep{pastor2003liquidity}, respectively. Columns 6, 12, and 18 add both CIV and LIQ to the Fama-French 5 factor model.  Below risk premium estimates, we report Newey and West $t$-statistics with 12 lags that adjust for errors in variables as in \cite{shanken1992estimation}. $ \bar{R}^2 $ at the bottom of the table is the adjusted R-squared.

The main takeaway from Table \ref{tab:FMB_decile} is that portfolios that sort on common bad fears have significant risk premia estimates across all specifications we consider. We can see that the point estimate for the risk premia ranges from -0.41 to 0.47 with $t$-statistics indicating significance at 1\% levels. Annualizing these estimates suggest risk prices ranging from -5.64\% to -4.92\%; which is similar to those we present in Table \ref{tab:single_sorts}. Risk premia estimates from the Fama-French 5 factor model have varying degrees of significance with far more variability in the point estimates for estimates. The same holds true for the additional controls we consider. Model fit ranges from 0.756 to 0.982 and the intercepts in all cross-sectional regressions are statistically insignificant. 

Tables \ref{tab:5x5VIX} and \ref{tab:5x5VIX2} report analogous results for double sorts that control for innovations in the VIX index, and market fears (total, good and bad), respectively. We can see from Table \ref{tab:5x5VIX} similar conclusions hold. Common bad fears prices the 25 portfolios that sort on innovations to the VIX and loadings to common bad fears. Estimates of the risk premia range from -0.22 to -0.17 which imply annual risk prices of -2.64\% to -2.04\%. This is similar in magnitude to the risk adjusted returns for spread portfolios in Table \ref{tab:single_sortsVIX}. Observe that there are varying degrees of significance for the Fama-French 5-factors and also the additional controls we consider. The adjusted R-squared statistics fall relative to Table \ref{tab:FMB_decile} and the intercepts are mostly significant at conventional levels. Similar conclusions hold for Table \ref{tab:5x5VIX2} where we control for market fears. Again, risk premia estimates for common bad fears range from -0.23 to -0.1. Here the two specifications in columns 15 and 18 are insignificant at conventional levels; although column 15 is significant at the 10\% level. For significant estimates for common bad fear risk premia estimates the annual prices of risk range from -2.76\% to -2.40\%.

In the Appendix, we show corresponding results that control for portfolios that control for: i) the market risk premium; ii) market capitalization; and iii) trading volume. In general, these results yield similar conclusions to those we report here. The two notable differences are that portfolios loading on common fears and common bad fears result in significant risk-adjusted returns when controlling for the market risk premium, and Fama-MacBeth analysis controlling for trading volume have significant exposures to common bad fears at the 10\% level. Long--short portfolios loading on common bad fears generate risk-adjusted returns that range from -0.26\% per month (-3.12\% annualized) to -0.21\% per month (-2.52\% annualized). The annualized risk premia estimates from corresponding Fama-MacBeth regressions range from -3.24\% to -2.04\%.

\begin{table}[!hp]
  \centering
  \caption{\textbf{Fama-MacBeth Analysis: 5x5 Common Fears / VIX  Portfolios} \\
  \small{This table shows the Fama-MacBeth two-pass regression analysis for the 25 portfolios we construct on loadings to: common fears, $\beta^{\Delta\cf}$,  and innovations to the VIX index ($\Delta$\vix) in columns 1--6; common good fears, $\beta^{\Delta\cfg} $ and $\Delta$\vix, in columns 7--12; and common bad fears, $\beta^{\Delta\cfb} $, and $\Delta$\vix in columns 13--18. $\lambda_{\cf}$ denotes the corresponding risk premium estimates for common fears, common good fears and common bad fears.  Results in columns 1, 7, 13 control for the Fama-French 5 factor model. MKT is the market risk premium; SMB is small-minus-big; HML is high-minus-low; RMW is robust-minus-weak; CMA is conservative minus aggressive). Meanwhile columns 2--5, 7--12, and 14--17 control for the variance risk premium (VRP), momentum (MOM), common idiosyncratic volatility (CIV) \citep{herskovic2016common}, and liquidity (LIQ) \citep{pastor2003liquidity}, respectively. Columns 6, 12, 18 add all aforementioned factors to the Fama-French 5 factor model.  Below risk premia estimates we report Newey and West $t$-statistics with 12 lags that adjust for errors in variables as in \cite{shanken1992estimation}. $ \bar{R}^2 $ is the adjusted R-squared.}
  }
    \adjustbox{max height=9.70in ,max width=\textwidth, keepaspectratio}{
    \begin{tabular}{lrrrrrrrrrrrrrrrrrr}
    \toprule
      \midrule
    \midrule
          & \multicolumn{6}{c}{5 X 5 $\beta^{\Delta\cf} $ / $\beta^{\Delta\text{VIX}} $ portfolios} & \multicolumn{6}{c}{5 X 5 $\beta^{\Delta\cfg} $ / $\beta^{\Delta\text{VIX}} $ portfolios} & \multicolumn{6}{c}{5 X 5 $\beta^{\Delta\cfb} $ / $\beta^{\Delta\text{VIX}} $ portfolios} \\
    \midrule
          & 1     & 2     & 3     & 4     & 5     & 6     & 7     & 8     & 9     & 10    & 11    & 12    & 13    & 14    & 15    & 16    & 17    & 18 \\
    \midrule
    $\lambda_{0}$ & 0.67  & 0.90  & 0.89  & 0.75  & 0.83  & 1.08  & 1.17  & 1.23  & 1.21  & 1.16  & 1.21  & 1.24  & 0.65  & 0.66  & 0.83  & 0.67  & 0.84  & 0.91 \\
    $t$-stat & 2.11  & 2.95  & 2.83  & 2.67  & 2.78  & 3.58  & 5.27  & 5.94  & 5.62  & 4.98  & 5.90  & 5.76  & 2.01  & 2.07  & 2.57  & 2.16  & 2.70  & 2.99 \\
    $\lambda_{\cf}$ & -0.20 & -0.16 & -0.14 & -0.18 & -0.15 & -0.11 & 0.15  & 0.16  & 0.12  & 0.14  & 0.18  & 0.14  & \textbf{-0.19} & \textbf{-0.22} & \textbf{-0.17} & \textbf{-0.20} & \textbf{-0.18} & \textbf{-0.19} \\
    $t$-stat & -1.83 & -1.50 & -1.36 & -1.75 & -1.46 & -1.15 & 2.08  & 2.23  & 1.74  & 1.94  & 2.68  & 2.05  & \textbf{-2.34} & \textbf{-2.78} & \textbf{-2.02} & \textbf{-2.40} & \textbf{-2.17} & \textbf{-2.47} \\
    $\lambda_{MKT}$ & 0.75  & 0.49  & 0.52  & 0.64  & 0.45  & 0.25  & -0.14 & -0.22 & -0.15 & -0.14 & -0.16 & -0.16 & 0.52  & 0.49  & 0.31  & 0.46  & 0.17  & 0.05 \\
    $t$-stat & 1.64  & 1.07  & 1.18  & 1.40  & 0.96  & 0.54  & -0.48 & -0.70 & -0.50 & -0.45 & -0.52 & -0.49 & 1.15  & 1.08  & 0.78  & 1.10  & 0.36  & 0.10 \\
    $\lambda_{SMB}$ & -0.55 & -0.93 & -0.60 & -0.55 & -0.56 & -0.90 & 0.23  & 0.15  & 0.14  & 0.23  & 0.06  & -0.03 & -0.04 & -0.17 & -0.02 & 0.00  & 0.03  & 0.07 \\
    $t$-stat & -1.44 & -2.49 & -1.57 & -1.43 & -1.47 & -2.35 & 0.56  & 0.37  & 0.34  & 0.57  & 0.15  & -0.06 & -0.13 & -0.57 & -0.07 & 0.00  & 0.08  & 0.21 \\
    $\lambda_{HML}$ & -1.61 & -1.64 & -1.45 & -1.62 & -1.30 & -1.48 & 0.79  & 0.57  & 0.54  & 0.79  & 0.40  & 0.18  & -0.67 & -0.70 & -0.59 & -0.63 & -0.67 & -0.57 \\
    $t$-stat & -2.54 & -2.57 & -2.33 & -2.54 & -2.19 & -2.29 & 2.11  & 1.39  & 1.41  & 2.08  & 0.86  & 0.38  & -1.47 & -1.53 & -1.36 & -1.39 & -1.46 & -1.33 \\
    $\lambda_{RMW}$ & -0.46 & -0.27 & -0.37 & -0.35 & 0.02  & -0.03 & -0.15 & -0.17 & -0.09 & -0.16 & -0.15 & -0.14 & 0.15  & 0.17  & 0.19  & 0.24  & 0.45  & 0.62 \\
    $t$-stat & -1.07 & -0.65 & -0.85 & -0.86 & 0.06  & -0.07 & -0.57 & -0.64 & -0.33 & -0.57 & -0.56 & -0.47 & 0.50  & 0.56  & 0.64  & 0.76  & 1.55  & 1.84 \\
    $\lambda_{CMA}$ & -0.78 & -1.28 & -0.86 & -0.80 & -0.90 & -1.30 & -0.03 & -0.05 & -0.05 & -0.03 & -0.25 & -0.20 & -0.49 & -0.54 & -0.50 & -0.45 & -0.61 & -0.53 \\
    $t$-stat & -2.39 & -3.44 & -2.72 & -2.57 & -2.74 & -3.48 & -0.14 & -0.25 & -0.22 & -0.13 & -0.92 & -0.72 & -1.62 & -1.81 & -1.63 & -1.42 & -2.04 & -1.70 \\
    $\lambda_{VRP}$ &       & 0.39  &       &       &       & 0.35  &       & 0.16  &       &       &       & 0.12  &       & 0.24  &       &       &       & 0.11 \\
    $t$-stat &       & 3.08  &       &       &       & 2.55  &       & 1.04  &       &       &       & 0.71  &       & 2.76  &       &       &       & 0.92 \\
    $\lambda_{MOM}$ &       &       & -0.08 &       &       & 0.05  &       &       & -1.57 &       &       & -1.51 &       &       & -0.41 &       &       & -0.20 \\
    $t$-stat &       &       & -0.12 &       &       & 0.08  &       &       & -1.73 &       &       & -1.67 &       &       & -0.77 &       &       & -0.30 \\
    $\lambda_{CIV}$ &       &       &       & -0.19 &       & -0.20 &       &       &       & -0.17 &       & 0.30  &       &       &       & -0.14 &       & -0.24 \\
    $t$-stat &       &       &       & -1.04 &       & -1.12 &       &       &       & -0.17 &       & 0.31  &       &       &       & -0.75 &       & -1.25 \\
    $\lambda_{LIQ}$ &       &       &       &       & -0.21 & -0.07 &       &       &       &       & -0.20 & -0.17 &       &       &       &       & -0.24 & -0.21 \\
    $t$-stat &       &       &       &       & -2.52 & -0.73 &       &       &       &       & -1.67 & -1.36 &       &       &       &       & -2.22 & -1.24 \\
    \midrule
    $ \bar{R}^2 $ & 0.530 & 0.721 & 0.532 & 0.511 & 0.610 & 0.688 & 0.589 & 0.661 & 0.610 & 0.566 & 0.670 & 0.685 & 0.374 & 0.429 & 0.361 & 0.346 & 0.461 & 0.388 \\
      \midrule
    \midrule
    \bottomrule
    \end{tabular}%
    }
  \label{tab:5x5VIX}%
\end{table}%

\begin{table}[!hp]
  \centering
  \caption{\textbf{Fama-MacBeth Analysis: 5x5 Common Fears / Market Fears  Portfolios} \\
  \small{This table shows the Fama-MacBeth two-pass regression analysis for the 25 p portfolios we construct on loadings to: common fears, $\beta^{\Delta\cf}$,  and market fears, $\beta^{\Delta\mf}$ ,in columns 1--6; common good fears, $\beta^{\Delta\cfg} $ and good market fears, $\beta^{\Delta\mfg}$, in columns 7--12; and common bad fears, $\beta^{\Delta\cfb} $, and bad market fears,  $\beta^{\Delta\mfb}$, in columns 13--18. $\lambda_{\cf}$ denotes the corresponding risk premium estimates for common fears, common good fears and common bad fears.  Results in columns 1, 7, 13 control for the Fama-French 5 factor model. MKT is the market risk premium; SMB is small-minus-big; HML is high-minus-low; RMW is robust-minus-weak; CMA is conservative minus aggressive). Meanwhile columns 2--5, 7--12, and 14--17 control for the variance risk premium (VRP), momentum (MOM), common idiosyncratic volatility (CIV) \citep{herskovic2016common}, and liquidity (LIQ) \citep{pastor2003liquidity}, respectively. Columns 6, 12, 18 add all aforementioned factors to the Fama-French 5 factor model.  Below risk premia estimates we report Newey and West $t$-statistics with 12 lags that adjust for errors in variables as in \cite{shanken1992estimation}. $ \bar{R}^2 $ is the adjusted R-squared.}
  }
      \adjustbox{max height=9.70in ,max width=\textwidth, keepaspectratio}{
    \begin{tabular}{lrrrrrrrrrrrrrrrrrr}
    \toprule
         \midrule
    \midrule
          & \multicolumn{6}{c}{5 X 5 $\beta^{\Delta\cf} $ / $\beta^{\Delta\mf}$ portfolios} & \multicolumn{6}{c}{5 X 5 $\beta^{\Delta\cfg} $ / $\beta^{\Delta\mfg} $ portfolios} & \multicolumn{6}{c}{5 X 5 $\beta^{\Delta\cfb} $ / $\beta^{\Delta\mfb}$ portfolios} \\
    \midrule
          & 1     & 2     & 3     & 4     & 5     & 6     & 7     & 8     & 9     & 10    & 11    & 12    & 13    & 14    & 15    & 16    & 17    & 18 \\
    \midrule
    $\lambda_{0}$ & 0.61  & 0.61  & 1.16  & 0.70  & 0.74  & 1.19  & 1.37  & 1.29  & 1.35  & 1.41  & 1.26  & 1.26  & 0.96  & 0.96  & 1.39  & 1.05  & 1.01  & 1.19 \\
    $t$-stat & 2.87  & 2.94  & 5.40  & 3.50  & 3.87  & 5.60  & 4.95  & 4.26  & 5.09  & 4.88  & 4.18  & 4.02  & 4.95  & 4.90  & 7.82  & 5.56  & 5.20  & 5.48 \\
    $\lambda_{\cf}$ & -0.14 & -0.14 & -0.02 & -0.11 & -0.09 & -0.01 & 0.16  & 0.16  & 0.14  & 0.16  & 0.16  & 0.12  & \textbf{-0.22} & \textbf{-0.20} & -0.15 & \textbf{-0.23} & \textbf{-0.23} & -0.10 \\
    $t$-stat & -1.48 & -1.47 & -0.29 & -1.24 & -1.13 & -0.10 & 1.64  & 1.59  & 1.40  & 1.65  & 1.64  & 1.26  & \textbf{-2.75} & \textbf{-2.27} & -1.79 & \textbf{-2.82} & \textbf{-2.84} & -0.99 \\
    $\lambda_{MKT}$ & 0.71  & 0.72  & 0.13  & 0.63  & 0.52  & 0.05  & -0.65 & -0.53 & -0.59 & -0.69 & -0.48 & -0.43 & 0.22  & 0.30  & -0.25 & 0.13  & 0.14  & 0.18 \\
    $t$-stat & 1.72  & 1.73  & 0.35  & 1.53  & 1.24  & 0.14  & -2.10 & -1.65 & -2.02 & -2.22 & -1.52 & -1.38 & 0.53  & 0.70  & -0.68 & 0.30  & 0.31  & 0.39 \\
    $\lambda_{SMB}$ & -0.42 & -0.47 & -0.38 & -0.48 & -0.46 & -0.35 & 0.88  & 0.69  & 0.77  & 0.86  & 0.67  & 0.44  & -0.28 & -0.64 & -0.19 & -0.34 & -0.27 & -0.84 \\
    $t$-stat & -1.30 & -1.59 & -1.19 & -1.50 & -1.45 & -1.22 & 2.39  & 1.74  & 2.25  & 2.40  & 1.71  & 1.11  & -1.04 & -1.73 & -0.68 & -1.18 & -0.99 & -1.75 \\
    $\lambda_{HML}$ & -0.73 & -0.75 & -0.51 & -0.65 & -0.84 & -0.60 & 1.03  & 1.02  & 0.94  & 0.99  & 0.70  & 0.72  & -0.98 & -1.16 & -0.41 & -0.95 & -0.96 & -0.83 \\
    $t$-stat & -1.66 & -1.62 & -1.22 & -1.51 & -1.79 & -1.33 & 3.01  & 2.96  & 3.08  & 2.99  & 1.57  & 1.93  & -2.10 & -2.28 & -0.89 & -2.10 & -2.11 & -1.68 \\
    $\lambda_{RMW}$ & 0.06  & 0.08  & -0.10 & -0.02 & 0.15  & -0.03 & 0.15  & 0.26  & 0.18  & 0.15  & 0.34  & 0.39  & -0.03 & 0.16  & -0.21 & -0.08 & 0.04  & -0.11 \\
    $t$-stat & 0.25  & 0.32  & -0.43 & -0.06 & 0.59  & -0.12 & 0.54  & 0.95  & 0.65  & 0.58  & 1.12  & 1.24  & -0.12 & 0.63  & -0.75 & -0.26 & 0.17  & -0.43 \\
    $\lambda_{CMA}$ & -0.53 & -0.56 & -0.49 & -0.36 & -0.62 & -0.57 & 0.19  & 0.20  & 0.12  & 0.18  & -0.01 & 0.01  & -0.95 & -1.13 & -0.69 & -0.96 & -0.99 & -0.90 \\
    $t$-stat & -1.79 & -1.83 & -1.68 & -1.25 & -1.98 & -1.81 & 0.88  & 0.89  & 0.59  & 0.82  & -0.02 & 0.02  & -2.80 & -2.96 & -2.00 & -2.79 & -2.66 & -2.25 \\
    $\lambda_{VRP}$ &       & 0.07  &       &       &       & 0.04  &       & 0.12  &       &       &       & 0.16  &       & 0.27  &       &       &       & 0.43 \\
    $t$-stat &       & 0.49  &       &       &       & 0.30  &       & 0.88  &       &       &       & 1.12  &       & 1.63  &       &       &       & 1.74 \\
    $\lambda_{MOM}$ &       &       & -2.15 &       &       & -1.89 &       &       & -0.77 &       &       & -1.05 &       &       & -1.70 &       &       & -1.51 \\
    $t$-stat &       &       & -3.54 &       &       & -3.15 &       &       & -0.95 &       &       & -1.31 &       &       & -2.37 &       &       & -2.12 \\
    $\lambda_{CIV}$ &       &       &       & -0.35 &       & -0.22 &       &       &       & -0.09 &       & -0.08 &       &       &       & -0.23 &       & -0.14 \\
    $t$-stat &       &       &       & -2.20 &       & -0.16 &       &       &       & -0.83 &       & -0.77 &       &       &       & -1.18 &       & -0.82 \\
    $\lambda_{LIQ}$ &       &       &       &       & -0.20 & -0.18 &       &       &       &       & -0.20 & -0.10 &       &       &       &       & -0.02 & 0.16 \\
    $t$-stat &       &       &       &       & -1.75 & -1.71 &       &       &       &       & -1.56 & -0.83 &       &       &       &       & -0.21 & 1.30 \\
    \midrule
    $ \bar{R}^2 $ & 0.165 & 0.121 & 0.453 & 0.219 & 0.253 & 0.406 & 0.441 & 0.497 & 0.415 & 0.419 & 0.469 & 0.440 & 0.570 & 0.616 & 0.695 & 0.562 & 0.556 & 0.715 \\
         \midrule
    \midrule
    \bottomrule
    \end{tabular}%
    }
  \label{tab:5x5VIX2}%
\end{table}%

\newpage
\subsection{Alternative Test Assets}

It is natural to question whether common fears, common good fears, and common bad fears carry similar pricing implications to other test assets. We consider a variety of anomaly portfolios that contain double sorts (5 x 5) on: size and investment (ME/INV); size and the market risk premium (ME/MKT); size and book-to-market (ME/BM); size and operating profit (ME/OP); operating profit/investment (OP/INV); book-to-market and investment (BM/INV); book-to-market and operating profit (BM/OP); size and residual variance (ME/RESVAR); size and variance (ME/VAR); size and accruals (ME/AC); and size and momentum (ME/MOM).

\begin{figure}[!hp]
	\centering
		\scalebox{0.56}{\includegraphics{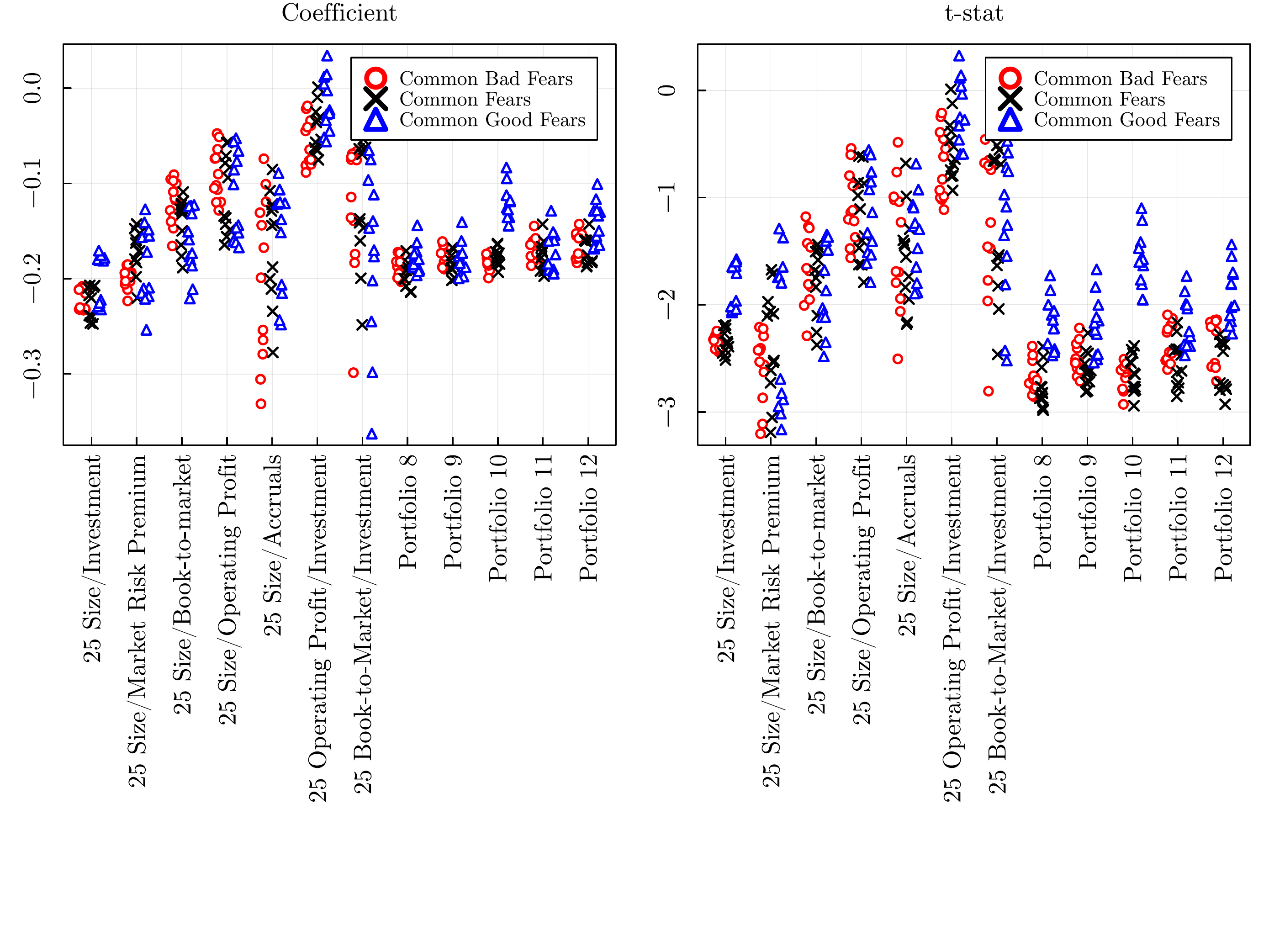}}\\
		\caption{\textbf{Common Fears, Common Good Fears, and Common Bad Fears Risk Premium Estimates and $t$-statistics: Alternative Test Assets}\\ \small{Notes: This figure shows the point estimates (LHS plots) and corresponding $t$-statistics (RHS plots) for the risk premia associated to common fears (yellow dots), common good fears (blue dots), and common bad fears (red dots). The first seven set of test assets are: size and investment (ME/INV); size and the market risk premium (ME/MKT); size and book-to-market (ME/BM); size and operating profit (ME/OP); size and accruals (ME/AC); operating profit/investment (OP/INV); and book-to-market and investment (BM/INV). Portfolio 8 takes ME/INV, ME/MKT, ME/BM, and decile portfolios loading on common fears. Portfolios 9 and 10 consider the same anomaly portfolios but then add the respective double sorted portfolios on common investor fears and VIX, and common investor fears and market fears respectively. Portfolio 11 considers the following anomaly portfolios: ME/INV, ME/MKT, ME/BM, ME/OP, size/residual variance, size/variance, and adds the decile portfolios we sort on common fears. Portfolio 12 considers the same anomaly portfolios as Portfolio 11, but adds the 25 double sorted portfolios on common fears and VIX.}}
      \label{fig:boxplot}
\end{figure}

Figure \ref{fig:boxplot} shows plots for the risk premia estimates (LHS plot) and corresponding $t$-statistics (RHS plot) to common fears, common good fears, and common bad fears respectively for a battery of test assets and asset pricing models.\footnote{For the sake of brevity, we refrain from reporting results here for: BM/OP; ME/RESVAR; ME/VAR; and ME/MOM. Common fears do not price i) and iv). In subsequent analysis below, we include the test assets  in ii) and iii). All results are available upon request.} We first consider seven anomaly portfolios in isolation (ME/INV, ME/MKT, ME/BM, ME/OP, ME/AC, OP/INV, and BM/INV), and then add these anomaly portfolios to portfolios that we sort on loadings to common fears (Portfolios 8--12). For each set of test assets, we consider 12 pricing models. The first six consider: the Fama-French 3 factor model; add our additional controls (VRP, MOM, CIV, LIQ) to the Fama-French 3 factor model; and then all additional controls together. The next six do the same for the Fama-French 5 factor model.

From these plots, we can see that our main results hold for alternative test assets. these risk premia estimates are almost never positive (the exception is 25 OP/INV portfolios and common good fears); even for those that are statistically insignificant. Although common fears and common good fears are able to price the cross sectional variation in some test assets, it is common bad fears that predominantly yield significant negative risk premium estimates across all specifications. For risk premium estimates that are statistically significant, common fears annualized risk premia estimates range from -3.36\% to -1.68\%. Common good fears annualized risk premia estimates range from -3.00\% to -1.80\% , and common bad fears annualized risk premia estimates  range from -3.96\% to -1.68\%. Test assets with the minimum risk premia point estimates are 25 portfolios on ME/INV, 25 portfolios on ME/MKT, 25 portfolios on BM/INV. Those with the maximum risk premia estimates are when we add multiple anomaly portfolios to portfolios we sort on common fears loadings (e.g. 25 portfolios sorted on: ME/INV, ME/BM, ME/MKT, ME/OP, ME/RESVAR, and ME/VAR).

\begin{table}[!hp]
  \centering
  \caption{\textbf{Fama-MacBeth Analysis; Alternative Test Assets: Size/Investment, Size/Book to Market, Size/Market Risk Premium} \\
  \small{This table shows the Fama-MacBeth two-pass regression analysis for the 25 portfolios that sort on: i) size/investment (ME/INV); ii) size/book-to-market (ME/BM); iii) size/market risk premium (ME/MKT) and the 25 respective portfolios we construct on loadings to: common fears, $\beta^{\Delta\cf}$,  and market fears, $\beta^{\Delta\mf}$ ,in columns 1--5; common good fears, $\beta^{\Delta\cfg} $ and good market fears, $\beta^{\Delta\mfg}$, in columns 6--10; and common bad fears, $\beta^{\Delta\cfb} $, and bad market fears,  $\beta^{\Delta\mfb}$, in columns 11--15. $\lambda_{\cf}$ denotes the corresponding risk premium estimates for common fears, common good fears and common bad fears. All results control initially for the Fama-French 5 factors.  MKT is the market risk premium; SMB is small-minus-big; HML is high-minus-low; RMW is robust-minus-weak; CMA is conservative minus aggressive). Results in columns 1--4, 6--9, 11--14 use the Fama-French 5 factors plus: the variance risk premium (VRP); momentum (MOM), common idiosyncratic volatility (CIV) \citep{herskovic2016common}; and liquidity (LIQ) \citep{pastor2003liquidity}. Columns 5, 10, 15 add all aforementioned additional controls to the Fama-French 5 factor model. Below risk premia estimates we report Newey and West $t$-statistics with 12 lags that adjust for errors in variables as in \cite{shanken1992estimation}. $ \bar{R}^2 $ is the adjusted R-squared.}
  }
      \adjustbox{max height=9.70in ,max width=\textwidth, keepaspectratio}{
    \begin{tabular}{lrrrrrrrrrrrrrrr}
    \toprule
    \midrule
    \midrule
    & \multicolumn{15}{c}{ 25 ME/INV,25 ME/BM, 25 ME/MKT,  Plus 25:}\\
          & \multicolumn{5}{c}{ $\beta^{\Delta\cf} / \beta^{\Delta\mf}$ } & \multicolumn{5}{c}{$\beta^{\Delta\cfg} / \beta^{\Delta\mfg}$ } & \multicolumn{5}{c}{$\beta^{\Delta\cfb} / \beta^{\Delta\mfb}$ } \\
    \midrule
          & 1     & 2     & 3     & 4     & 5     & 6     & 7     & 8     & 9     & 10    & 11    & 12    & 13    & 14    & 15 \\
    \midrule
    $\lambda_{0}$ & 0.56  & 0.66  & 0.62  & 0.63  & 0.71  & 0.59  & 0.78  & 0.64  & 0.61  & 0.76  & 0.55  & 0.67  & 0.59  & 0.66  & 0.74 \\
    $t$-stat & 2.22  & 2.76  & 2.74  & 2.78  & 3.10  & 2.17  & 3.02  & 2.42  & 2.30  & 2.87  & 2.18  & 2.68  & 2.51  & 2.92  & 3.22 \\
    $\lambda_{\cf}$ & \textbf{-0.19} & \textbf{-0.17} & \textbf{-0.18} & \textbf{-0.18} & \textbf{-0.16} & -0.13 & -0.11 & -0.12 & -0.09 & -0.08 & \textbf{-0.20} & \textbf{-0.18} & \textbf{-0.18} & \textbf{-0.18} & \textbf{-0.17} \\
    $t$-stat & \textbf{-2.65} & \textbf{-2.42} & \textbf{-2.54} & \textbf{-2.45} & \textbf{-2.38} & -1.60 & -1.41 & -1.47 & -1.21 & -1.10 & \textbf{-2.78} & \textbf{-2.51} & \textbf{-2.61} & \textbf{-2.58} & \textbf{-2.55} \\
    $\lambda_{MKT}$ & 0.48  & 0.38  & 0.42  & 0.39  & 0.31  & 0.45  & 0.24  & 0.39  & 0.42  & 0.26  & 0.48  & 0.37  & 0.44  & 0.36  & 0.28 \\
    $t$-stat & 1.40  & 1.16  & 1.24  & 1.10  & 0.98  & 1.40  & 0.78  & 1.24  & 1.29  & 0.86  & 1.36  & 1.14  & 1.28  & 0.96  & 0.87 \\
    $\lambda_{SMB}$ & 0.15  & 0.16  & 0.16  & 0.17  & 0.18  & 0.16  & 0.19  & 0.17  & 0.17  & 0.20  & 0.17  & 0.18  & 0.18  & 0.19  & 0.20 \\
    $t$-stat & 0.93  & 1.00  & 0.97  & 1.01  & 1.07  & 0.95  & 1.20  & 1.01  & 1.03  & 1.21  & 1.02  & 1.09  & 1.05  & 1.12  & 1.17 \\
    $\lambda_{HML}$ & -0.26 & -0.27 & -0.25 & -0.24 & -0.24 & -0.18 & -0.23 & -0.16 & -0.17 & -0.22 & -0.28 & -0.28 & -0.26 & -0.25 & -0.26 \\
    $t$-stat & -1.15 & -1.19 & -1.12 & -1.08 & -1.08 & -0.79 & -1.03 & -0.73 & -0.78 & -0.96 & -1.24 & -1.28 & -1.19 & -1.16 & -1.18 \\
    $\lambda_{RMW}$ & 0.24  & 0.31  & 0.27  & 0.26  & 0.31  & 0.17  & 0.39  & 0.21  & 0.19  & 0.38  & 0.31  & 0.40  & 0.34  & 0.31  & 0.37 \\
    $t$-stat & 1.06  & 1.43  & 1.18  & 1.11  & 1.44  & 0.82  & 1.99  & 1.02  & 0.93  & 1.93  & 1.41  & 2.04  & 1.53  & 1.39  & 1.91 \\
    $\lambda_{CMA}$ & -0.14 & -0.15 & -0.15 & -0.13 & -0.14 & -0.14 & -0.15 & -0.15 & -0.15 & -0.16 & -0.13 & -0.15 & -0.15 & -0.14 & -0.14 \\
    $t$-stat & -0.97 & -1.06 & -1.00 & -0.91 & -0.97 & -1.01 & -1.06 & -1.08 & -1.09 & -1.15 & -0.96 & -1.08 & -1.04 & -0.98 & -1.03 \\
    $\lambda_{VRP}$ & 0.14  &       &       &       & 0.03  & 0.13  &       &       &       & 0.05  & 0.14  &       &       &       & 0.03 \\
    $t$-stat & 1.23  &       &       &       & 0.37  & 1.12  &       &       &       & 0.57  & 1.26  &       &       &       & 0.45 \\
    $\lambda_{MOM}$ &       & -0.35 &       &       & -0.39 &       & -1.40 &       &       & -1.28 &       & -0.49 &       &       & -0.45 \\
    $t$-stat &       & -0.65 &       &       & -0.74 &       & -2.36 &       &       & -2.21 &       & -0.91 &       &       & -0.83 \\
    $\lambda_{CIV}$ &       &       & -0.92 &       & 0.09  &       &       & -1.11 &       & 0.02  &       &       & -0.79 &       & -0.08 \\
    $t$-stat &       &       & -1.00 &       & 0.10  &       &       & -1.40 &       & 0.03  &       &       & -0.75 &       & -0.07 \\
    $\lambda_{LIQ}$ &       &       &       & -0.19 & -0.22 &       &       &       & -0.21 & -0.23 &       &       &       & -0.19 & -0.20 \\
    $t$-stat &       &       &       & -1.74 & -2.26 &       &       &       & -1.99 & -2.45 &       &       &       & -1.59 & -1.88 \\
    \midrule
    $ \bar{R}^2 $ & 0.468 & 0.459 & 0.455 & 0.556 & 0.551 & 0.353 & 0.413 & 0.344 & 0.466 & 0.510 & 0.515 & 0.506 & 0.498 & 0.584 & 0.578 \\
        \midrule
    \midrule
    \bottomrule
    \end{tabular}%
    }
  \label{tab:FMB_ALT}%
\end{table}%

We end this section by showing the risk premia estimates for a set of test assets that contains anomaly portfolios and also portfolios we sort on loadings to common and market fears. Table \ref{tab:FMB_ALT} shows the risk premia estimates where our test assets are the 25 portfolios on i) ME/INV; ii) ME/BM; and ii) ME/MKT plus the respective 25 portfolios that we sort on common fears/market fears in columns 1--5; common good fears/good market fears in columns 6--10; and common bad fears/bad market fears in columns 10--15. For these test assets, we can see that common fears and common bad fears are priced with $t$-statistics indicating significance at 1\% levels. the estimates range from -0.16\% per month (-1.92\% annualized) to -0.20\% per month (-2.40\% annualized). As for common good fears, the risk premia estimates are negative, but insignificant. Additional factor risk premia we control for range from being statistically insignificant to marginally significance. In some cases, the sign of the risk premia estimates change \citep[see e.g. the CIV factor of][]{herskovic2016common}. Pricing models examining common fears and common bad fears having adjusted R-squared values that range from 0.459 to 0.578, with those looking at common good fears ranging from 0.352 to 0.51.

Table \ref{tab:FMB_ALT2} in the Appendix presents Fama-MacBeth results for additional test assets. Here, we add 25 portfolios sorted on  ME/OP, ME/RESVAR, and ME/VAR to those we consider in Table \ref{tab:FMB_ALT}. These results yield qualitatively similar conclusions with annualized risk premium estimates for common fears and common bad fears both ranging between -1.92\% and -1.68\% respectively.

\section{Robustness Analysis}\label{sec:Robust}

\subsection{Accounting for Market Fears in Fama-MacBeth Analysis}

Another natural question is whether the pricing implications for common firm-level investor fears hold when controlling for market fears within our Fama-MacBeth analysis. We therefore construct a factor mimicking portfolio to resemble market fears, good market fears, and bad market fears in the exact manner as we do for our measures of common firm-level investor fears. We present the Fama-MacBeth estimates for 25 portfolios we sort on loadings to: i) common fears, $\beta^{\Delta\cf}$,  and market fears, $\beta^{\Delta\mf}$; ii) common good fears, $\beta^{\Delta\cfg} $ and good market fears, $\beta^{\Delta\mfg}$; and iii) common bad fears, $\beta^{\Delta\cfb} $, and bad market fears,  $\beta^{\Delta\mfb}$ in Table \ref{tab:5x5CFMF}.

Two noteworthy points emerge from Table \ref{tab:5x5CFMF}. First, Our main results concerning common fears and common bad fears hold even in light of allowing for a market fears pricing factor. Three of the five specifications pricing bad common fears are statistically significant at 1\% levels with another two being marginally signficiant at 10\% levels. For statistically significant bad common fears risk premia, the annualized risk premium estimates are between -2.64\% and -2.52\%. This magnitude if economically meaningful and similar to those we present in Table \ref{tab:5x5VIX2}.

Second, the risk premia estimates for market fears, good market fears and bad market fears are all positive except for the results in column 6. Second within these specifications, both good market fears and common good fears appear to command positive risk-premia that are statistically significant. The magnitude of these premia are similar at annualized values ranging from 1.92\% to 2.40\%. This result relates to those in \cite{kilic2019good} where good variance risk-premia appear to have a positive relationship with future returns.

\begin{table}[!hp]
  \centering
  \caption{\textbf{Fama-MacBeth Analysis: 5x5 Common Fears / Market Fears  Portfolios: Accounting for Market Fears} \\
  \small{This table shows the Fama-MacBeth two-pass regression analysis for the 25 p portfolios we construct on loadings to: common fears, $\beta^{\Delta\cf}$,  and market fears, $\beta^{\Delta\mf}$ ,in columns 1--6; common good fears, $\beta^{\Delta\cfg} $ and good market fears, $\beta^{\Delta\mfg}$, in columns 7--12; and common bad fears, $\beta^{\Delta\cfb} $, and bad market fears,  $\beta^{\Delta\mfb}$, in columns 13--18. $\lambda_{\cf}$ denotes the corresponding risk premium estimates for common fears, common good fears and common bad fears. $\lambda_mf$ denotes the risk premium estimates for market fears, good market fears, and bad market fears. Results in columns 1, 7, 13 control for the Fama-French 5 factor model. MKT is the market risk premium; SMB is small-minus-big; HML is high-minus-low; RMW is robust-minus-weak; CMA is conservative minus aggressive). Meanwhile columns 2--5, 8--11, and 14--17 control for the variance risk premium (VRP), momentum (MOM), common idiosyncratic volatility (CIV) \citep{herskovic2016common}, and liquidity (LIQ) \citep{pastor2003liquidity}, respectively. Columns 6, 12, 18 add all aforementioned factors to the Fama-French 5 factor model.  Below risk premia estimates we report Newey and West $t$-statistics with 12 lags that adjust for errors in variables as in \cite{shanken1992estimation}. $ \bar{R}^2 $ is the adjusted R-squared.}
  }
      \adjustbox{max height=9.70in ,max width=\textwidth, keepaspectratio}{
    \begin{tabular}{lrrrrrrrrrrrrrrrrrr}
    \toprule
         \midrule
    \midrule
          & \multicolumn{6}{c}{5 X 5 $\beta^{\Delta\cf} $ / $\beta^{\Delta\mf} $ portfolios} & \multicolumn{6}{c}{5 X 5 $\beta^{\Delta\cfg} $ / $\beta^{\Delta\mfg}$ portfolios} & \multicolumn{6}{c}{5 X 5 $\beta^{\Delta\cfb} $ / $\beta^{\Delta\mfb}$ portfolios} \\
    \midrule
          & 1     & 2     & 3     & 4     & 5     & 6     & 7     & 8     & 9     & 10    & 11    & 12    & 13    & 14    & 15    & 16    & 17    & 18 \\
    \midrule
    $\lambda_{0}$ & 0.53  & 0.54  & 1.15  & 0.70  & 0.72  & 1.22  & 1.46  & 1.37  & 1.45  & 1.43  & 1.36  & 1.26  & 1.01  & 1.06  & 1.42  & 1.07  & 1.14  & 1.22 \\
    $t$-stat & 2.39  & 2.48  & 5.45  & 3.48  & 3.54  & 5.75  & 5.30  & 4.60  & 5.49  & 4.95  & 4.41  & 4.03  & 5.17  & 5.55  & 7.76  & 5.64  & 5.58  & 5.55 \\
    $\lambda_{\cf}$ & -0.15 & -0.14 & -0.03 & -0.11 & -0.10 & 0.00  & 0.20  & 0.21  & 0.19  & 0.21  & 0.20  & 0.15  & \textbf{-0.21} & -0.16 & -0.14 & \textbf{-0.22} & \textbf{-0.22} & -0.07 \\
    $t$-stat & -1.60 & -1.59 & -0.32 & -1.31 & -1.23 & -0.05 & 2.03  & 2.06  & 1.99  & 2.07  & 2.01  & 1.62  & \textbf{-2.67} & -1.80 & -1.73 & \textbf{-2.74} & \textbf{-2.75} & -0.71 \\
    $\lambda_{\mf}$ & 0.14  & 0.13  & 0.02  & 0.01  & 0.06  & -0.02 & 0.17  & 0.19  & 0.17  & 0.18  & 0.17  & 0.20  & 0.08  & 0.18  & 0.06  & 0.07  & 0.13  & 0.15 \\
    $t$-stat & 1.20  & 1.17  & 0.14  & 0.10  & 0.44  & -0.13 & 2.70  & 3.04  & 2.82  & 2.52  & 2.70  & 2.83  & 1.31  & 2.25  & 0.94  & 1.00  & 1.69  & 1.92 \\
    $\lambda_{MKT}$ & 0.72  & 0.73  & 0.14  & 0.63  & 0.53  & 0.04  & -0.72 & -0.58 & -0.70 & -0.69 & -0.58 & -0.35 & 0.09  & 0.07  & -0.34 & 0.04  & -0.14 & -0.02 \\
    $t$-stat & 1.73  & 1.75  & 0.36  & 1.51  & 1.22  & 0.11  & -2.31 & -1.80 & -2.34 & -2.21 & -1.72 & -1.17 & 0.23  & 0.18  & -0.95 & 0.11  & -0.32 & -0.04 \\
    $\lambda_{SMB}$ & -0.27 & -0.31 & -0.37 & -0.48 & -0.43 & -0.41 & 0.93  & 0.68  & 0.87  & 0.95  & 0.74  & 0.41  & -0.08 & -0.44 & -0.02 & -0.14 & 0.05  & -0.51 \\
    $t$-stat & -0.79 & -0.97 & -1.06 & -1.44 & -1.29 & -1.34 & 2.48  & 1.71  & 2.53  & 2.57  & 1.82  & 1.05  & -0.24 & -1.20 & -0.07 & -0.40 & 0.16  & -1.03 \\
    $\lambda_{HML}$ & -0.76 & -0.77 & -0.51 & -0.65 & -0.84 & -0.59 & 0.94  & 0.91  & 0.90  & 0.97  & 0.65  & 0.66  & -0.90 & -1.11 & -0.36 & -0.89 & -0.82 & -0.77 \\
    $t$-stat & -1.71 & -1.67 & -1.24 & -1.54 & -1.81 & -1.35 & 2.88  & 2.71  & 2.96  & 2.96  & 1.51  & 1.79  & -1.88 & -2.17 & -0.77 & -1.89 & -1.81 & -1.56 \\
    $\lambda_{RMW}$ & 0.01  & 0.03  & -0.11 & -0.02 & 0.14  & -0.02 & -0.01 & 0.10  & 0.01  & -0.04 & 0.17  & 0.21  & -0.12 & 0.10  & -0.28 & -0.14 & -0.01 & -0.08 \\
    $t$-stat & 0.06  & 0.11  & -0.44 & -0.06 & 0.54  & -0.07 & -0.05 & 0.38  & 0.02  & -0.14 & 0.51  & 0.63  & -0.44 & 0.37  & -0.96 & -0.48 & -0.05 & -0.30 \\
    $\lambda_{CMA}$ & -0.54 & -0.55 & -0.49 & -0.36 & -0.62 & -0.56 & 0.23  & 0.24  & 0.19  & 0.25  & 0.05  & 0.06  & -0.91 & -1.17 & -0.67 & -0.92 & -0.98 & -0.93 \\
    $t$-stat & -1.81 & -1.82 & -1.69 & -1.29 & -1.98 & -1.83 & 1.00  & 1.06  & 0.89  & 1.05  & 0.19  & 0.24  & -2.67 & -3.04 & -1.93 & -2.63 & -2.64 & -2.33 \\
    $\lambda_{VRP}$ &       & 0.05  &       &       &       & 0.04  &       & 0.12  &       &       &       & 0.16  &       & 0.37  &       &       &       & 0.46 \\
    $t$-stat &       & 0.40  &       &       &       & 0.32  &       & 0.89  &       &       &       & 1.13  &       & 1.99  &       &       &       & 1.85 \\
    $\lambda_{MOM}$ &       &       & -2.13 &       &       & -1.93 &       &       & -0.51 &       &       & -1.12 &       &       & -1.64 &       &       & -1.32 \\
    $t$-stat &       &       & -3.66 &       &       & -3.41 &       &       & -0.67 &       &       & -1.37 &       &       & -2.32 &       &       & -1.95 \\
    $\lambda_{CIV}$ &       &       &       & -0.35 &       & -0.35 &       &       &       & 0.63  &       & 0.81  &       &       &       & -1.71 &       & 0.36 \\
    $t$-stat &       &       &       & -2.06 &       & -0.25 &       &       &       & 0.48  &       & 0.64  &       &       &       & -0.85 &       & 0.20 \\
    $\lambda_{LIQ}$ &       &       &       &       & -0.20 & -0.18 &       &       &       &       & -0.20 & -0.13 &       &       &       &       & -0.08 & 0.07 \\
    $t$-stat &       &       &       &       & -1.69 & -1.79 &       &       &       &       & -1.60 & -1.10 &       &       &       &       & -0.76 & 0.59 \\
    \midrule
    $ \bar{R}^2 $ & 0.136 & 0.086 & 0.421 & 0.173 & 0.210 & 0.365 & 0.560 & 0.702 & 0.535 & 0.542 & 0.581 & 0.675 & 0.579 & 0.711 & 0.701 & 0.562 & 0.593 & 0.784 \\
         \midrule
    \midrule
    \bottomrule
    \end{tabular}%
    }
  \label{tab:5x5CFMF}%
\end{table}%

Tables \ref{tab:FMB_ALT_CFMF} and \ref{tab:FMB_ALT_CFMF2} report analogous results to those in Table \ref{tab:5x5CFMF}, but include additional anomaly portfolios as test assets. The former adds the respective 25 portfolios on: ME/INV; ME/BM; and ME/MKT to the double sorts on common firm-level investor fears/market fears portfolios. The latter adds the respective 25 portfolios on: ME/INV; ME/BM; ME/MKT; ME/OP; ME/RESVAR; ME/VAR to the double sorts on common firm-level investor fears/market fears portfolios. These results show that common bad fears bear significant negative risk premia that conform with those within our main results. The significant positive premia for common good fears and good market fears we observe in Table \ref{tab:5x5CFMF} disappears.

\subsection{Three-pass Regression Analysis}

All results we present until now suffer from omitted variable bias. We therefore investigate the pricing implications of common firm-level investor fears using the three-pass regressions approach in \cite{giglio2021asset}. This procedure is valid even if we do not specify or observe all factors within a pricing model and relies on PCA of the test assets to recover the factor space and additional regressions to proxy risk premia. In what follows, we present results from pricing models that contain a measure of common firm-level investor fears, and the market risk premium.

There are two benefits we exploit of the \cite{giglio2021asset} approach. First, as long as the test assets remain the same, the risk-premia estimates and model fit do not change as you start adding additional pricing factors to the specification. This means we can consider multiple dimensions of robustness to our main results with brevity. Second, we are able to test the null hypothesis that factors we consider are weak pricing factors. The benefit of this within the three-pass regression procedure is that we are able to understand whether the test assets capture well variation in the pricing factor itself whilst being able to recover the risk-premia estimates of potentially strong factors and interpret inference reliably.

Table \ref{tab:3PASS} shows results for three-pass regressions that use a measure of common firm-level investor fears and the market risk premium. Each panel assesses the sensitivities to how we define the factor tracking common firm-level investor fears. In Panel A, we show results for tradable factors of common firm-level investor fears which are factor mimicking portfolios as with our main results. Panels B and C consider nontradable measures of common firm-level investor fears. Panel B uses the innovations in the factros we extract using PCA as we outline in Section \ref{sec:MFIV}. Panel C uses innovations in an equally weighted average of firm level implied volatilities in a similar vein to how \cite{herskovic2016common} measure their CIV factor. We report risk-premia estimates with their corresponding $t$-statistics below. We also report the $p$-values from the Wald test in \cite{giglio2021asset} with the null hypothesis that the factor is weak. Rejection of the null indicates the factor is a strong pricing factor. We also report the adjusted R-squared for each cross-sectional regression and the number of factors from test assets to recover the factor space.

The columns assess the sensitivity of our estimates to changes in the test assets. Columns 1--3 show results from decile portfolios that sort on common, common good, and common bad fears. Columns 4--6 contain the 25 portfolios that sort on: ME/INV; ME/BM; ME/MKT; ME/OP; ME/RESVAR; and ME/VAR. Columns 7--9 add the respective portfolios in columns 1--3 to those in 4--6. Columns 10--12 add all decile beta portfolios to columns 4--6 and then uses common fears, common good fears and common bad fears in columns 10, 11, and 12 respectively.  Columns 13--15 add the respective decile beta portfolios on common fears,  as well as double sorts on common fears and market fears to the test assets in columns 4--6. Risk premium estimates in columns 1, 4, 7, 10, 13 use common fears, in columns 2, 5, 8, 11, 14 use common good fears, and in columns 3, 6, 9, 12, 15 use common bad fears.

Risk premium estimates for common firm-level investor fears are statistically significant across all specifications of how we define the risk factor, and across the breadth of test assets. The majority of the time the risk premium relating to common bad fears is larger in absolute value than those from common good fears or common fears. The factor mimicking portfolios we construct to track common fears, common good fears, and common bad fears in Panel A have comparable point estimates to risk premia estimates using innovations in the extracted factor using PCA in Panel B. The annualized risk premia estimates range from -2.76\% to -1.44\% and -3.72\% to -1.44\% in Panels A and B respectively. Those risk premia in Panel C are larger in absolute value with annualized estimates ranging from -7.68\% to -3.96\%\footnote{The results from Fama-MacBeth two-pass regressions using innovations of an equally weighted average of implied volatilities having varying degrees of statistical significance the magnitude of coefficient estimates are qualitatively similar.}. The risk premia estimates associated to the market fluctuate between 0.68\% and 1.44\% per month and model fit ranges from 0.37 to 0.85. The Wald tests reject the null that common firm-level investor fears are weak asset pricing factors, as do those for the market risk premium. As we increase the number of test assets beyond decile portfolios, the number of factors to recover the factor space increases from one to three.

As an additional exercise, we estimate results using all pricing factors we consider above and are available upon request. These results show that only the HML factor of \cite{fama2015five} five factors and the LIQ factor of \cite{pastor2003liquidity} are statistically significant; with the former in most cases only marginally significant. These results also suggest that MOM, VRP, and CIV of \cite{herskovic2016common} are statistically and economically insignificant. The Wald tests indicate that we cannot reject the null that MOM and CIV are weak pricing factors.

\begin{table}[!hp]
  \centering
  \caption{\textbf{\cite{giglio2021asset} Three-pass Regression Analysis} \\
  \small{This table shows results from the three-pass regression analysis of \cite{giglio2021asset}.  Columns 1--3 show results from decile portfolios that sort on common, common good, and common bad fears. Columns 4--6 contain the 25 portfolios that sort on: size/investment; size/book-to-market; size/market risk premium; size/operating profit; size/residual variance; and size/total variance. Columns 7--9 add the respective portfolios in columns 1--3 to those in 4--6. Columns 10--12 add all decile beta portfolios to columns 4--6 and then uses common fears, common good fears and common bad fears in columns 10, 11, and 12 respectively.  Columns 13--15 add the respective decile beta portfolios on common fears,  as well as double sorts on common fears and market fears to the test assets in columns 4--6. $\lambda_{\cf}$ denotes the corresponding risk premium estimates for common fears, common good fears and common bad fears. Risk premium estimates in columns 1, 4, 7, 10, 13 use common fears, in columns 2, 5, 8, 11, 14 use common good fears, and in columns 3, 6, 9, 12, 15 use common bad fears. Panel A constructs the common fears factor from factor mimicking portfolios; Panel B takes the innovations to the extracted common factor; and Panel C takes an equally weighted average of firm-level fears to construct the common factor. $t$-statistics are reported below risk premia estimates. The Wald (p-value) entries are p-values from the null hypothesis that the asset pricing factor is a weak pricing factor.  rejection of the null implies the factor is a strong pricing factor.  $ \bar{R}^2 $ is the adjusted R-squared and No. Factors is the number of PCA factors the model requires to recover the factor space.}
  }
        \adjustbox{max height=9.70in ,max width=\textwidth, keepaspectratio}{
    \begin{tabular}{lrrrrrrrrrrrrrrr}
    \toprule
    \midrule
    \midrule
          & \multicolumn{15}{c}{\textbf{A:} Factor Mimicking Portfolios} \\
          & 1     & 2     & 3     & 4     & 5     & 6     & 7     & 8     & 9     & 10    & 11    & 12    & 13    & 14    & 15 \\
    \midrule
    $\lambda_{0}$ & 0.02  & 0.29  & -0.29 & 0.33  & 0.33  & 0.33  & 0.33  & 0.33  & 0.33  & 0.30  & 0.30  & 0.30  & 0.37  & 0.40  & 0.34 \\
    $t$-stat & 0.08  & 2.63  & -0.80 & 3.65  & 3.65  & 3.65  & 3.79  & 3.91  & 3.64  & 3.36  & 3.36  & 3.36  & 4.42  & 5.36  & 3.82 \\
    $\lambda_{\cf}$ & \textbf{-0.19} & \textbf{-0.16} & \textbf{-0.23} & \textbf{-0.12} & \textbf{-0.14} & \textbf{-0.12} & \textbf{-0.12} & \textbf{-0.14} & \textbf{-0.12} & \textbf{-0.13} & \textbf{-0.15} & \textbf{-0.12} & \textbf{-0.12} & \textbf{-0.13} & \textbf{-0.12} \\
    $t$-stat & \textbf{-3.00} & \textbf{-2.73} & \textbf{-2.77} & \textbf{-2.25} & \textbf{-2.52} & \textbf{-2.07} & \textbf{-2.25} & \textbf{-2.54} & \textbf{-2.07} & \textbf{-2.37} & \textbf{-2.69} & \textbf{-2.14} & \textbf{-2.19} & \textbf{-2.38} & \textbf{-2.06} \\
    Wald (p-value) & 0.00  & 0.00  & 0.00  & 0.00  & 0.00  & 0.00  & 0.00  & 0.00  & 0.00  & 0.00  & 0.00  & 0.00  & 0.00  & 0.00  & 0.00 \\
    $\lambda_{MKT}$ & 1.11  & 0.80  & 1.44  & 0.70  & 0.70  & 0.70  & 0.70  & 0.70  & 0.70  & 0.75  & 0.75  & 0.75  & 0.68  & 0.65  & 0.72 \\
    $t$-stat & 3.25  & 2.77  & 3.12  & 2.28  & 2.28  & 2.28  & 2.29  & 2.30  & 2.30  & 2.45  & 2.45  & 2.45  & 2.25  & 2.13  & 2.35 \\
    Wald (p-value) & 0.00  & 0.00  & 0.00  & 0.00  & 0.00  & 0.00  & 0.00  & 0.00  & 0.00  & 0.00  & 0.00  & 0.00  & 0.00  & 0.00  & 0.00 \\
    \midrule
    $ \bar{R}^2 $ & 0.76  & 0.85  & 0.62  & 0.43  & 0.43  & 0.43  & 0.42  & 0.44  & 0.40  & 0.41  & 0.41  & 0.41  & 0.40  & 0.42  & 0.37 \\
    No. Factors  & 1.00  & 1.00  & 1.00  & 3.00  & 3.00  & 3.00  & 3.00  & 3.00  & 3.00  & 3.00  & 3.00  & 3.00  & 3.00  & 3.00  & 3.00 \\
    \midrule
    \midrule
          & \multicolumn{15}{c}{\textbf{B:} Innovations in extracted PCA} \\
          & 1     & 2     & 3     & 4     & 5     & 6     & 7     & 8     & 9     & 10    & 11    & 12    & 13    & 14    & 15 \\
    \midrule
    $\lambda_{0}$ & 0.02  & 0.29  & -0.29 & 0.33  & 0.33  & 0.33  & 0.33  & 0.33  & 0.33  & 0.30  & 0.30  & 0.30  & 0.37  & 0.40  & 0.34 \\
    $t$-stat & 0.08  & 2.63  & -0.80 & 3.65  & 3.65  & 3.65  & 3.79  & 3.91  & 3.64  & 3.36  & 3.36  & 3.36  & 4.42  & 5.36  & 3.82 \\
    $\lambda_{\cf}$ & \textbf{-0.24} & \textbf{-0.15} & \textbf{-0.31} & \textbf{-0.16} & \textbf{-0.14} & \textbf{-0.16} & \textbf{-0.15} & \textbf{-0.14} & \textbf{-0.16} & \textbf{-0.16} & \textbf{-0.14} & \textbf{-0.16} & \textbf{-0.15} & \textbf{-0.12} & \textbf{-0.16} \\
    $t$-stat & \textbf{-2.74} & \textbf{-2.48} & \textbf{-2.72} & \textbf{-1.97} & \textbf{-2.14} & \textbf{-1.97} & \textbf{-1.96} & \textbf{-2.11} & \textbf{-1.97} & \textbf{-2.02} & \textbf{-2.21} & \textbf{-2.02} & \textbf{-1.89} & \textbf{-1.95} & \textbf{-1.94} \\
    Wald (p-value) & 0.00  & 0.00  & 0.00  & 0.00  & 0.00  & 0.00  & 0.00  & 0.00  & 0.00  & 0.00  & 0.00  & 0.00  & 0.00  & 0.00  & 0.00 \\
    $\lambda_{MKT}$ & 1.11  & 0.80  & 1.44  & 0.70  & 0.70  & 0.70  & 0.70  & 0.70  & 0.70  & 0.75  & 0.75  & 0.75  & 0.68  & 0.65  & 0.72 \\
    $t$-stat & 3.25  & 2.77  & 3.12  & 2.28  & 2.28  & 2.28  & 2.29  & 2.30  & 2.30  & 2.45  & 2.45  & 2.45  & 2.25  & 2.13  & 2.35 \\
    Wald (p-value) & 0.00  & 0.00  & 0.00  & 0.00  & 0.00  & 0.00  & 0.00  & 0.00  & 0.00  & 0.00  & 0.00  & 0.00  & 0.00  & 0.00  & 0.00 \\
    \midrule
    $ \bar{R}^2 $ & 0.76  & 0.85  & 0.62  & 0.43  & 0.43  & 0.43  & 0.42  & 0.44  & 0.40  & 0.41  & 0.41  & 0.41  & 0.40  & 0.42  & 0.37 \\
    No. Factors  & 1.00  & 1.00  & 1.00  & 3.00  & 3.00  & 3.00  & 3.00  & 3.00  & 3.00  & 3.00  & 3.00  & 3.00  & 3.00  & 3.00  & 3.00 \\
    \midrule
     \midrule
          & \multicolumn{15}{c}{\textbf{C:} Innovations of equally weighted average of implied volatilities} \\
          & 1     & 2     & 3     & 4     & 5     & 6     & 7     & 8     & 9     & 10    & 11    & 12    & 13    & 14    & 15 \\
    \midrule
    $\lambda_{0}$ & 0.02  & 0.29  & -0.29 & 0.33  & 0.33  & 0.33  & 0.33  & 0.33  & 0.33  & 0.30  & 0.30  & 0.30  & 0.37  & 0.40  & 0.34 \\
    $t$-stat & 0.08  & 2.63  & -0.80 & 3.65  & 3.65  & 3.65  & 3.79  & 3.91  & 3.64  & 3.36  & 3.36  & 3.36  & 4.42  & 5.36  & 3.82 \\
    $\lambda_{\cf}$ & \textbf{-0.49} & \textbf{-0.33} & \textbf{-0.64} & \textbf{-0.40} & \textbf{-0.37} & \textbf{-0.40} & \textbf{-0.39} & \textbf{-0.37} & \textbf{-0.40} & \textbf{-0.42} & \textbf{-0.38} & \textbf{-0.42} & \textbf{-0.38} & \textbf{-0.33} & \textbf{-0.40} \\
    $t$-stat & \textbf{-3.28} & \textbf{-2.76} & \textbf{-3.10} & \textbf{-3.22} & \textbf{-3.04} & \textbf{-3.22} & \textbf{-3.20} & \textbf{-2.98} & \textbf{-3.21} & \textbf{-3.35} & \textbf{-3.06} & \textbf{-3.35} & \textbf{-3.03} & \textbf{-2.65} & \textbf{-3.16} \\
    Wald (p-value) & 0.00  & 0.00  & 0.00  & 0.00  & 0.00  & 0.00  & 0.00  & 0.00  & 0.00  & 0.00  & 0.00  & 0.00  & 0.00  & 0.00  & 0.00 \\
    $\lambda_{MKT}$ & 1.11  & 0.80  & 1.44  & 0.70  & 0.70  & 0.70  & 0.70  & 0.70  & 0.70  & 0.75  & 0.75  & 0.75  & 0.68  & 0.65  & 0.72 \\
    $t$-stat & 3.25  & 2.77  & 3.12  & 2.28  & 2.28  & 2.28  & 2.29  & 2.30  & 2.30  & 2.45  & 2.45  & 2.45  & 2.25  & 2.13  & 2.35 \\
    Wald (p-value) & 0.00  & 0.00  & 0.00  & 0.00  & 0.00  & 0.00  & 0.00  & 0.00  & 0.00  & 0.00  & 0.00  & 0.00  & 0.00  & 0.00  & 0.00 \\
    \midrule
    $ \bar{R}^2 $ & 0.76  & 0.85  & 0.62  & 0.43  & 0.43  & 0.43  & 0.42  & 0.44  & 0.40  & 0.41  & 0.41  & 0.41  & 0.40  & 0.42  & 0.37 \\
    No. Factors  & 1.00  & 1.00  & 1.00  & 3.00  & 3.00  & 3.00  & 3.00  & 3.00  & 3.00  & 3.00  & 3.00  & 3.00  & 3.00  & 3.00  & 3.00 \\
    \midrule
    \midrule
    \bottomrule
    \end{tabular}%
    }
  \label{tab:3PASS}%
\end{table}%

\section{Conclusion}\label{sec:Conc}

This paper shows strong co-movement within firm-level investor fears using individual equity options. Using the model-free approach of \cite{bakshi2000spanning,Bakshi2003} to extract implied variances, we define investor fears as uncertainty regarding future price movements of the underlying stocks. We decompose such fears into good and bad fears that link to upward and downward anticipated future price movements respectively. We provide empirical evidence in favour of commonalities within firm-level investor fears containing different information to those we infer from index options.

Stocks with a higher loading to common firm-level investor fears earn lower returns. Our results indicate there are statistically significant and economically meaningful risk premia to common firm-level investor fears; particularly those relating to common bad fears. Sorting stocks on their common bad fear beta yields an annualized risk-adjusted return spread of around 5\%. Fama-MacBeth regressions further substantiate this result with common bad fears risk premium estimates ranging from -5.63\% to -4.92\%. Meaningful common firm-level investor fear risk premia are present after controlling for market fears from index options, as well as a variety of other controls, although the annualized magnitude declines to around 3\% in absolute value. Common fears are important in understanding return differences within anomaly portfolios with similar risk premium estimates. We also show that common firm-level investor fear risk premia survive the three-pass regression procedure in \cite{giglio2021asset}.

Several important implications emerge from our analysis. First, the common component within in firm-level implied variances is different to that within implied variances on index options. This means that investor beliefs common to individual stocks is inherently different to beliefs about the market. Second, the common component within firm-level investor fears constitutes a priced source of risk. We show that the risk premium associated to the common component within firm-level fears from put options, common bad fears, is substantially larger than the corresponding premia that links to call options, common good fears. This suggests that not only do market participants react differently to good and bad anticipated future outcomes, but also that they will accept substantially lower returns in equilibrium on assets that hedge against common bad fears.

\newpage
\appendix
\section*{Appendix}
\setcounter{table}{0}
\renewcommand{\thetable}{A\arabic{table}}

\section{Discretization Procedure of Model-Free Implied Variance}

Considering total implied variance in Equation (\ref{eq:MFIV}), the discretization is $$\sigma^2_{i,t}  =\frac{2}{T}\sum_{i=1}^{n}\frac{\Delta K_i}{K_i^2}e^{rT}Q(K_i)-\frac{1}{T}\left[\frac{F}{K_0}-1\right]^2,$$ where $T$ is time to expiration, $F$ is the forward index level derived from the put-call parity as $F= e^{r^fT}[C(K,T)-P(K,T)]+K$ with the risk-free rate $r^f$, $K_0$ is the reference price, the first exercise price less or equal to the forward level $F(K_0 \leq F)$, and $K_i$ is the $i$th OTM strike price available on a specific date (call if $K_i>K_0$, put if $K_i<K_0$, and both call and put if $K_i = K_0$). $Q(K_i)$ is the average bid-ask of OTM options with exercise price equal to $K_i$. If $K_i=K_0$, it will be equal to the average between the at-the-money (ATM) call and put price, relative to the strike price, and $\Delta(K_i)$ is the sum divided by two of the two nearest prices to the exercise price $K_0$, namely, $\frac{(K_{i+1}-K_{i-1})}{2}$ for $2\leq i \leq n-1$.
For further details see the CBOE VIX white paper available at \url{https://cdn.cboe.com/api/global/us_indices/governance/Volatility_Index_Methodology_Cboe_Volatility_Index.pdf}. Note the standard CBOE methodology considers an interpolation between the two closest to 30-days expiration dates. In our data construction, we take into account only one expiration date closest to 30-days due to options data availability with respect to firm-level stocks.

\begin{landscape}
\section{Additional Results}

\subsection{Double Portfolio Sorts}
\begin{table}[!hp]
  \centering
  \caption{\textbf{Conditional Double Portfolio Sorts from Loadings on Common Fears and $\Delta$VIX} \\
  \small{Notes: This table shows value-weighted portfolios from conditional double sorts that we sort on loadings to: i) common fears and innovations to the VIX index in Panel A; ii) common good fears and innovations to the VIX index in Panel B; and iii) common bad fears and innovations to the VIX index in Panel C.  In Panel D we report double sorts on common bad fears and common good fears.}}
      \adjustbox{max height=12in ,max width= 9.70in, keepaspectratio}{
    \begin{tabular}{rlrrrrrrrrlrrrrrrr}
    \toprule
        \midrule
    \midrule
    \multicolumn{1}{l}{\textbf{A:} Common Fears} &       & \multicolumn{1}{l}{Low $\beta^{\Delta\cf}$} &       &       &        \multicolumn{2}{r}{High $\beta^{\Delta\cf}$}  & &       & \multicolumn{1}{l}{\textbf{B:} Common Good Fears} &       & \multicolumn{1}{l}{Low $\beta^{\Delta\cfg}$} &       &       &       \multicolumn{2}{r}{High $\beta^{\Delta\cfg}$}  & &  \\
          &       &$i$= 1     & 2     & 3     & 4     & 5     & 5-$j$   & \multicolumn{1}{l}{$t$-stat (5-$j$)} &       &       & $i$=1     & 2     & 3     & 4     & 5     & 5-$j$   & \multicolumn{1}{l}{$t$-stat (5-$j$)} \\
    \midrule
    \multicolumn{1}{l}{Low $\beta^{\Delta\text{VIX}} $} & \multicolumn{1}{r}{$j$=1} & 1.28  & 1.13  & 0.84  & 0.87  & 0.76  &\textbf{ -0.52} & \textbf{-1.99} & \multicolumn{1}{l}{Low $\beta^{\Delta\text{VIX}} $} & \multicolumn{1}{r}{$j$=1} & 1.17  & 0.88  & 0.98  & 0.83  & 0.76  & -0.41 & -1.42 \\
          & \multicolumn{1}{r}{2} & 1.33  & 1.07  & 0.94  & 0.76  & 0.80  & \textbf{-0.53} & \textbf{-2.14} &       & \multicolumn{1}{r}{2} & 1.23  & 1.17  & 1.11  & 0.90  & 0.80  & -0.43 & -1.52 \\
          & \multicolumn{1}{r}{3} & 1.11  & 1.11  & 1.15  & 0.99  & 0.88  & -0.23 & -1.02 &       & \multicolumn{1}{r}{3} & 1.12  & 1.03  & 0.94  & 0.98  & 0.75  & -0.37 & -1.38 \\
          & \multicolumn{1}{r}{4} & 1.06  & 0.85  & 0.89  & 0.95  & 0.70  & \textbf{-0.36} & \textbf{-1.87} &       & \multicolumn{1}{r}{4} & 1.10  & 0.87  & 0.97  & 0.84  & 0.81  & -0.29 & -1.23 \\
    \multicolumn{1}{l}{High $\beta^{\Delta\text{VIX}} $} & \multicolumn{1}{r}{5} & 1.45  & 1.16  & 0.99  & 1.05  & 0.80  & \textbf{-0.65} & \textbf{-2.66} & \multicolumn{1}{l}{High $\beta^{\Delta\text{VIX}} $} & \multicolumn{1}{r}{5} & 1.19  & 1.26  & 1.26  & 1.10  & 0.82  & -0.38 & -1.51 \\
          & 5-$i$   & 0.17  & 0.02  & 0.15  & 0.18  & 0.04  &       &       &       & 5-$i$   & 0.02  & 0.38  & 0.29  & 0.27  & 0.06  &       &  \\
          & $t$-stat (5-$i$) & 0.53  & 0.09  & 0.57  & 0.64  & 0.14  &       &       &       & $t$-stat (5-$i$) & 0.09  & 1.57  & 1.10  & 0.92  & 0.22  &       &  \\
    \midrule
    \midrule
    \multicolumn{1}{l}{\textbf{D:} Common Good} &       & \multicolumn{1}{l}{Low $\beta^{\Delta\cfg}$} &       &       &       \multicolumn{2}{r}{High $\beta^{\Delta\cfg}$}  & &      & \multicolumn{1}{l}{\textbf{C:} Common Bad Fears} &       & \multicolumn{1}{l}{Low $\beta^{\Delta\cfb}$} &       &       &       \multicolumn{2}{r}{High $\beta^{\Delta\cfb}$}  & &  \\
 \multicolumn{1}{l}{ vs. Bad Fears}      &   &$i$= 1     & 2     & 3     & 4     & 5     & 5-$j$   & \multicolumn{1}{l}{$t$-stat (5-$j$)} &       &       & $i$=1     & 2     & 3     & 4     & 5     & 5-$j$   & \multicolumn{1}{l}{$t$-stat (5-$j$)} \\
    \midrule
    \multicolumn{1}{l}{Low $\beta^{\Delta\cfb}$} & \multicolumn{1}{r}{$j$=1} & 1.50  & 1.19  & 1.19  & 1.28  & 1.12  & -0.38 & -1.37 & \multicolumn{1}{l}{Low $\beta^{\Delta\text{VIX}} $} & \multicolumn{1}{r}{$j$=1} & 1.32  & 1.26  & 0.99  & 0.81  & 0.77  & \textbf{-0.55} & \textbf{-2.10} \\
          & \multicolumn{1}{r}{2} & 1.19  & 1.23  & 1.18  & 0.84  & 0.97  & -0.22 & -0.81 &       & \multicolumn{1}{r}{2} & 1.19  & 1.06  & 0.83  & 0.88  & 0.77  & \textbf{-0.42} & \textbf{-1.97} \\
          & \multicolumn{1}{r}{3} & 0.77  & 1.01  & 0.99  & 0.92  & 0.91  & 0.14  & 0.55  &       & \multicolumn{1}{r}{3} & 1.10  & 1.18  & 1.06  & 1.12  & 0.84  & -0.27 & -1.38 \\
          & \multicolumn{1}{r}{4} & 1.02  & 0.99  & 0.86  & 0.86  & 0.77  & -0.24 & -0.91 &       & \multicolumn{1}{r}{4} & 1.10  & 1.00  & 0.78  & 0.75  & 0.84  & -0.25 & -1.34 \\
    \multicolumn{1}{l}{High $\beta^{\Delta\cfb}$} & \multicolumn{1}{r}{5} & 0.93  & 0.72  & 0.95  & 0.82  & 0.72  & -0.21 & -0.80 & \multicolumn{1}{l}{High $\beta^{\Delta\text{VIX}} $} & \multicolumn{1}{r}{5} & 1.42  & 0.80  & 1.14  & 0.98  & 0.87  & \textbf{-0.55} & \textbf{-2.36} \\
          & 5-$i$  & \textbf{-0.57} &\textbf{-0.47} & -0.23 & \textbf{-0.46} & -0.40 &       &       &       & 5-$i$   & 0.10  & -0.45 & 0.15  & 0.17  & 0.10  &       &  \\
          & $t$-stat (5-$i$) & \textbf{-2.12} & \textbf{-2.20} & -1.18 & \textbf{-2.00} & -1.68 &       &       &       & $t$-stat (5-$i$) & 0.30  & -1.50 & 0.50  & 0.59  & 0.36  &       &  \\
              \midrule
    \midrule
    \bottomrule
    \end{tabular}%
    }
  \label{tab:double_sorts}%
\end{table}%

\end{landscape}

\begin{landscape}

\begin{table}[!hp]
  \centering
  \caption{\textbf{Unconditional Double Portfolio Sorts from Loadings on Common Fears and Market Fears} \\
  \small{Notes: This table shows value-weighted portfolios from unconditional double sorts that we sort on loadings to: i) common fears and innovations to market fears in Panel A; ii) common good fears and innovations to good market fears in Panel B; and iii) common bad fears and innovations to bad market fears in Panel C.  In Panel D we report double sorts on common bad fears and common good fears.}}
      \adjustbox{max height=12in ,max width= 9.70in, keepaspectratio}{
    \begin{tabular}{rlrrrrrrrrlrrrrrrr}
    \toprule
    \midrule
    \midrule
    \multicolumn{2}{l}{A: Common Fears} & \multicolumn{1}{l}{Low $\beta^{\Delta\cf}$} &       &       &       & \multicolumn{2}{l}{High $\beta^{\Delta\cf}$} &       & \multicolumn{2}{l}{B: Common Good Fears} & \multicolumn{1}{l}{Low $\beta^{\Delta\cfg}$} &       &       &       & \multicolumn{2}{l}{High $\beta^{\Delta\cfg}$} &  \\
          &       & 1     & 2     & 3     & 4     & 5     & 5-$j$ & \multicolumn{1}{l}{$t$-stat (5-$j$)} &       &       & 1     & 2     & 3     & 4     & 5     & 5-$j$ & \multicolumn{1}{l}{$t$-stat (5-$j$)} \\
    \midrule
    \multicolumn{1}{l}{low $\beta^{\Delta\mf}$} & \multicolumn{1}{r}{1} & 1.30  & 0.98  & 0.85  & 0.72  & 0.66  & -0.64 & \textbf{-2.26} & \multicolumn{1}{l}{low $\beta^{\Delta\mfg}$} & \multicolumn{1}{r}{1} & 1.02  & 0.97  & 0.90  & 0.86  & 0.59  & -0.43 & -1.57 \\
          & \multicolumn{1}{r}{2} & 1.32  & 1.06  & 1.01  & 1.05  & 0.83  & -0.49 & \textbf{-2.10} &       & \multicolumn{1}{r}{2} & 1.22  & 1.15  & 0.97  & 1.11  & 0.80  & -0.42 & -1.50 \\
          & \multicolumn{1}{r}{3} & 1.10  & 1.11  & 1.09  & 1.01  & 0.70  & -0.40 & -1.65 &       & \multicolumn{1}{r}{3} & 1.18  & 1.04  & 0.88  & 0.79  & 0.83  & -0.35 & -1.26 \\
          & \multicolumn{1}{r}{4} & 1.04  & 0.98  & 0.72  & 0.86  & 0.86  & -0.18 & -0.85 &       & \multicolumn{1}{r}{4} & 0.93  & 1.02  & 0.91  & 1.07  & 0.83  & -0.10 & -0.36 \\
    \multicolumn{1}{l}{High $\beta^{\Delta\mf}$} & \multicolumn{1}{r}{5} & 1.50  & 1.14  & 1.04  & 0.98  & 1.03  & -0.48 & \textbf{-1.96} & \multicolumn{1}{l}{High $\beta^{\Delta\mfg}$} & \multicolumn{1}{r}{5} & 1.27  & 1.36  & 1.29  & 1.05  & 1.00  & -0.28 & -0.92 \\
          & 5-$i$ & 0.20  & 0.16  & 0.19  & 0.26  & 0.36  &       &       &       & 5-$i$ & 0.25  & 0.39  & 0.39  & 0.19  & 0.41  &       &  \\
          & $t$-stat (5-$i$) & 0.67  & 0.59  & 0.70  & 0.95  & 1.37  &       &       &       & $t$-stat (5-$i$) & 1.04  & 1.65  & 1.70  & 0.76  & 1.57  &       &  \\
    \midrule
    \midrule
    \multicolumn{2}{l}{D: Common Good } & \multicolumn{1}{l}{Low $\beta^{\Delta\cfg}$} &       &       &       & \multicolumn{2}{l}{High $\beta^{\Delta\cfg}$} &       & \multicolumn{2}{l}{C: Common Bad Fears} & \multicolumn{1}{l}{Low $\beta^{\Delta\cfb}$} &       &       &       & \multicolumn{2}{l}{High $\beta^{\Delta\cfb}$} &  \\
    \multicolumn{1}{l}{vs Bad} &       & 1     & 2     & 3     & 4     & 5     & 5-$j$ & \multicolumn{1}{l}{$t$-stat (5-$j$)} &       &       & 1     & 2     & 3     & 4     & 5     & 5-$j$ & \multicolumn{1}{l}{$t$-stat (5-$j$)} \\
    \midrule
    \multicolumn{1}{l}{Low $\beta^{\Delta\cfb}$} & \multicolumn{1}{r}{1} & 1.43  & 1.35  & 1.14  & 1.38  & 0.99  & -0.44 & -1.55 & \multicolumn{1}{l}{low $\beta^{\Delta\mfb}$} & \multicolumn{1}{r}{1} & 1.31  & 1.02  & 0.91  & 0.82  & 0.80  & -0.51 & \textbf{-1.94} \\
          & \multicolumn{1}{r}{2} & 1.01  & 1.06  & 1.19  & 1.10  & 0.95  & -0.06 & -0.27 &       & \multicolumn{1}{r}{2} & 1.29  & 1.20  & 1.09  & 0.98  & 0.71  & -0.58 & \textbf{-2.49} \\
          & \multicolumn{1}{r}{3} & 0.76  & 1.04  & 1.03  & 0.96  & 0.90  & 0.13  & 0.53  &       & \multicolumn{1}{r}{3} & 1.09  & 1.04  & 1.02  & 0.98  & 0.78  & -0.31 & -1.60 \\
          & \multicolumn{1}{r}{4} & 0.96  & 0.89  & 1.02  & 0.94  & 0.78  & -0.18 & -0.68 &       & \multicolumn{1}{r}{4} & 1.03  & 0.92  & 0.78  & 0.89  & 0.79  & -0.24 & -1.17 \\
    \multicolumn{1}{l}{High $\beta^{\Delta\cfb}$} & \multicolumn{1}{r}{5} & 0.82  & 0.85  & 0.88  & 0.78  & 0.73  & -0.09 & -0.35 & \multicolumn{1}{l}{High $\beta^{\Delta
    \mfb}$} & \multicolumn{1}{r}{5} & 1.40  & 0.95  & 1.05  & 0.86  & 0.98  & -0.42 & -1.72 \\
          & 5-$i$ & -0.60 & -0.50 & -0.26 & -0.60 & -0.26 &       &       &       & 5-$i$ & 0.09  & -0.07 & 0.15  & 0.04  & 0.19  &       &  \\
          & $t$-stat (5-$i$) & \textbf{-2.10} & \textbf{-2.05} & -1.05 & \textbf{-2.60} & -0.98 &       &       &       & $t$-stat (5-$i$) & 0.31  & -0.25 & 0.51  & 0.15  & 0.65  &       &  \\
    \midrule
    \midrule
    \bottomrule
    \end{tabular}%
    }
  \label{tab:double_sorts2}%
\end{table}%

\end{landscape}

\subsection{Portfolios Controlling for the Market Risk Premium, Market Capitalization and Trading Volume}

\begin{table}[!hp]
  \centering
  \caption{\textbf{Portfolio Sorts from Loadings on Common Fears, Common Good Fears, and Common Bad Fears Controlling for the Market Risk Premium}\\
  \small{Notes: This table shows value-weighted portfolios that we sort on loadings to: i) common fears in Panel A; ii) common good fears in Panel B; and iii) common bad fears in Panel C whilst controlling for the market risk premium.  In each Panel, we report monthly excess returns for quintile portfolios and the long-short portfolio that goes long the portfolio of stocks with a high loading on common fears and short the portfolio of stocks with low loadings to common fears. We also report the risk adjusted returns from the Fama-French 5 factor model and the Fama-French 5-factor model accounting for Momentum.}}
    \begin{tabular}{lrrrrrr}
    \toprule
    \midrule
    \midrule
       & \multicolumn{1}{l}{low $\beta^{\Delta\cf}$} &       &       &       \multicolumn{2}{r}{High $\beta^{\Delta\cf}$} &  \\
    \midrule
    \textbf{A:} Common Fears  & 1     & 2     & 3     & 4     & 5     & 5-1 \\
    \midrule
    mean (\%) & 1.23  & 0.99  & 1.03  & 0.89  & 0.83  & \textbf{-0.39} \\
    $t$-stat & 3.77  & 3.36  & 3.72  & 3.46  & 3.44  & \textbf{-2.73} \\
   $\alpha^{\text{FF5}}$  & 0.42  & 0.24  & 0.29  & 0.24  & 0.21  & -0.20 \\
    $t$-stat & 5.85  & 3.20  & 5.04  & 3.93  & 3.25  & -1.76 \\
   $\alpha^{\text{FF5 + MOM}}$ & 0.42  & 0.23  & 0.28  & 0.23  & 0.20  & \textbf{-0.22} \\
    $t$-stat & 5.82  & 3.18  & 4.98  & 3.83  & 3.55  & \textbf{-2.01} \\
    \midrule
    \midrule
    \textbf{B:} Common Good Fears & 1     & 2     & 3     & 4     & 5     & 5-1 \\
    \midrule
    mean (\%) & 1.17  & 1.04  & 1.01  & 0.95  & 0.83  & \textbf{-0.34} \\
    $t$-stat & 3.60  & 3.50  & 3.65  & 3.62  & 3.39  & \textbf{-2.23} \\
    $\alpha^{\text{FF5}}$   & 0.40  & 0.28  & 0.29  & 0.26  & 0.17  & -0.24 \\
    $t$-stat & 6.37  & 4.00  & 4.76  & 4.71  & 1.91  & -1.89 \\
    $\alpha^{\text{FF5 + MOM}}$ & 0.40  & 0.28  & 0.29  & 0.25  & 0.16  & \textbf{-0.24} \\
    $t$-stat & 6.25  & 3.91  & 4.71  & 4.89  & 1.96  & \textbf{-2.02} \\
    \midrule
    \midrule
    \textbf{C:} Common Bad Fears & 1     & 2     & 3     & 4     & 5     & 5-1 \\
    \midrule
    mean (\%) & 1.20  & 0.99  & 1.02  & 0.93  & 0.80  & \textbf{-0.39} \\
    $t$-stat & 3.76  & 3.30  & 3.78  & 3.59  & 3.25  & \textbf{-2.79} \\
     $\alpha^{\text{FF5}}$ & 0.39  & 0.22  & 0.29  & 0.28  & 0.19  & -0.20 \\
    $t$-stat & 6.41  & 4.04  & 5.25  & 4.32  & 2.42  & -1.84 \\
    $\alpha^{\text{FF5 + MOM}}$ & 0.39  & 0.22  & 0.29  & 0.27  & 0.18  & \textbf{-0.21} \\
    $t$-stat & 6.43  & 4.04  & 4.71  & 4.31  & 2.42  & \textbf{-2.06} \\
    \midrule
    \midrule
    \bottomrule
    \end{tabular}%
  \label{tab:single_sortsMKT}%
\end{table}%

\begin{table}[!hp]
  \centering
  \caption{\textbf{Portfolio Sorts from Loadings on Common Fears, Common Good Fears, and Common Bad Fears Controlling for Market Capitalization}\\
  \small{Notes: This table shows value-weighted portfolios that we sort on loadings to: i) common fears in Panel A; ii) common good fears in Panel B; and iii) common bad fears in Panel C whilst controlling for market capitalization.  In each Panel, we report monthly excess returns for quintile portfolios and the long-short portfolio that goes long the portfolio of stocks with a high loading on common fears and short the portfolio of stocks with low loadings to common fears. We also report the risk adjusted returns from the Fama-French 5 factor model and the Fama-French 5-factor model accounting for Momentum.}}
    \begin{tabular}{lrrrrrr}
    \toprule
    \midrule
    \midrule
       & \multicolumn{1}{l}{low $\beta^{\Delta\cf}$} &       &       &       \multicolumn{2}{r}{High $\beta^{\Delta\cf}$} &  \\
    \midrule
    \textbf{A:} Common Fears  & 1     & 2     & 3     & 4     & 5     & 5-1 \\
    \midrule
    mean (\%) & 1.21  & 1.09  & 0.97  & 0.94  & 0.78  & \textbf{-0.43} \\
    $t$-stat & 3.46  & 3.54  & 3.68  & 3.75  & 3.38  & \textbf{-2.42} \\
    $\alpha^{\text{FF5}}$  & 0.40  & 0.33  & 0.30  & 0.31  & 0.16  & \textbf{-0.23} \\
    $t$-stat & 5.09  & 6.36  & 4.38  & 4.27  & 2.29  & \textbf{-1.95} \\
    $\alpha^{\text{FF5 + MOM}}$ & 0.40  & 0.33  & 0.29  & 0.30  & 0.15  & \textbf{-0.25} \\
    $t$-stat & 5.29  & 6.30  & 4.71  & 4.35  & 2.46  & \textbf{-2.44} \\
    \midrule
    \midrule
    \textbf{B:} Common Good Fears & 1     & 2     & 3     & 4     & 5     & 5-1 \\
    \midrule
    mean (\%) & 1.11  & 1.12  & 1.01  & 0.93  & 0.82  & -0.28 \\
    $t$-stat & 3.09  & 3.51  & 3.62  & 3.78  & 3.68  & -1.38 \\
    $\alpha^{\text{FF5}}$  & 0.33  & 0.35  & 0.30  & 0.30  & 0.20  & -0.13 \\
    $t$-stat & 3.51  & 5.97  & 5.04  & 4.34  & 1.99  & -0.73 \\
    $\alpha^{\text{FF5 + MOM}}$ & 0.33  & 0.35  & 0.30  & 0.29  & 0.18  & -0.15 \\
    $t$-stat & 3.63  & 6.08  & 5.01  & 4.37  & 2.15  & -0.94 \\
    \midrule
    \midrule
    \textbf{C:} Common Bad Fears & 1     & 2     & 3     & 4     & 5     & 5-1 \\
    \midrule
    mean (\%) & 1.23  & 1.06  & 0.98  & 0.96  & 0.79  & \textbf{-0.44} \\
    $t$-stat & 3.61  & 3.63  & 3.73  & 3.68  & 3.21  & \textbf{-2.78} \\
    $\alpha^{\text{FF5}}$ & 0.41  & 0.33  & 0.31  & 0.30  & 0.16  & \textbf{-0.26} \\
    $t$-stat & 5.57  & 5.81  & 5.29  & 4.32  & 2.21  & \textbf{-2.17} \\
    $\alpha^{\text{FF5 + MOM}}$ & 0.42  & 0.33  & 0.30  & 0.29  & 0.14  & \textbf{-0.27} \\
    $t$-stat & 5.75  & 5.59  & 5.20  & 4.54  & 2.21  & \textbf{-2.49} \\
    \midrule
    \midrule
    \bottomrule
    \end{tabular}%
  \label{tab:single_sortsMKCAP}%
\end{table}%

\begin{table}[!hp]
  \centering
  \caption{\textbf{Portfolio Sorts from Loadings on Common Fears, Common Good Fears, and Common Bad Fears Controlling for Trading Volume}\\
  \small{Notes: This table shows value-weighted portfolios that we sort on loadings to: i) common fears in Panel A; ii) common good fears in Panel B; and iii) common bad fears in Panel C whilst controlling for trading volume.  In each Panel, we report monthly excess returns for quintile portfolios and the long-short portfolio that goes long the portfolio of stocks with a high loading on common fears and short the portfolio of stocks with low loadings to common fears. We also report the risk adjusted returns from the Fama-French 5 factor model and the Fama-French 5-factor model accounting for Momentum.}}
    \begin{tabular}{lrrrrrr}
    \toprule
    \midrule
    \midrule
       & \multicolumn{1}{l}{low $\beta^{\Delta\cf}$} &       &       &       \multicolumn{2}{r}{High $\beta^{\Delta\cf}$} &  \\
    \midrule
    \textbf{A:} Common Fears  & 1     & 2     & 3     & 4     & 5     & 5-1 \\
    \midrule
    mean (\%) & 1.21  & 1.14  & 0.94  & 0.95  & 0.78  & \textbf{-0.43} \\
    $t$-stat & 3.44  & 3.74  & 3.67  & 3.81  & 3.37  & \textbf{-2.42} \\
    $\alpha^{\text{FF5}}$  & 0.40  & 0.38  & 0.28  & 0.32  & 0.16  & \textbf{-0.23} \\
    $t$-stat & 4.97  & 6.01  & 4.78  & 4.71  & 2.25  & \textbf{-1.87} \\
    $\alpha^{\text{FF5 + MOM}}$ & 0.40  & 0.38  & 0.28  & 0.31  & 0.15  & \textbf{-0.25} \\
    $t$-stat & 5.08  & 5.96  & 4.61  & 5.05  & 2.48  & \textbf{-2.28} \\
    \midrule
    \midrule
    \textbf{B:} Common Good Fears & 1     & 2     & 3     & 4     & 5     & 5-1 \\
    \midrule
    mean (\%) & 1.10  & 1.12  & 1.07  & 0.94  & 0.79  & -0.31 \\
    $t$-stat & 3.09  & 3.47  & 3.89  & 3.93  & 3.67  & -1.44 \\
    $\alpha^{\text{FF5}}$  & 0.32  & 0.35  & 0.36  & 0.33  & 0.20  & -0.12 \\
    $t$-stat & 2.95  & 5.06  & 6.91  & 5.35  & 1.94  & -0.65 \\
   $\alpha^{\text{FF5 + MOM}}$ & 0.32  & 0.36  & 0.35  & 0.33  & 0.18  & -0.14 \\
    $t$-stat & 3.07  & 5.11  & 6.99  & 5.32  & 2.11  & -0.87 \\
    \midrule
    \midrule
    \textbf{C:} Common Bad Fears & 1     & 2     & 3     & 4     & 5     & 5-1 \\
    \midrule
    mean (\%) & 1.21  & 1.10  & 0.97  & 0.92  & 0.79  & \textbf{-0.42} \\
    $t$-stat & 3.60  & 3.82  & 3.74  & 3.56  & 3.28  & \textbf{-2.66} \\
    $\alpha^{\text{FF5}}$  & 0.41  & 0.40  & 0.30  & 0.27  & 0.17  & \textbf{-0.24} \\
    $t$-stat & 5.51  & 5.51  & 4.84  & 3.98  & 2.45  & \textbf{-2.17} \\
    $\alpha^{\text{FF5 + MOM}}$ & 0.41  & 0.40  & 0.29  & 0.26  & 0.15  & \textbf{-0.25} \\
    $t$-stat & 5.53  & 5.22  & 4.85  & 4.37  & 2.59  & \textbf{-2.45} \\
    \midrule
    \midrule
    \bottomrule
    \end{tabular}
  \label{tab:single_sortsVOL}%
\end{table}%

\begin{table}[!hp]
  \centering
  \caption{\textbf{Fama-MacBeth Analysis: 5x5 Common Fears / Market Risk Premium Portfolios} \\
  \small{This table shows the Fama-MacBeth two-pass regression analysis for the 25 portfolios we construct on loadings to: common fears, $\beta^{\Delta\cf}$,  and the market risk premium in columns 1--6; common good fears, $\beta^{\Delta\cfg} $ and the market risk premium, in columns 7--12; and common bad fears, $\beta^{\Delta\cfb} $, and the market risk premium in columns 13--18. $\lambda_{\cf}$ denotes the corresponding risk premium estimates for common fears, common good fears and common bad fears. Results in columns 1, 7, 13 control for the Fama-French 5 factor model. MKT is the market risk premium; SMB is small-minus-big; HML is high-minus-low; RMW is robust-minus-weak; CMA is conservative minus aggressive). Meanwhile columns 2--5, 8--11, and 14--17 control for the variance risk premium (VRP), momentum (MOM), common idiosyncratic volatility (CIV) \citep{herskovic2016common}, and liquidity (LIQ) \citep{pastor2003liquidity}, respectively. Columns 6, 12, 18 add all aforementioned factors to the Fama-French 5 factor model.  Below risk premia estimates we report Newey and West $t$-statistics with 12 lags that adjust for errors in variables as in \cite{shanken1992estimation}. $ \bar{R}^2 $ is the adjusted R-squared.}
  }
      \adjustbox{max height=9.70in ,max width=\textwidth, keepaspectratio}{
    \begin{tabular}{lrrrrrrrrrrrrrrrrrr}
    \toprule
        \midrule
    \midrule
          & \multicolumn{6}{c}{5 X 5 $\beta^{\Delta\cf} $ / $\beta^{\text{MKT}} $ portfolios} & \multicolumn{6}{c}{5 X 5 $\beta^{\Delta\cfg} $ / $\beta^{\text{MKT}} $ portfolios} & \multicolumn{6}{c}{5 X 5 $\beta^{\Delta\cfb} $ / $\beta^{\text{MKT}} $ portfolios} \\
    \midrule
          & 1     & 2     & 3     & 4     & 5     & 6     & 7     & 8     & 9     & 10    & 11    & 12    & 13    & 14    & 15    & 16    & 17    & 18 \\
    \midrule
    $\lambda_{0}$ & 0.81  & 0.71  & 1.31  & 0.89  & 0.92  & 1.37  & 1.26  & 1.26  & 1.22  & 1.26  & 1.26  & 1.21  & 0.79  & 0.73  & 1.48  & 1.02  & 0.89  & 1.39 \\
    $t$-stat & 4.26  & 3.10  & 5.40  & 4.38  & 4.96  & 5.38  & 4.81  & 4.70  & 4.78  & 4.78  & 4.87  & 4.75  & 3.88  & 3.37  & 6.36  & 4.98  & 4.40  & 5.46 \\
    $\lambda_{\cf}$ & \textbf{-0.19} & \textbf{-0.19} & \textbf{-0.18} & \textbf{-0.19} & \textbf{-0.14} & \textbf{-0.14} & 0.15  & 0.15  & 0.13  & 0.15  & 0.15  & 0.12  & \textbf{-0.16} & \textbf{-0.19} & \textbf{-0.15} & -0.12 & \textbf{-0.15} & -0.15 \\
    $t$-stat & \textbf{-2.56} & \textbf{-2.57} & \textbf{-2.47} & \textbf{-2.56} & \textbf{-2.15} & \textbf{-2.09} & 1.39  & 1.30  & 1.35  & 1.39  & 1.39  & 1.23  & \textbf{-2.16} & \textbf{-2.36} & \textbf{-1.97} & -1.56 & \textbf{-1.99} & -1.77 \\
    $\lambda_{MKT}$ & 0.21  & 0.37  & -0.37 & 0.09  & 0.03  & -0.54 & -0.56 & -0.56 & -0.54 & -0.56 & -0.56 & -0.54 & 0.38  & 0.58  & -0.37 & 0.08  & 0.23  & -0.22 \\
    $t$-stat & 0.50  & 0.96  & -0.73 & 0.21  & 0.06  & -1.10 & -1.50 & -1.48 & -1.49 & -1.49 & -1.51 & -1.47 & 0.90  & 1.44  & -0.76 & 0.18  & 0.51  & -0.43 \\
    $\lambda_{SMB}$ & 0.23  & 0.02  & 0.34  & 0.26  & 0.21  & 0.48  & 0.87  & 0.85  & 0.93  & 0.87  & 0.88  & 0.92  & -0.13 & -0.64 & -0.07 & -0.11 & -0.18 & -0.31 \\
    $t$-stat & 0.58  & 0.05  & 0.81  & 0.64  & 0.53  & 1.16  & 2.19  & 2.20  & 2.18  & 2.23  & 2.27  & 2.20  & -0.37 & -2.40 & -0.19 & -0.31 & -0.54 & -1.11 \\
    $\lambda_{HML}$ & -0.46 & -0.51 & -0.72 & -0.57 & -0.60 & -0.76 & 1.08  & 1.08  & 1.11  & 1.08  & 1.11  & 1.15  & -0.17 & -0.31 & -0.44 & -0.49 & -0.21 & -0.57 \\
    $t$-stat & -1.23 & -1.32 & -1.81 & -1.45 & -1.52 & -1.84 & 2.17  & 2.15  & 2.17  & 2.18  & 2.11  & 2.34  & -0.48 & -0.83 & -1.17 & -1.30 & -0.58 & -1.42 \\
    $\lambda_{RMW}$ & 0.11  & 0.18  & 0.25  & 0.20  & 0.32  & 0.36  & 0.76  & 0.76  & 0.76  & 0.75  & 0.79  & 0.80  & 0.03  & -0.08 & 0.13  & 0.21  & 0.15  & 0.10 \\
    $t$-stat & 0.38  & 0.66  & 0.93  & 0.76  & 1.19  & 1.36  & 2.90  & 2.96  & 2.90  & 2.77  & 2.78  & 2.91  & 0.09  & -0.25 & 0.47  & 0.75  & 0.54  & 0.30 \\
    $\lambda_{CMA}$ & -0.29 & -0.28 & -0.28 & -0.30 & -0.36 & -0.35 & -0.12 & -0.11 & -0.08 & -0.13 & -0.12 & -0.07 & -0.36 & -0.36 & -0.35 & -0.49 & -0.39 & -0.39 \\
    $t$-stat & -1.56 & -1.50 & -1.50 & -1.63 & -2.02 & -1.90 & -0.34 & -0.33 & -0.23 & -0.37 & -0.34 & -0.20 & -1.65 & -1.65 & -1.61 & -2.21 & -1.86 & -1.80 \\
    $\lambda_{VRP}$ &       & 0.16  &       &       &       & -0.04 &       & -0.02 &       &       &       & -0.01 &       & 0.33  &       &       &       & 0.21 \\
    $t$-stat &       & 1.00  &       &       &       & -0.32 &       & -0.15 &       &       &       & -0.09 &       & 2.10  &       &       &       & 1.42 \\
    $\lambda_{MOM}$ &       &       & -1.50 &       &       & -0.95 &       &       & -0.29 &       &       & -0.31 &       &       & -2.75 &       &       & -2.41 \\
    $t$-stat &       &       & -2.08 &       &       & -1.52 &       &       & -0.41 &       &       & -0.45 &       &       & -2.94 &       &       & -2.87 \\
    $\lambda_{CIV}$ &       &       &       & -0.23 &       & -0.11 &       &       &       & 0.00  &       & -0.01 &       &       &       & -0.48 &       & -0.25 \\
    $t$-stat &       &       &       & -1.10 &       & -0.57 &       &       &       & 0.00  &       & -0.01 &       &       &       & -2.22 &       & -1.24 \\
    $\lambda_{LIQ}$ &       &       &       &       & -0.21 & -0.18 &       &       &       &       & -0.01 & 0.00  &       &       &       &       & -0.16 & 0.03 \\
    $t$-stat &       &       &       &       & -1.49 & -1.22 &       &       &       &       & -0.10 & -0.02 &       &       &       &       & -1.33 & 0.25 \\
    \midrule
    $ \bar{R}^2 $ & 0.367 & 0.363 & 0.518 & 0.361 & 0.487 & 0.499 & 0.640 & 0.620 & 0.628 & 0.620 & 0.621 & 0.557 & 0.232 & 0.370 & 0.549 & 0.265 & 0.276 & 0.498 \\
    \midrule
    \midrule
    \bottomrule
    \end{tabular}%
    }
  \label{tab:5x5MKT}%
\end{table}%

\begin{table}[!hp]
  \centering
  \caption{\textbf{Fama-MacBeth Analysis: 5x5 Common Fears / Market Capitalization Portfolios} \\
  \small{This table shows the Fama-MacBeth two-pass regression analysis for the 25 portfolios we construct on loadings to: common fears, $\beta^{\Delta\cf}$,  and market capitalization (MKCAP) in columns 1--6; common good fears, $\beta^{\Delta\cfg} $ and market MKCAP in columns 7--12; and common bad fears, $\beta^{\Delta\cfb} $, and MKCAP in columns 13--18. $\lambda_{\cf}$ denotes the corresponding risk premium estimates for common fears, common good fears and common bad fears. Results in columns 1, 7, 13 control for the Fama-French 5 factor model. MKT is the market risk premium; SMB is small-minus-big; HML is high-minus-low; RMW is robust-minus-weak; CMA is conservative minus aggressive). Meanwhile columns 2--5, 8--11, and 14--17 control for the variance risk premium (VRP), momentum (MOM), common idiosyncratic volatility (CIV) \citep{herskovic2016common}, and liquidity (LIQ) \citep{pastor2003liquidity}, respectively. Columns 6, 12, 18 add all aforementioned factors to the Fama-French 5 factor model.  Below risk premia estimates we report Newey and West $t$-statistics with 12 lags that adjust for errors in variables as in \cite{shanken1992estimation}. $ \bar{R}^2 $ is the adjusted R-squared.}
  }
      \adjustbox{max height=9.70in ,max width=\textwidth, keepaspectratio}{
    \begin{tabular}{lrrrrrrrrrrrrrrrrrr}
    \toprule
        \midrule
    \midrule
          & \multicolumn{6}{c}{5 X 5 $\beta^{\Delta\cf} $ / $\beta^{\text{MKCAP}} $ portfolios} & \multicolumn{6}{c}{5 X 5 $\beta^{\Delta\cfg} $ / $\beta^{\text{MKCAP}} $ portfolios} & \multicolumn{6}{c}{5 X 5 $\beta^{\Delta\cfb} $ / $\beta^{\text{MKCAP}} $ portfolios} \\
    \midrule
          & 1     & 2     & 3     & 4     & 5     & 6     & 7     & 8     & 9     & 10    & 11    & 12    & 13    & 14    & 15    & 16    & 17    & 18 \\
    \midrule
    $\lambda_{0}$ & 0.20  & 0.20  & 0.40  & 0.20  & 0.20  & 0.45  & 0.26  & 0.36  & 0.51  & 0.43  & 0.25  & 0.73  & 0.00  & -0.10 & 0.37  & 0.00  & 0.03  & 0.19 \\
    $t$-stat & 0.76  & 0.75  & 1.55  & 0.79  & 0.75  & 1.42  & 0.94  & 1.34  & 1.27  & 1.60  & 0.93  & 1.82  & 0.01  & -0.27 & 1.10  & 0.00  & 0.09  & 0.51 \\
    $\lambda_{\cf}$ & \textbf{-0.23} & \textbf{-0.23} & \textbf{-0.20} & \textbf{-0.23} & \textbf{-0.23} & \textbf{-0.20} & -0.19 & -0.12 & -0.11 & -0.13 & -0.23 & -0.06 & \textbf{-0.27} & \textbf{-0.26} & \textbf{-0.20} & \textbf{-0.26} & \textbf{-0.30} & \textbf{-0.20} \\
    $t$-stat & \textbf{-2.87} & \textbf{-3.18} & \textbf{-2.62} & \textbf{-2.87} & \textbf{-2.80} & \textbf{-2.58} & -1.53 & -0.83 & -0.59 & -1.05 & -1.87 & -0.36 & \textbf{-2.74} & \textbf{-2.63} & \textbf{-2.18} & \textbf{-2.79} & \textbf{-3.10} & \textbf{-2.39} \\
    $\lambda_{MKT}$ & 0.96  & 0.96  & 0.74  & 0.95  & 0.96  & 0.69  & 0.88  & 0.77  & 0.63  & 0.65  & 0.81  & 0.31  & 1.15  & 1.25  & 0.76  & 1.15  & 1.08  & 0.94 \\
    $t$-stat & 2.46  & 2.45  & 2.06  & 2.45  & 2.42  & 1.80  & 2.28  & 1.90  & 1.40  & 1.67  & 2.02  & 0.63  & 2.31  & 2.45  & 1.56  & 2.33  & 2.14  & 1.86 \\
    $\lambda_{SMB}$ & 0.19  & 0.19  & 0.18  & 0.20  & 0.19  & 0.19  & 0.24  & 0.18  & 0.27  & 0.30  & 0.24  & 0.30  & 0.26  & 0.21  & 0.22  & 0.26  & 0.26  & 0.18 \\
    $t$-stat & 1.00  & 0.94  & 0.95  & 1.00  & 1.00  & 0.96  & 1.34  & 0.98  & 1.49  & 1.63  & 1.32  & 1.50  & 1.31  & 1.05  & 1.15  & 1.28  & 1.30  & 0.95 \\
    $\lambda_{HML}$ & -0.10 & -0.09 & -0.28 & -0.06 & -0.10 & -0.38 & -0.14 & -0.15 & -0.35 & -0.18 & -0.23 & -0.52 & -0.21 & -0.07 & -0.44 & -0.16 & -0.18 & -0.23 \\
    $t$-stat & -0.25 & -0.24 & -0.67 & -0.14 & -0.25 & -0.68 & -0.59 & -0.64 & -1.05 & -0.77 & -0.90 & -1.36 & -0.73 & -0.26 & -1.34 & -0.48 & -0.63 & -0.60 \\
    $\lambda_{RMW}$ & 0.02  & 0.02  & 0.16  & 0.03  & 0.02  & 0.20  & -0.13 & 0.04  & -0.09 & -0.03 & -0.11 & 0.08  & 0.35  & 0.20  & 0.48  & 0.36  & 0.47  & 0.33 \\
    $t$-stat & 0.08  & 0.05  & 0.51  & 0.11  & 0.08  & 0.61  & -0.62 & 0.17  & -0.44 & -0.13 & -0.52 & 0.35  & 1.12  & 0.62  & 1.53  & 1.12  & 1.40  & 0.90 \\
    $\lambda_{CMA}$ & -0.41 & -0.40 & -0.46 & -0.35 & -0.41 & -0.52 & -0.30 & -0.22 & -0.15 & -0.24 & -0.33 & -0.08 & -0.42 & -0.21 & -0.48 & -0.38 & -0.33 & -0.25 \\
    $t$-stat & -1.32 & -1.17 & -1.50 & -0.99 & -1.30 & -1.32 & -0.72 & -0.55 & -0.33 & -0.59 & -0.78 & -0.18 & -1.84 & -0.82 & -2.21 & -1.51 & -1.49 & -0.97 \\
    $\lambda_{VRP}$ &       & 0.04  &       &       &       & 0.01  &       & 0.17  &       &       &       & 0.13  &       & 0.39  &       &       &       & 0.37 \\
    $t$-stat &       & 0.18  &       &       &       & 0.06  &       & 1.16  &       &       &       & 0.85  &       & 2.28  &       &       &       & 2.19 \\
    $\lambda_{MOM}$ &       &       & -0.77 &       &       & -0.79 &       &       & -1.02 &       &       & -1.22 &       &       & -1.05 &       &       & -1.17 \\
    $t$-stat &       &       & -1.06 &       &       & -1.10 &       &       & -1.11 &       &       & -1.29 &       &       & -1.67 &       &       & -1.80 \\
    $\lambda_{CIV}$ &       &       &       & -0.82 &       & -0.11 &       &       &       & -0.23 &       & -0.18 &       &       &       & -0.68 &       & -0.59 \\
    $t$-stat &       &       &       & -0.73 &       & -0.08 &       &       &       & -2.13 &       & -1.67 &       &       &       & -0.40 &       & -0.35 \\
    $\lambda_{LIQ}$ &       &       &       &       & 0.04  & -0.01 &       &       &       &       & -0.09 & -0.13 &       &       &       &       & -0.08 & 0.00 \\
    $t$-stat &       &       &       &       & 0.33  & -0.05 &       &       &       &       & -0.87 & -1.04 &       &       &       &       & -0.87 & 0.03 \\
    \midrule
    $ \bar{R}^2 $ & 0.742 & 0.728 & 0.748 & 0.730 & 0.728 & 0.700 & 0.717 & 0.734 & 0.734 & 0.773 & 0.751 & 0.830 & 0.632 & 0.733 & 0.687 & 0.612 & 0.645 & 0.731 \\
    \midrule
    \midrule
    \bottomrule
    \end{tabular}%
    }
  \label{tab:5x5MKCAP}%
\end{table}%

\begin{table}[!hp]
  \centering
  \caption{\textbf{Fama-MacBeth Analysis: 5x5 Common Fears / Trading Volume Portfolios} \\
  \small{This table shows the Fama-MacBeth two-pass regression analysis for the 25 portfolios we construct on loadings to: common fears, $\beta^{\Delta\cf}$,  and trading volume (VOLUM) in columns 1--6; common good fears, $\beta^{\Delta\cfg} $ and VOLUM, in columns 7--12; and common bad fears, $\beta^{\Delta\cfb} $, and VOLUM in columns 13--18. $\lambda_{\cf}$ denotes the corresponding risk premium estimates for common fears, common good fears and common bad fears. Results in columns 1, 7, 13 control for the Fama-French 5 factor model. MKT is the market risk premium; SMB is small-minus-big; HML is high-minus-low; RMW is robust-minus-weak; CMA is conservative minus aggressive). Meanwhile columns 2--5, 8--11, and 14--17 control for the variance risk premium (VRP), momentum (MOM), common idiosyncratic volatility (CIV) \citep{herskovic2016common}, and liquidity (LIQ) \citep{pastor2003liquidity}, respectively. Columns 6, 12, 18 add all aforementioned factors to the Fama-French 5 factor model.  Below risk premia estimates we report Newey and West $t$-statistics with 12 lags that adjust for errors in variables as in \cite{shanken1992estimation}. $ \bar{R}^2 $ is the adjusted R-squared.}
  }
      \adjustbox{max height=9.70in ,max width=\textwidth, keepaspectratio}{
    \begin{tabular}{lrrrrrrrrrrrrrrrrrr}
    \toprule
        \midrule
    \midrule
          & \multicolumn{6}{c}{5 X 5 $\beta^{\Delta\cf} $ / $\beta^{\text{VOLUM}} $ portfolios} & \multicolumn{6}{c}{5 X 5 $\beta^{\Delta\cfg} $ / $\beta^{\text{VOLUM}} $ portfolios} & \multicolumn{6}{c}{5 X 5 $\beta^{\Delta\cfb} $ / $\beta^{\text{VOLUM}} $ portfolios} \\
    \midrule
          & 1     & 2     & 3     & 4     & 5     & 6     & 7     & 8     & 9     & 10    & 11    & 12    & 13    & 14    & 15    & 16    & 17    & 18 \\
    \midrule
    $\lambda_{0}$ & 0.36  & 0.33  & 0.80  & 0.33  & 0.41  & 0.88  & 0.28  & 0.17  & 0.51  & 0.31  & 0.30  & 0.41  & 0.42  & 0.33  & 0.54  & 0.48  & 0.40  & 0.42 \\
    $t$-stat & 1.40  & 1.26  & 2.98  & 1.29  & 1.50  & 2.99  & 1.02  & 0.53  & 1.53  & 1.22  & 1.19  & 1.17  & 1.46  & 1.16  & 1.58  & 1.65  & 1.40  & 1.23 \\
    $\lambda_{\cf}$ & -0.15 & -0.10 & -0.08 & -0.16 & -0.13 & -0.03 & -0.19 & \textbf{-0.29} & -0.13 & -0.18 & -0.18 & -0.23 & -0.19 & -0.21 & -0.17 & -0.17 & -0.21 & -0.20 \\
    $t$-stat & -1.70 & -1.10 & -0.92 & -1.80 & -1.48 & -0.34 & -1.58 & \textbf{-2.43} & -0.96 & -1.52 & -1.51 & -1.89 & -1.61 & -1.79 & -1.55 & -1.44 & -1.81 & -1.82 \\
    $\lambda_{MKT}$ & 0.83  & 0.88  & 0.32  & 0.89  & 0.77  & 0.28  & 0.89  & 0.98  & 0.63  & 0.84  & 0.85  & 0.70  & 0.68  & 0.76  & 0.56  & 0.62  & 0.65  & 0.63 \\
    $t$-stat & 1.99  & 2.16  & 0.83  & 2.19  & 1.67  & 0.68  & 2.48  & 2.70  & 1.29  & 2.20  & 2.14  & 1.42  & 1.47  & 1.71  & 1.22  & 1.31  & 1.38  & 1.47 \\
    $\lambda_{SMB}$ & 0.32  & 0.26  & 0.32  & 0.31  & 0.32  & 0.28  & 0.17  & 0.06  & 0.23  & 0.18  & 0.17  & 0.12  & 0.40  & 0.38  & 0.40  & 0.41  & 0.42  & 0.40 \\
    $t$-stat & 1.56  & 1.34  & 1.58  & 1.54  & 1.56  & 1.41  & 0.88  & 0.31  & 1.11  & 0.88  & 0.89  & 0.58  & 2.09  & 1.98  & 2.06  & 2.11  & 2.17  & 2.02 \\
    $\lambda_{HML}$ & -0.04 & -0.12 & -0.48 & 0.00  & -0.10 & -0.61 & 0.13  & 0.05  & -0.16 & 0.09  & 0.09  & -0.26 & -0.40 & -0.46 & -0.48 & -0.40 & -0.47 & -0.53 \\
    $t$-stat & -0.09 & -0.23 & -0.88 & 0.00  & -0.19 & -0.97 & 0.49  & 0.19  & -0.41 & 0.31  & 0.28  & -0.57 & -0.78 & -0.87 & -0.92 & -0.76 & -0.88 & -0.98 \\
    $\lambda_{RMW}$ & -0.92 & -0.87 & -0.52 & -1.17 & -0.94 & -0.85 & -0.14 & -0.07 & 0.09  & -0.12 & -0.19 & 0.18  & -0.39 & -0.17 & -0.37 & -0.45 & -0.14 & -0.07 \\
    $t$-stat & -2.14 & -2.09 & -1.35 & -2.50 & -2.13 & -1.92 & -0.30 & -0.14 & 0.16  & -0.26 & -0.40 & 0.36  & -0.85 & -0.38 & -0.79 & -0.97 & -0.28 & -0.14 \\
    $\lambda_{CMA}$ & 0.14  & 0.11  & -0.17 & 0.18  & 0.13  & -0.20 & -0.01 & -0.16 & -0.06 & -0.04 & -0.02 & -0.21 & -0.47 & -0.46 & -0.49 & -0.41 & -0.44 & -0.41 \\
    $t$-stat & 0.41  & 0.31  & -0.52 & 0.54  & 0.37  & -0.61 & -0.03 & -0.36 & -0.13 & -0.08 & -0.03 & -0.47 & -1.54 & -1.51 & -1.61 & -1.30 & -1.46 & -1.39 \\
    $\lambda_{VRP}$ &       & 0.13  &       &       &       & 0.11  &       & 0.20  &       &       &       & 0.25  &       & 0.15  &       &       &       & 0.15 \\
    $t$-stat &       & 1.12  &       &       &       & 0.91  &       & 1.29  &       &       &       & 1.49  &       & 1.27  &       &       &       & 1.26 \\
    $\lambda_{MOM}$ &       &       & -1.41 &       &       & -1.65 &       &       & -1.14 &       &       & -1.15 &       &       & -0.34 &       &       & -0.29 \\
    $t$-stat &       &       & -1.76 &       &       & -2.04 &       &       & -1.25 &       &       & -1.39 &       &       & -0.42 &       &       & -0.34 \\
    $\lambda_{CIV}$ &       &       &       & 1.17  &       & 1.83  &       &       &       & -0.50 &       & -0.45 &       &       &       & -1.46 &       & -0.80 \\
    $t$-stat &       &       &       & 0.98  &       & 1.35  &       &       &       & -0.33 &       & -0.34 &       &       &       & -0.97 &       & -0.54 \\
    $\lambda_{LIQ}$ &       &       &       &       & -0.03 & -0.07 &       &       &       &       & -0.01 & 0.01  &       &       &       &       & -0.09 & -0.05 \\
    $t$-stat &       &       &       &       & -0.25 & -0.68 &       &       &       &       & -0.08 & 0.11  &       &       &       &       & -1.00 & -0.62 \\
\midrule
    $ \bar{R}^2 $ & 0.716 & 0.749 & 0.781 & 0.716 & 0.708 & 0.834 & 0.681 & 0.717 & 0.707 & 0.665 & 0.669 & 0.715 & 0.656 & 0.680 & 0.642 & 0.650 & 0.674 & 0.648 \\
    \midrule
    \midrule
    \bottomrule
    \end{tabular}%
    }
  \label{tab:5x5VOLUM}%
\end{table}%

\newpage
\subsection{Adding Alternative Alternative Test Assets II}

\begin{table}[!hp]
  \centering
  \caption{\textbf{Fama-MacBeth Analysis; Alternative Test Assets: Size/Investment, Size,/Book to Market, Size/Market Risk Premium, Size/Operating Profit, Size/Residual Variance, Size/Total Variance} \\
  \small{This table shows the Fama-MacBeth two-pass regression analysis for the 25 portfolios that sort on: i) size/investment (ME/INV); ii) size/book-to-market (ME/BM); iii) size/market risk premium (ME/MKT); iv) size/operating profit (ME/OP); v) size/residual variance (ME/RES VAR); and vi) size/total variance (ME/VAR), and the 25 respective portfolios we construct on loadings to: common fears, $\beta^{\Delta\cf}$,  and market fears, $\beta^{\Delta\mf}$ ,in columns 1--5; common good fears, $\beta^{\Delta\cfg} $ and good market fears, $\beta^{\Delta\mfg}$, in columns 6--10; and common bad fears, $\beta^{\Delta\cfb} $, and bad market fears,  $\beta^{\Delta\mfb}$, in columns 11--15. $\lambda_{\cf}$ denotes the corresponding risk premium estimates for common fears, common good fears and common bad fears. All results control initially for the Fama-French 5 factors.  MKT is the market risk premium; SMB is small-minus-big; HML is high-minus-low; RMW is robust-minus-weak; CMA is conservative minus aggressive). Results in columns 1--4, 6--9, 11--14 use the Fama-French 5 factors plus: the variance risk premium (VRP); momentum (MOM), common idiosyncratic volatility (CIV) \citep{herskovic2016common}; and liquidity (LIQ) \citep{pastor2003liquidity}. Columns 5, 10, 15 add all aforementioned additional controls to the Fama-French 5 factor model. Below risk premia estimates we report Newey and West $t$-statistics with 12 lags that adjust for errors in variables as in \cite{shanken1992estimation}. $ \bar{R}^2 $ is the adjusted R-squared.}
  }
      \adjustbox{max height=9.70in ,max width=\textwidth, keepaspectratio}{
    \begin{tabular}{lrrrrrrrrrrrrrrr}
    \toprule
    \midrule
    \midrule
    & \multicolumn{15}{c}{ 25 ME/INV, 25 ME/BM, 25 ME/MKT, 25 ME/OP 25 ME/RES VAR, 25 ME/VAR,  Plus 25:}\\
          & \multicolumn{5}{c}{ $\beta^{\Delta\cf} / \beta^{\Delta\mf}$ } & \multicolumn{5}{c}{$\beta^{\Delta\cfg} / \beta^{\Delta\mfg}$ } & \multicolumn{5}{c}{$\beta^{\Delta\cfb} / \beta^{\Delta\mfb}$ } \\
    \midrule
          & 1     & 2     & 3     & 4     & 5     & 6     & 7     & 8     & 9     & 10    & 11    & 12    & 13    & 14    & 15 \\
    \midrule
    $\lambda_{0}$ & 0.73  & 0.75  & 0.72  & 0.75  & 0.80  & 0.67  & 0.72  & 0.68  & 0.66  & 0.72  & 0.77  & 0.78  & 0.76  & 0.83  & 0.85 \\
    $t$-stat & 2.66  & 3.61  & 3.06  & 2.75  & 3.80  & 2.51  & 3.09  & 2.68  & 2.49  & 2.92  & 2.75  & 3.65  & 3.00  & 2.95  & 4.11 \\
    $\lambda_{\cf}$ & \textbf{-0.16} & \textbf{-0.16} & \textbf{-0.16} & \textbf{-0.16} & \textbf{-0.14} & -0.13 & -0.13 & -0.13 & -0.12 & -0.10 & \textbf{-0.16} & \textbf{-0.15} & \textbf{-0.15} & \textbf{-0.16} & \textbf{-0.14} \\
    $t$-stat & \textbf{-2.43} & \textbf{-2.34} & \textbf{-2.37} & \textbf{-2.35} & \textbf{-2.28} & -1.81 & -1.80 & -1.70 & -1.55 & -1.44 & \textbf{-2.25} & \textbf{-2.14} & \textbf{-2.16} & \textbf{-2.21} & \textbf{-2.15} \\
    $\lambda_{MKT}$ & 0.29  & 0.27  & 0.30  & 0.27  & 0.21  & 0.36  & 0.30  & 0.35  & 0.36  & 0.30  & 0.25  & 0.24  & 0.26  & 0.18  & 0.16 \\
    $t$-stat & 0.72  & 0.81  & 0.80  & 0.65  & 0.68  & 1.05  & 0.94  & 1.06  & 1.06  & 1.02  & 0.58  & 0.74  & 0.68  & 0.41  & 0.50 \\
    $\lambda_{SMB}$ & 0.17  & 0.17  & 0.17  & 0.17  & 0.17  & 0.18  & 0.17  & 0.18  & 0.18  & 0.18  & 0.18  & 0.18  & 0.18  & 0.18  & 0.19 \\
    $t$-stat & 1.00  & 1.02  & 1.01  & 1.00  & 1.04  & 1.05  & 1.04  & 1.06  & 1.05  & 1.08  & 1.04  & 1.06  & 1.04  & 1.06  & 1.11 \\
    $\lambda_{HML}$ & -0.25 & -0.25 & -0.25 & -0.22 & -0.23 & -0.19 & -0.20 & -0.19 & -0.18 & -0.19 & -0.26 & -0.26 & -0.26 & -0.23 & -0.23 \\
    $t$-stat & -1.09 & -1.07 & -1.09 & -0.98 & -1.00 & -0.84 & -0.88 & -0.84 & -0.82 & -0.83 & -1.14 & -1.11 & -1.15 & -1.02 & -1.02 \\
    $\lambda_{RMW}$ & 0.22  & 0.22  & 0.21  & 0.22  & 0.21  & 0.20  & 0.21  & 0.20  & 0.21  & 0.21  & 0.23  & 0.23  & 0.23  & 0.22  & 0.22 \\
    $t$-stat & 1.32  & 1.45  & 1.38  & 1.33  & 1.43  & 1.34  & 1.45  & 1.42  & 1.41  & 1.47  & 1.41  & 1.55  & 1.46  & 1.37  & 1.50 \\
    $\lambda_{CMA}$ & -0.19 & -0.19 & -0.19 & -0.18 & -0.17 & -0.19 & -0.18 & -0.19 & -0.18 & -0.18 & -0.19 & -0.19 & -0.19 & -0.18 & -0.17 \\
    $t$-stat & -1.42 & -1.41 & -1.42 & -1.31 & -1.24 & -1.46 & -1.35 & -1.45 & -1.42 & -1.36 & -1.39 & -1.38 & -1.39 & -1.30 & -1.28 \\
    $\lambda_{VRP}$ & 0.04  &       &       &       & -0.05 & 0.04  &       &       &       & -0.07 & 0.03  &       &       &       & -0.06 \\
    $t$-stat & 0.39  &       &       &       & -0.54 & 0.33  &       &       &       & -0.78 & 0.29  &       &       &       & -0.76 \\
    $\lambda_{MOM}$ &       & 0.03  &       &       & -0.15 &       & -0.15 &       &       & -0.24 &       & 0.06  &       &       & -0.08 \\
    $t$-stat &       & 0.05  &       &       & -0.28 &       & -0.32 &       &       & -0.49 &       & 0.12  &       &       & -0.15 \\
    $\lambda_{CIV}$ &       &       & -0.04 &       & 0.18  &       &       & -0.26 &       & 0.08  &       &       & 0.20  &       & 0.25 \\
    $t$-stat &       &       & -0.03 &       & 0.20  &       &       & -0.28 &       & 0.10  &       &       & 0.16  &       & 0.26 \\
    $\lambda_{LIQ}$ &       &       &       & -0.21 & -0.26 &       &       &       & -0.21 & -0.26 &       &       &       & -0.22 & -0.27 \\
    $t$-stat &       &       &       & -2.06 & -2.86 &       &       &       & -2.01 & -2.94 &       &       &       & -2.04 & -2.72 \\
    \midrule
    $ \bar{R}^2 $ & 0.492 & 0.492 & 0.492 & 0.606 & 0.624 & 0.434 & 0.441 & 0.434 & 0.552 & 0.580 & 0.511 & 0.510 & 0.511 & 0.623 & 0.634 \\
        \midrule
    \midrule
    \bottomrule
    \end{tabular}%
    }
  \label{tab:FMB_ALT2}%
\end{table}%

\newpage
\subsection{Additional Results Controlling for Market Fears}

\begin{table}[!hp]
  \centering
  \caption{\textbf{Fama-MacBeth Analysis; Alternative Test Assets: Size/Investment, Size, Book to Market, Size/Market Risk Premium: Accounting for Market Fears} \\
  \small{This table shows the Fama-MacBeth two-pass regression analysis for the 25 portfolios that sort on: i) size/investment (ME/INV); ii) size/book-to-market (ME/BM); iii) size/market risk premium (ME/MKT) and the 25 respective portfolios we construct on loadings to: common fears, $\beta^{\Delta\cf}$,  and market fears, $\beta^{\Delta\mf}$ ,in columns 1--5; common good fears, $\beta^{\Delta\cfg} $ and good market fears, $\beta^{\Delta\mfg}$, in columns 6--10; and common bad fears, $\beta^{\Delta\cfb} $, and bad market fears,  $\beta^{\Delta\mfb}$, in columns 11--15. $\lambda_{\cf}$ denotes the corresponding risk premium estimates for common fears, common good fears and common bad fears. $\lambda_mf$ denotes the risk premium estimates for market fears, good market fears, and bad market fears. All results control initially for the Fama-French 5 factors.  MKT is the market risk premium; SMB is small-minus-big; HML is high-minus-low; RMW is robust-minus-weak; CMA is conservative minus aggressive). Results in columns 1--4, 6--9, 11--14 use the Fama-French 5 factors plus: the variance risk premium (VRP); momentum (MOM), common idiosyncratic volatility (CIV) \citep{herskovic2016common}; and liquidity (LIQ) \citep{pastor2003liquidity}. Columns 5, 10, 15 add all aforementioned additional controls to the Fama-French 5 factor model. Below risk premia estimates we report Newey and West $t$-statistics with 12 lags that adjust for errors in variables as in \cite{shanken1992estimation}. $ \bar{R}^2 $ is the adjusted R-squared.}
  }
      \adjustbox{max height=9.70in ,max width=\textwidth, keepaspectratio}{
    \begin{tabular}{lrrrrrrrrrrrrrrr}
    \toprule
    \midrule
    \midrule
    & \multicolumn{15}{c}{ 25 ME/INV,25 ME/BM, 25 ME/MKT, 25 Plus:}\\
          & \multicolumn{5}{c}{ $\beta^{\Delta\cf} / \beta^{\Delta\mf}$ } & \multicolumn{5}{c}{$\beta^{\Delta\cfg} / \beta^{\Delta\mfg}$ } & \multicolumn{5}{c}{$\beta^{\Delta\cfb} / \beta^{\Delta\mfb}$ } \\
    \midrule
          & 1     & 2     & 3     & 4     & 5     & 6     & 7     & 8     & 9     & 10    & 11    & 12    & 13    & 14    & 15 \\
    \midrule
    $\lambda_{0}$ & 0.45  & 0.65  & 0.49  & 0.51  & 0.70  & 0.54  & 0.73  & 0.60  & 0.55  & 0.66  & 0.56  & 0.72  & 0.57  & 0.62  & 0.73 \\
    $t$-stat & 1.96  & 3.10  & 2.56  & 2.44  & 3.68  & 1.89  & 2.82  & 2.27  & 1.96  & 2.42  & 2.18  & 2.93  & 2.46  & 2.69  & 3.16 \\
    $\lambda_{\cf}$ & \textbf{-0.16} & -0.14 & \textbf{-0.15} & -0.13 & -0.11 & -0.14 & -0.13 & -0.13 & -0.11 & -0.11 & \textbf{-0.20} & \textbf{-0.18} & \textbf{-0.19} & \textbf{-0.19} & \textbf{-0.18} \\
    $t$-stat & \textbf{-2.07} & -1.88 & \textbf{-2.06} & -1.80 & -1.58 & -1.82 & -1.71 & -1.75 & -1.49 & -1.52 & \textbf{-2.66} & \textbf{-2.30} & \textbf{-2.51} & \textbf{-2.51} & \textbf{-2.37} \\
    $\lambda_{\mf}$ & 0.10  & 0.08  & 0.09  & 0.11  & 0.11  & 0.01  & 0.00  & 0.00  & 0.03  & 0.03  & 0.08  & 0.03  & 0.05  & 0.07  & 0.06 \\
    $t$-stat & 1.35  & 1.08  & 1.05  & 1.57  & 1.57  & 0.27  & 0.10  & 0.02  & 0.76  & 0.73  & 2.03  & 0.49  & 0.95  & 1.42  & 1.42 \\
    $\lambda_{MKT}$ & 0.60  & 0.39  & 0.56  & 0.53  & 0.33  & 0.51  & 0.30  & 0.44  & 0.49  & 0.36  & 0.47  & 0.30  & 0.46  & 0.39  & 0.28 \\
    $t$-stat & 1.64  & 1.18  & 1.54  & 1.35  & 1.00  & 1.50  & 0.95  & 1.33  & 1.42  & 1.20  & 1.37  & 0.98  & 1.38  & 1.12  & 0.94 \\
    $\lambda_{SMB}$ & 0.15  & 0.17  & 0.15  & 0.16  & 0.18  & 0.16  & 0.19  & 0.17  & 0.18  & 0.19  & 0.17  & 0.19  & 0.17  & 0.18  & 0.20 \\
    $t$-stat & 0.91  & 1.03  & 0.92  & 0.96  & 1.10  & 0.99  & 1.18  & 1.06  & 1.09  & 1.18  & 1.03  & 1.15  & 1.05  & 1.13  & 1.19 \\
    $\lambda_{HML}$ & -0.27 & -0.30 & -0.26 & -0.25 & -0.28 & -0.22 & -0.25 & -0.20 & -0.20 & -0.23 & -0.27 & -0.29 & -0.27 & -0.25 & -0.30 \\
    $t$-stat & -1.16 & -1.31 & -1.15 & -1.12 & -1.22 & -0.93 & -1.06 & -0.88 & -0.88 & -1.01 & -1.19 & -1.29 & -1.20 & -1.14 & -1.32 \\
    $\lambda_{RMW}$ & 0.25  & 0.40  & 0.27  & 0.23  & 0.36  & 0.24  & 0.40  & 0.26  & 0.28  & 0.40  & 0.20  & 0.38  & 0.24  & 0.22  & 0.33 \\
    $t$-stat & 1.10  & 2.00  & 1.18  & 1.03  & 1.84  & 1.16  & 1.92  & 1.27  & 1.30  & 1.90  & 0.89  & 1.95  & 1.02  & 0.93  & 1.78 \\
    $\lambda_{CMA}$ & -0.14 & -0.15 & -0.13 & -0.13 & -0.15 & -0.08 & -0.12 & -0.09 & -0.11 & -0.13 & -0.16 & -0.17 & -0.17 & -0.18 & -0.17 \\
    $t$-stat & -1.01 & -1.10 & -0.95 & -0.90 & -1.12 & -0.54 & -0.86 & -0.62 & -0.74 & -0.94 & -1.15 & -1.22 & -1.20 & -1.23 & -1.22 \\
    $\lambda_{VRP}$ & 0.05  &       &       &       & 0.01  & 0.13  &       &       &       & 0.09  & 0.12  &       &       &       & 0.11 \\
    $t$-stat & 0.43  &       &       &       & 0.14  & 1.13  &       &       &       & 0.92  & 1.08  &       &       &       & 1.30 \\
    $\lambda_{MOM}$ &       & -1.02 &       &       & -0.99 &       & -1.16 &       &       & -1.03 &       & -0.84 &       &       & -0.95 \\
    $t$-stat &       & -1.76 &       &       & -1.76 &       & -2.03 &       &       & -1.89 &       & -1.52 &       &       & -1.77 \\
    $\lambda_{CIV}$ &       &       & -0.82 &       & 0.74  &       &       & -1.07 &       & 0.27  &       &       & 0.08  &       & 0.97 \\
    $t$-stat &       &       & -0.76 &       & 0.81  &       &       & -1.36 &       & 0.35  &       &       & 0.09  &       & 1.02 \\
    $\lambda_{LIQ}$ &       &       &       & -0.19 & -0.25 &       &       &       & -0.23 & -0.22 &       &       &       & -0.10 & -0.11 \\
    $t$-stat &       &       &       & -1.70 & -2.73 &       &       &       & -2.25 & -2.53 &       &       &       & -1.05 & -1.35 \\
    \midrule
    $ \bar{R}^2 $ & 0.398 & 0.451 & 0.403 & 0.478 & 0.527 & 0.356 & 0.417 & 0.345 & 0.471 & 0.504 & 0.515 & 0.529 & 0.497 & 0.536 & 0.562 \\
        \midrule
    \midrule
    \bottomrule
    \end{tabular}%
    }
  \label{tab:FMB_ALT_CFMF}%
\end{table}%

\begin{table}[!hp]
  \centering
  \caption{\textbf{Fama-MacBeth Analysis; Alternative Test Assets: Size/Investment, Size,/Book to Market, Size/Market Risk Premium, Size/Operating Profit, Size/Residual Variance, Size/Total Variance: Accounting for Market Fears} \\
  \small{This table shows the Fama-MacBeth two-pass regression analysis for the 25 portfolios that sort on: i) size/investment (ME/INV); ii) size/book-to-market (ME/BM); iii) size/market risk premium (ME/MKT); iv) size/operating profit (ME/OP); v) size/residual variance (ME/RES VAR); and vi) size/total variance (ME/VAR), and the 25 respective portfolios we construct on loadings to: common fears, $\beta^{\Delta\cf}$,  and market fears, $\beta^{\Delta\mf}$ ,in columns 1--5; common good fears, $\beta^{\Delta\cfg} $ and good market fears, $\beta^{\Delta\mfg}$, in columns 6--10; and common bad fears, $\beta^{\Delta\cfb} $, and bad market fears,  $\beta^{\Delta\mfb}$, in columns 11--15. $\lambda_{\cf}$ denotes the corresponding risk premium estimates for common fears, common good fears and common bad fears. $\lambda_mf$ denotes the risk premium estimates for market fears, good market fears, and bad market fears. All results control initially for the Fama-French 5 factors.  MKT is the market risk premium; SMB is small-minus-big; HML is high-minus-low; RMW is robust-minus-weak; CMA is conservative minus aggressive). Results in columns 1--4, 6--9, 11--14 use the Fama-French 5 factors plus: the variance risk premium (VRP); momentum (MOM), common idiosyncratic volatility (CIV) \citep{herskovic2016common}; and liquidity (LIQ) \citep{pastor2003liquidity}. Columns 5, 10, 15 add all aforementioned additional controls to the Fama-French 5 factor model. Below risk premia estimates we report Newey and West $t$-statistics with 12 lags that adjust for errors in variables as in \cite{shanken1992estimation}. $ \bar{R}^2 $ is the adjusted R-squared.}
  }
      \adjustbox{max height=9.70in ,max width=\textwidth, keepaspectratio}{
    \begin{tabular}{lrrrrrrrrrrrrrrr}
    \toprule
    \midrule
    \midrule
    & \multicolumn{15}{c}{ 25 ME/INV, 25 ME/BM, 25 ME/MKT, 25 ME/OP 25 ME/RES VAR, 25 ME/VAR,  Plus:}\\
          & \multicolumn{5}{c}{ $\beta^{\Delta\cf} / \beta^{\Delta\mf}$ } & \multicolumn{5}{c}{$\beta^{\Delta\cfg} / \beta^{\Delta\mfg}$ } & \multicolumn{5}{c}{$\beta^{\Delta\cfb} / \beta^{\Delta\mfb}$ } \\
    \midrule
          & 1     & 2     & 3     & 4     & 5     & 6     & 7     & 8     & 9     & 10    & 11    & 12    & 13    & 14    & 15 \\
    \midrule
    $\lambda_{0}$ & 0.66  & 0.71  & 0.65  & 0.67  & 0.79  & 0.67  & 0.74  & 0.68  & 0.66  & 0.70  & 0.70  & 0.76  & 0.68  & 0.73  & 0.80 \\
    $t$-stat & 2.69  & 3.67  & 3.26  & 2.84  & 4.24  & 2.44  & 3.11  & 2.64  & 2.39  & 2.78  & 2.62  & 3.62  & 2.90  & 2.79  & 3.81 \\
    $\lambda_{\cf}$ & -0.13 & -0.13 & -0.13 & -0.11 & -0.09 & -0.13 & -0.14 & -0.14 & -0.12 & -0.11 & \textbf{-0.15} & \textbf{-0.14} & \textbf{-0.14} & \textbf{-0.15} & \textbf{-0.14} \\
    $t$-stat & -1.85 & -1.83 & -1.85 & -1.60 & -1.39 & -1.88 & -1.82 & -1.82 & -1.67 & -1.58 & \textbf{-2.11} & \textbf{-2.01} & \textbf{-2.02} & \textbf{-2.06} & \textbf{-2.05} \\
    $\lambda_{\mf}$ & 0.04  & 0.06  & 0.05  & 0.09  & 0.05  & -0.01 & -0.02 & -0.01 & 0.00  & 0.00  & 0.06  & 0.06  & 0.05  & 0.07  & 0.05 \\
    $t$-stat & 0.77  & 0.89  & 0.72  & 1.27  & 0.91  & -0.34 & -0.51 & -0.37 & 0.09  & 0.13  & 1.18  & 0.93  & 0.74  & 1.16  & 1.07 \\
    $\lambda_{MKT}$ & 0.37  & 0.32  & 0.38  & 0.36  & 0.23  & 0.36  & 0.28  & 0.35  & 0.36  & 0.32  & 0.32  & 0.25  & 0.34  & 0.28  & 0.21 \\
    $t$-stat & 0.91  & 0.89  & 0.97  & 0.85  & 0.69  & 0.95  & 0.86  & 1.01  & 0.97  & 1.05  & 0.78  & 0.79  & 0.91  & 0.68  & 0.68 \\
    $\lambda_{SMB}$ & 0.17  & 0.16  & 0.17  & 0.17  & 0.17  & 0.18  & 0.17  & 0.18  & 0.18  & 0.18  & 0.18  & 0.18  & 0.18  & 0.19  & 0.19 \\
    $t$-stat & 1.02  & 0.99  & 1.00  & 0.99  & 1.04  & 1.09  & 1.07  & 1.09  & 1.07  & 1.11  & 1.09  & 1.09  & 1.10  & 1.13  & 1.13 \\
    $\lambda_{HML}$ & -0.26 & -0.27 & -0.26 & -0.23 & -0.24 & -0.22 & -0.23 & -0.22 & -0.20 & -0.21 & -0.26 & -0.26 & -0.26 & -0.23 & -0.26 \\
    $t$-stat & -1.12 & -1.17 & -1.14 & -1.04 & -1.06 & -0.95 & -0.98 & -0.95 & -0.88 & -0.92 & -1.13 & -1.15 & -1.15 & -1.04 & -1.13 \\
    $\lambda_{RMW}$ & 0.21  & 0.22  & 0.21  & 0.20  & 0.21  & 0.21  & 0.22  & 0.21  & 0.23  & 0.22  & 0.19  & 0.21  & 0.19  & 0.18  & 0.18 \\
    $t$-stat & 1.33  & 1.40  & 1.34  & 1.24  & 1.41  & 1.40  & 1.54  & 1.49  & 1.52  & 1.53  & 1.24  & 1.34  & 1.21  & 1.14  & 1.28 \\
    $\lambda_{CMA}$ & -0.17 & -0.18 & -0.17 & -0.16 & -0.16 & -0.13 & -0.13 & -0.13 & -0.14 & -0.14 & -0.20 & -0.20 & -0.20 & -0.20 & -0.20 \\
    $t$-stat & -1.31 & -1.33 & -1.30 & -1.23 & -1.20 & -0.99 & -0.97 & -0.99 & -1.06 & -1.09 & -1.48 & -1.48 & -1.48 & -1.49 & -1.46 \\
    $\lambda_{VRP}$ & 0.00  &       &       &       & -0.08 & 0.04  &       &       &       & -0.05 & 0.03  &       &       &       & -0.04 \\
    $t$-stat & -0.01 &       &       &       & -1.00 & 0.34  &       &       &       & -0.51 & 0.37  &       &       &       & -0.44 \\
    $\lambda_{MOM}$ &       & -0.11 &       &       & -0.25 &       & -0.17 &       &       & -0.18 &       & -0.09 &       &       & -0.22 \\
    $t$-stat &       & -0.21 &       &       & -0.49 &       & -0.35 &       &       & -0.36 &       & -0.18 &       &       & -0.42 \\
    $\lambda_{CIV}$ &       &       & -0.14 &       & 0.44  &       &       & -0.19 &       & 0.30  &       &       & 0.53  &       & 0.94 \\
    $t$-stat &       &       & -0.12 &       & 0.47  &       &       & -0.19 &       & 0.37  &       &       & 0.44  &       & 0.95 \\
    $\lambda_{LIQ}$ &       &       &       & -0.21 & -0.27 &       &       &       & -0.21 & -0.26 &       &       &       & -0.16 & -0.19 \\
    $t$-stat &       &       &       & -2.21 & -3.06 &       &       &       & -2.17 & -2.97 &       &       &       & -1.82 & -2.43 \\
    \midrule
    $ \bar{R}^2 $ & 0.455 & 0.458 & 0.453 & 0.555 & 0.593 & 0.433 & 0.442 & 0.433 & 0.547 & 0.569 & 0.516 & 0.518 & 0.517 & 0.588 & 0.598 \\
        \midrule
    \midrule
    \bottomrule
    \end{tabular}%
    }
  \label{tab:FMB_ALT_CFMF2}%
\end{table}%

\bibliographystyle{elsarticle-harv}
\bibliography{BBE_CEV}

\begin{thebibliography}{44}
\expandafter\ifx\csname natexlab\endcsname\relax\def\natexlab#1{#1}\fi
\expandafter\ifx\csname url\endcsname\relax
  \def\url#1{\texttt{#1}}\fi
\expandafter\ifx\csname urlprefix\endcsname\relax\def\urlprefix{URL }\fi

\bibitem[{Acemoglu et~al.(2012)Acemoglu, Carvalho, Ozdaglar, and
  Tahbaz-Salehi}]{acemoglu2012network}
Acemoglu, D., Carvalho, V.~M., Ozdaglar, A., Tahbaz-Salehi, A., 2012. The
  network origins of aggregate fluctuations. Econometrica 80~(5), 1977--2016.

\bibitem[{An et~al.(2014)An, Ang, Bali, and Cakici}]{an2014joint}
An, B.-J., Ang, A., Bali, T.~G., Cakici, N., 2014. The joint cross section of
  stocks and options. Journal of Finance 69~(5), 2279--2337.

\bibitem[{Ang et~al.(2006{\natexlab{a}})Ang, Chen, and Xing}]{ang2006downside}
Ang, A., Chen, J., Xing, Y., 2006{\natexlab{a}}. Downside risk. Review of
  Financial Studies 19~(4), 1191--1239.

\bibitem[{Ang et~al.(2006{\natexlab{b}})Ang, Hodrick, Xing, and
  Zhang}]{ang2006cross}
Ang, A., Hodrick, R.~J., Xing, Y., Zhang, X., 2006{\natexlab{b}}. The
  cross-section of volatility and expected returns. Journal of Finance 61~(1),
  259--299.

\bibitem[{Bakshi et~al.(1997)Bakshi, Cao, and Chen}]{bakshi1997}
Bakshi, G., Cao, C., Chen, Z., 1997. Empirical performance of alternative
  option pricing models. Journal of Finance 52~(5), 2003--2049.

\bibitem[{Bakshi et~al.(2003)Bakshi, Kapadia, and Madan}]{Bakshi2003}
Bakshi, G., Kapadia, N., Madan, D., 2003. Stock return characteristics, skew
  laws, and the differential pricing of individual equity options. Review of
  Financial Studies 16~(1), 101--143.

\bibitem[{Bakshi and Madan(2000)}]{bakshi2000spanning}
Bakshi, G., Madan, D., 2000. Spanning and derivative-security valuation.
  Journal of Financial Economics 55~(2), 205--238.

\bibitem[{Bali and Hovakimian(2009)}]{bali2009volatility}
Bali, T.~G., Hovakimian, A., 2009. Volatility spreads and expected stock
  returns. Management Science 55~(11), 1797--1812.

\bibitem[{Barun{\'\i}k et~al.(2022)Barun{\'\i}k, Bevilacqua, and
  Tunaru}]{barunik2022}
Barun{\'\i}k, J., Bevilacqua, M., Tunaru, R., 2022. Asymmetric network
  connectedness of fears. Review of Economics and Statistics 104~(6),
  1304--1316.

\bibitem[{Bloom(2009)}]{bloom2009impact}
Bloom, N., 2009. The impact of uncertainty shocks. econometrica 77~(3),
  623--685.

\bibitem[{Bollerslev et~al.(2020)Bollerslev, Li, and Zhao}]{bollerslev2020good}
Bollerslev, T., Li, S.~Z., Zhao, B., 2020. Good volatility, bad volatility, and
  the cross section of stock returns. Journal of Financial and Quantitative
  Analysis 55~(3), 751--781.

\bibitem[{Breckenfelder and T{\'e}dongap(2012)}]{breckenfelder2012asymmetry}
Breckenfelder, J., T{\'e}dongap, R., 2012. Asymmetry matters: A high-frequency
  risk-reward trade-off. SSRN 1828283.

\bibitem[{Buraschi and Jiltsov(2006)}]{buraschi2006}
Buraschi, A., Jiltsov, A., 2006. Model uncertainty and option markets with
  heterogeneous beliefs. Journal of Finance 61~(6), 2841--2897.

\bibitem[{Campbell(1993)}]{campbell1993intertemporal}
Campbell, J.~Y., 1993. Intertemporal asset pricing without consumption data.
  American Economic Review 83~(3), 487.

\bibitem[{Campbell(1996)}]{campbell1996understanding}
Campbell, J.~Y., 1996. Understanding risk and return. Journal of Political
  Economy 104~(2), 298--345.

\bibitem[{Campbell et~al.(2018)Campbell, Giglio, Polk, and
  Turley}]{campbell2018intertemporal}
Campbell, J.~Y., Giglio, S., Polk, C., Turley, R., 2018. An intertemporal
  {CAPM} with stochastic volatility. Journal of Financial Economics 128~(2),
  207--233.

\bibitem[{Campbell et~al.(2001)Campbell, Lettau, Malkiel, and
  Xu}]{campbell2001have}
Campbell, J.~Y., Lettau, M., Malkiel, B.~G., Xu, Y., 2001. Have individual
  stocks become more volatile? an empirical exploration of idiosyncratic risk.
  Journal of Finance 56~(1), 1--43.

\bibitem[{Carr and Wu(2009)}]{carr2009variance}
Carr, P., Wu, L., 2009. Variance risk premiums. Review of Financial Studies
  22~(3), 1311--1341.

\bibitem[{Carr and Wu(2011)}]{carr2011}
Carr, P., Wu, L., 2011. A simple robust link between {A}merican puts and credit
  protection. Review of Financial Studies 24~(2), 473--505.

\bibitem[{Christoffersen et~al.(2018)Christoffersen, Fournier, and
  Jacobs}]{christoffersen2018factor}
Christoffersen, P., Fournier, M., Jacobs, K., 2018. The factor structure in
  equity options. Review of Financial Studies 31~(2), 595--637.

\bibitem[{Christoffersen et~al.(2012)Christoffersen, Jacobs, and
  Ornthanalai}]{christoffersen2012}
Christoffersen, P., Jacobs, K., Ornthanalai, C., 2012. Dynamic jump intensities
  and risk premiums: Evidence from {$S\&P500$} returns and options. Journal of
  Financial Economics 106~(3), 447--472.

\bibitem[{Cremers et~al.(2015)Cremers, Halling, and
  Weinbaum}]{cremers2015aggregate}
Cremers, M., Halling, M., Weinbaum, D., 2015. Aggregate jump and volatility
  risk in the cross-section of stock returns. Journal of Finance 70~(2),
  577--614.

\bibitem[{Cremers and Weinbaum(2010)}]{cremers2010deviations}
Cremers, M., Weinbaum, D., 2010. Deviations from put-call parity and stock
  return predictability. Journal of Financial and Quantitative Analysis 45~(2),
  335--367.

\bibitem[{Dew-Becker and Giglio(2023)}]{dew2023cross}
Dew-Becker, I., Giglio, S., 2023. Cross-sectional uncertainty and the business
  cycle: Evidence from 40 years of options data. American Economic Journal:
  Macroeconomics 15~(2), 65--96.

\bibitem[{Engle and Figlewski(2015)}]{engle2015modeling}
Engle, R., Figlewski, S., 2015. Modeling the dynamics of correlations among
  implied volatilities. Review of Finance 19~(3), 991--1018.

\bibitem[{Fama and French(2015)}]{fama2015five}
Fama, E.~F., French, K.~R., 2015. A five-factor asset pricing model. Journal of
  Financial Economics 116~(1), 1--22.

\bibitem[{Farago and T{\'e}dongap(2018)}]{farago2018downside}
Farago, A., T{\'e}dongap, R., 2018. Downside risks and the cross-section of
  asset returns. Journal of Financial Economics 129~(1), 69--86.

\bibitem[{Gabaix(2011)}]{gabaix2011granular}
Gabaix, X., 2011. The granular origins of aggregate fluctuations. Econometrica
  79~(3), 733--772.

\bibitem[{Giglio and Xiu(2021)}]{giglio2021asset}
Giglio, S., Xiu, D., 2021. Asset pricing with omitted factors. Journal of
  Political Economy 129~(7), 1947--1990.

\bibitem[{Han(2008)}]{Han2008}
Han, B., 2008. Investor sentiment and option prices. Review of Financial
  Studies 21~(1), 387--414.

\bibitem[{Herskovic et~al.(2016)Herskovic, Kelly, Lustig, and
  Van~Nieuwerburgh}]{herskovic2016common}
Herskovic, B., Kelly, B., Lustig, H., Van~Nieuwerburgh, S., 2016. The common
  factor in idiosyncratic volatility: Quantitative asset pricing implications.
  Journal of Financial Economics 119~(2), 249--283.

\bibitem[{Herskovic et~al.(2020)Herskovic, Kelly, Lustig, and
  Van~Nieuwerburgh}]{herskovic2020firm}
Herskovic, B., Kelly, B., Lustig, H., Van~Nieuwerburgh, S., 2020. Firm
  volatility in granular networks. Journal of Political Economy 128~(11),
  4097--4162.

\bibitem[{Jegadeesh and Titman(1993)}]{jegadeesh1993returns}
Jegadeesh, N., Titman, S., 1993. Returns to buying winners and selling losers:
  Implications for stock market efficiency. Journal of finance 48~(1), 65--91.

\bibitem[{Kahneman and Tversky(2013)}]{kahneman2013prospect}
Kahneman, D., Tversky, A., 2013. Prospect theory: An analysis of decision under
  risk. In: Handbook of the fundamentals of financial decision making: Part I.
  World Scientific, pp. 99--127.

\bibitem[{Kilic and Shaliastovich(2019)}]{kilic2019good}
Kilic, M., Shaliastovich, I., 2019. Good and bad variance premia and expected
  returns. Management Science 65~(6), 2522--2544.

\bibitem[{Martin and Wagner(2019)}]{martin2019expected}
Martin, I.~W., Wagner, C., 2019. What is the expected return on a stock?
  Journal of Finance 74~(4), 1887--1929.

\bibitem[{McCracken and Ng(2016)}]{mccracken2016fred}
McCracken, M.~W., Ng, S., 2016. {FRED-MD}: A monthly database for macroeconomic
  research. Journal of Business \& Economic Statistics 34~(4), 574--589.

\bibitem[{Merton(1973)}]{merton1973intertemporal}
Merton, R.~C., 1973. An intertemporal capital asset pricing model.
  Econometrica, 867--887.

\bibitem[{Muravyev et~al.(2022)Muravyev, Pearson, and
  Pollet}]{muravyev2022there}
Muravyev, D., Pearson, N.~D., Pollet, J.~M., 2022. Is there a risk premium in
  the stock lending market? evidence from equity options. Journal of Finance
  77~(3), 1787--1828.

\bibitem[{P{\'a}stor and Stambaugh(2003)}]{pastor2003liquidity}
P{\'a}stor, L., Stambaugh, R.~F., 2003. Liquidity risk and expected stock
  returns. Journal of Political Economy 111~(3), 642--685.

\bibitem[{Shanken(1992)}]{shanken1992estimation}
Shanken, J., 1992. On the estimation of beta-pricing models. Review of
  Financial Studies 5~(1), 1--33.

\bibitem[{Stock and Watson(2002{\natexlab{a}})}]{stock2002forecasting}
Stock, J.~H., Watson, M.~W., 2002{\natexlab{a}}. Forecasting using principal
  components from a large number of predictors. Journal of the American
  Statistical Association 97~(460), 1167--1179.

\bibitem[{Stock and Watson(2002{\natexlab{b}})}]{stock2002macroeconomic}
Stock, J.~H., Watson, M.~W., 2002{\natexlab{b}}. Macroeconomic forecasting
  using diffusion indexes. Journal of Business \& Economic Statistics 20~(2),
  147--162.

\bibitem[{Xing et~al.(2010)Xing, Zhang, and Zhao}]{xing2010does}
Xing, Y., Zhang, X., Zhao, R., 2010. What does the individual option volatility
  smirk tell us about future equity returns? Journal of Financial and
  Quantitative Analysis 45~(3), 641--662.

\end{thebibliography}

\end{document}